\documentclass[12pt]{article}
\usepackage{amsmath}
\usepackage{graphicx}
\usepackage{enumerate}
\usepackage{natbib}
\usepackage{url} % not crucial - just used below for the URL

\usepackage{bm}
\usepackage{amsfonts}
\usepackage{changes}
\usepackage{multirow}
\usepackage{subcaption}

\usepackage{color}

\newcommand{\blind}{0}

\newtheorem{theorem}{Theorem}

\newtheorem{lemma}{Lemma}%[section]
\newtheorem{condition}[theorem]{Condition}
\newtheorem{assumption}[theorem]{Assumption}
\newtheorem{proposition}[theorem]{Proposition}

\begin{document}

\def\spacingset#1{\renewcommand{\baselinestretch}%
{#1}\small\normalsize} \spacingset{1}

%%%%%%%%%%%%%%%%%%%%%%%%%%%%%%%%%%%%%%%%%%%%%%%%%%%%%%%%%%%%%%%%%%%%%%%%%%%%%%

\if0\blind
{
\title{A Two-stage Inference Procedure for Sample Local Average Treatment Effects in Randomized Experiments}
\author{Zhen Zhong\\
		Faculty of Business and Economics, The University of Hong Kong\\
		\\
		Per Johansson\\
		Department of Statistics, Uppsala
		University\\
		and Yau Mathematical Science Center,
		Tsinghua University\\
		\\
		Junni L. Zhang*\\
		National School of Development and Center for Statistical Science,
		\\Peking University}
  \maketitle
} \fi

\if1\blind
{
  \bigskip
  \bigskip
  \bigskip
  \begin{center}
    {\LARGE\bf A Two-stage Inference Procedure for Sample Local Average Treatment Effects in Randomized Experiments}
\end{center}
  \medskip
} \fi

\begin{abstract}
In a given randomized experiment, individuals are often volunteers and can differ in important ways from a population of interest. It is thus of interest to focus on the sample at hand. This paper focuses on inference about the sample local average treatment effect (LATE) in randomized experiments with non-compliance. We present a two-stage procedure that provides asymptotically correct coverage rate of the sample LATE in randomized experiments. The procedure uses a first-stage test to decide whether the instrument is strong or weak, and uses different confidence sets depending on the first-stage result. Proofs of the procedure is developed for the situation with and without regression adjustment and for two experimental designs (complete randomization and Mahalaonobis distance based rerandomization). Finite sample performance of the methods are studied using extensive Monte Carlo simulations and the methods are applied to data from a voter encouragement experiment.
\end{abstract}

\noindent%

\vfill

\newpage
\spacingset{1.9} % DON'T change the spacing!

\section{Introduction}
One requirement for the inference to an average treatment effect from a randomized experiment to be valid is that all experimental units comply to their treatment assignments. In practice, incomplete compliance to the assigned treatment is common.

One standard approach is to ignore the information on compliance behavior and to focus on the intention-to-treat (ITT) analysis. It is argued that the ITT-effect may be most policy-relevant, since one cannot in general force people to take a treatment. However, the ITT-effect may be misleading, for example, when a treatment with negative side effects may appear less effective simply because subjects adhere to it to a greater extent. The international guidelines for good clinical practice \citep{dixon1999international}, for example, suggest that the analysis of negative side effects should be according to treatment received.  Yet the ``as-treated'' analysis that compares those who receive treatment with those who receive control generally yields a biased estimate of the treatment effect, because the treatment and control groups are no longer similar.

Many scholars \citep[e.g.][]{McNamee_2009,Shrier_etal_2014,Shrier_etal_2017,Steele_etal_2015} argue that, when information on treatment received is available, the average treatment effect for those who comply to the assigned treatment, i.e., the complier average causal effect (CACE) or the local average treatment effect (LATE), should to be the main causal estimand.  In this literature the experimental units are most often regarded as being independently and randomly sampled from a super-population, and inference is made about the population LATE \citep[cf.][]{imbens_angrist_1994,Angrist_etal_1996}.  However, in general individuals in the experiments are volunteers which means that there is a risk that they differ in important ways from a population of interest. It is thus of interest to focus on the sample at hand and have easy to use and correct procedures for inference about the sample LATE.

There is a small but growing literature on the sample LATE  \citep[see e.g.][]{li2017general, kang2018inference, rambachan2020design, hong2020inference}. \cite{hong2020inference} show that with strong instruments (that is, in this setting, a large enough effect of treatment assignment on treatment received) the Wald estimator and the delta method for the estimation of standard errors provide correct inference to the sample LATE. A major concern with the standard Wald estimator is its poor performance when the  instrument instead is weak \citep{kang2018inference}. Under the assumption of homogeneous complier treatment effects, \cite{li2017general} provides AR-like \citep{anderson1949estimation} confidence set for the sample LATE, which remains valid, also, when the instrument is weak. 

We first consider complete randomization (CRE) without regression adjustment. We show the result in \cite{hong2020inference} under weaker assumptions. We derive the AR-like set with heterogeneous complier treatment effects, under weaker assumptions than in \cite{kang2018inference}. We then propose a two-stage procedure that provides asymptotically correct coverage rate of the sample LATE that in general yield substantially shorter interval than the AR-like confidence set. The procedure uses the Wald estimator with standard errors calculated using the delta method when a first-stage test concludes that the instrument is strong, and uses the AR-like confidence set when the first-stage test concludes that the instrument is weak.   We extend the procedure to another experimental design --- Mahalaonobis distance based rerandomization (ReM), and also to the situation with regression adjustment under either experimental design.

The next section defines the sample LATE. Section 3 presents the two experimental designs. Section 4 discusses inference under CRE without regression adjustment, Section 5 discusses inference under ReM without regression adjustment, and Section 6 discusses inference with regression adjustment. Section 7 presents a Monte Carlo study that compares the small
sample performance of different methods. Section 8 presents an application to a voter encouragement experiment. Section 9 concludes.

\section{The Sample LATE}\label{sec:SLATE}

Consider an experiment with $n$ units. For unit $i$, let $Z_i\in\{0,1\}$ denote the indicator of treatment assigned.  Let $\bm{Z}$ denote the vector of $Z_i$. We assume Stable Unit Treatment Value Assumption (SUTVA) \citep{Rubin_1978}, that is, there is only one version of treatment or control and there is no interference between units. For unit $i$, let $W_{i}(z)$, $z=0,1$, be the potential treatment received if assigned to treatment $z$. Let $\bm{W}(z)$, $z=0,1$, be the vector of $W_i(z)$. For unit $i$, define the four potential outcomes $Y_{i}(z,w)$ for which the treatment assigned and treatment received are fixed at $z=0,1$ and $w=0,1$. Only two of these potential outcomes, $Y_i(z)=Y_{i}(z,W_{i}(z))$, $z=0,1$, can possibly be observed.  Let $\bm{Y}(z)$, $z=0,1$, be the vector of $Y_i(z)$. We further assume the exclusion restriction assumption \citep{Angrist_etal_1996} that potential outcomes only depend on treatment received, with $Y_{i}(z,w)=Y_{i}(z^{\prime },w)$ for all $z,z^{\prime }$ and $w$.

There are four possible latent types of units: always-takers (\textit{at}) who would take treatment regardless of whether being assigned to treatment or control, with $W_i(1)=W_i(0)=1$; compliers (\textit{co}) who would take treatment when assigned to treatment and take control when assigned to control, with $W_i(1)=1$ and $W_i(0)=0$, or equivalently $W_i(1)-W_i(0)=1$; defiers (\textit{de}) who would take control when assigned to treatment and take treatment when assigned to control, with $W_i(1)=0$ and $W_i(0)=1$, or equivalently $W_i(1)-W_i(0)=-1$; and never-takers (\textit{nt}) who would take control regardless of whether being assigned to treatment or control, with $W_i(1)=W_i(0)=0$.

We focus on the finite population consisting of the $n$ experimental units. Hence $\bm{W}(0)$, $\bm{W}(1)$, $\bm{Y}(0)$ and $\bm{Y}(1)$ are treated as fixed, and only $\bm{Z}$ is treated as random. We make the following assumption:
\begin{assumption}\label{assu}
	(i) $W_i(1)\neq W_i(0)$ for some unit $i$; (ii) $W_{i}(1)\geq W_{i}(0)$. 
\end{assumption} 
Under Assumption \ref{assu}, there is at least one complier and there is no defier. Let $G_i$ denote the latent type for unit $i$.  Since $W_i(1)$ and $W_i(0)$ are fixed, $G_i$ is fixed.  The sample LATE is defined as
\begin{equation*}
	\tau=\frac{1}{n_{co}}\sum_{G_i=co}(Y_{i}(1)-Y_{i}(0)),
	\label{samp_LATE}
\end{equation*}
where $n_{co}=\sum_{i=1}^n (W_i(1)-W_i(0))$ is the number of compliers. 

The fraction of compliers is $\tau_W=n_{co}/n$, which is also the sample average effect of treatment assigned on $W$. Since potential outcomes only depend on treatment received, for always-takers and never-takers $Y_i(1)-Y_i(0)=0$.  Therefore, $\sum_{G_i=co}(Y_{i}(1)-Y_{i}(0))=\sum_{i=1}^n (Y_{i}(1)-Y_{i}(0))$. Define the sample average effect of treatment assigned on $Y$ as
\begin{equation*}
	\tau_{Y}=\frac{1}{n}\sum_{i=1}^{n}(Y_{i}(1)-Y_{i}(0)).
	\label{sample_ITT_Y}
\end{equation*}
We can get another form of the sample LATE:
\begin{equation}
	\tau=\frac{\tau_{Y}}{\tau_{W}}.
	\label{samp_LATE2}
\end{equation}
Assumption \ref{assu} implies that $\tau_W>0$, and therefore $\tau$ is finite.

\section{Experimental Designs}\label{sec:design}

Under complete randomization (CRE), $n_1$ out of $n$ units are randomly assigned to treatment and the remaining $n_0$ units are assigned to control. In a single experiment, an unfortunate treatment assignment may lead to unbalanced covariates between the treated and control groups. \cite{morgan2012rerandomization} proposed the Mahalanobis-distance based rerandomization (ReM) to improve covariate balance. 

For unit $i$, let $\boldsymbol{x}_i$ be the $K\times 1$ vector of covariates. The covariates are treated as fixed, and have been centered around zero, with $\overline{\boldsymbol{x}}\equiv 1/n\sum_{i=1}^n \boldsymbol{x}_i=\boldsymbol{0}$. Let $\boldsymbol{X}$ be the $n\times K$ matrix of covariates for all $n$ units. The finite population covariance matrix of the covariates is
\begin{equation*}
	\mathbb{S}_{\boldsymbol{x}\boldsymbol{x}}=\frac{1}{n-1}\sum_{i=1}^{n}\boldsymbol{x}_{i}
	\boldsymbol{x}_{i}^{\top}. \label{eq:Cov}
\end{equation*}

For any given treatment assignment vector $\boldsymbol{Z}$, the Mahalanobis distance is
\begin{equation*} M(\boldsymbol{Z},\boldsymbol{X)}=\frac{n_1n_0}{n}\left(\widehat{\tau}_{\boldsymbol{x}}^{\top}\mathbb{S}_{\boldsymbol{x}\boldsymbol{x}}^{-1}\widehat{\tau}_{\boldsymbol{x}}\right) \label{eq:MD}
\end{equation*}%
\noindent where
\begin{equation*}
	\widehat{\tau}_{\boldsymbol{x}}=\frac{1}{n_{1}}\sum_{i=1}^{n}Z_{i}%
	\boldsymbol{x}_{i}-\frac{1}{n_{0}}\sum_{i=1}^{n}(1-Z_{i})\boldsymbol{x}_{i}=%
	\overline{\boldsymbol{x}}_{1}-\overline{\boldsymbol{x}}_{0},
\end{equation*}
with $\overline{\boldsymbol{x}}_z$ being the mean of $\boldsymbol{x}_i$ for units with $Z_i=z$ ($z=0,1$). \cite{morgan2012rerandomization} suggested keeping randomly generating $\bm{Z}$ until 
\begin{equation*}
	M(\boldsymbol{Z},\boldsymbol{X)}\leq a,
\end{equation*}%
where $a$ is a positive constant. Asymptotically, the Mahalanobis distance follows a chi-square distribution with $K$ degrees of freedom, $\chi_K^2$.  Hence $a$ can be indirectly determined by setting $p_a=\Pr(\chi_K^2\leq a)$.  For example, with $p_a=0.01$, $a$ equals to the 0.01 quantile of a $\chi_K^2$ distribution. Note that when $a=\infty$, ReM reduces to CRE.

\section{Inference under CRE without Regression Adjustment}
\label{sec:CRE_noncompliance}

We introduce some notations for later use. For any set of quantities $\{Q_i:i=1,\cdots,n\}$, let $\overline{Q}=\frac{1}{n}\sum_{i=1}^{n}Q_i$ denote the mean for all $n$ units, let $\overline{Q}_z=\frac{1}{n_{z}}\sum_{i:\ Z_{i}=z}Q_{i}$ denote the mean for units assigned to treatment arm $z$, and let $\widehat{\tau}_Q = \overline{Q}_1 -\overline{Q}_0$ denote the difference-in-means. Define the finite population variance as $\mathbb{S}_{Q}^2=\frac{1}{n-1}\sum_{i=1}^{n}\left(Q_i-\overline{Q}\right)^2$, and the sample variance for units assigned to treatment arm $z$ as $S_{Q(z)}^2=\frac{1}{n_z-1}\sum_{i:Z_i=z}\left(Q_i-\overline{Q}_z\right)^2$. With another set of quantities $\{Q_i':i=1,\cdots,n\}$, define the sample covariance between $Q_i$ and $Q_i'$ for units assigned to treatment arm $z$ as $S_{Q(z),Q'(z)}=\frac{1}{n_z-1}\sum_{i:Z_i=z}\left(Q_i-\overline{Q}_z\right)\left(Q_i^{\prime}-\overline{Q^{\prime}}_z\right)$.

\subsection{Exiting Inference Procedures}

\subsubsection{The Wald Method}

For unit $i$, let the observed indicator of treatment received be $W_{i}=W_{i}(Z_{i})$, and let the observed outcome be $Y_{i}=Y_{i}(Z_{i})$.  Let $\widehat{\tau}_W=
\overline{W}_1 -\overline{W}_0$ and $\widehat{\tau}_Y=
\overline{Y}_1 -\overline{Y}_0$ denote the difference-in-means for $W_i$ and $Y_i$. The Wald estimator is defined as 
\begin{equation}
	\widehat{\tau}_{Wald}=\frac{\widehat{\tau}_{Y}}{\widehat{\tau}_{W}}=\frac{\overline{Y}_{1}-\overline{Y}_{0}}{\overline{W}_{1}-\overline{W}_{0}}.  \label{est_Wald}
\end{equation}

When the units are regarded as being independently and randomly sampled from a super population and treatment assignment is completely random, the Wald estimator is an asymptotically unbiased and normally distributed estimator for the population LATE, and the delta method can be used to estimate its asymptotic variance (e.g. \citealp{imbens2015causal}, Chapter 23). \cite{hong2020inference} obtained similar results when the estimand is the sample LATE.\footnote{\cite{hong2020inference} discussed stratified block randomization. When the number of strata is reduced to one, stratified block randomization reduces to CRE.} We next derive the asymptotic properties of the Wald method under a weaker set of assumptions than \cite{hong2020inference}.

Following \cite{li2017general} and \cite{ding2023first}, define the adjusted outcome $A_i=Y_i-\tau W_i$ with potential outcomes $A_i(z)=Y_i(z)-\tau W_i(z)$ for $z=0,1$. Under Assumption \ref{assu}, $\tau$ is finite, and therefore these outcomes are well defined. Let $\overline{A}(z)$ denote the mean of $A(z)$ for all $n$ units. Define $\tau_A=\frac{1}{n}\sum_{i=1}^n (A_i(1)-A_i(0))=\tau_Y-\tau \tau_W$. Applying~\eqref{samp_LATE2}, we have $\tau_A=0$. Let 
\begin{equation}
	\label{def_tauhat_A}
	\widehat{\tau}_{A}=\overline{A}_1-\overline{A}_0=\widehat{\tau}_Y-\tau\widehat{\tau}_W.
\end{equation}
Let $\mathbb{S}_{A(1)}^2$ and $\mathbb{S}_{A(0)}^2$ be the finite population variances for $A_i(1)$ and $A_i(0)$. Let $\mathbb{S}_{A(1)-A(0)}^2$ be the finite population variance for $A_i(1)-A_i(0)$. Define 
\begin{equation}
	\mathbb{V}_A=\frac{\mathbb{S}_{A(1)}^2}{n_1}+\frac{\mathbb{S}_{A(0)}^2}{n_0}-\frac{\mathbb{S}_{A(1)-A(0)}^2}{n}.\label{def_bV_A}
\end{equation}

We introduce a set of regularity conditions.
\setcounter{theorem}{0}
\begin{condition}
	\label{strength}
	$\liminf_{n\rightarrow\infty}\tau_W>0$.
\end{condition}
\begin{condition}
	\label{lindeberg-feller}
	As $n\rightarrow\infty$, (1) the proportion of units under treatment, $n_1/n$, has a limit in $(0,1)$; (2) $\limsup_{n\rightarrow\infty}n\mathbb{V}_A<\infty$; (3) for $z=0,1$,
	\begin{equation}
		\frac{\max_{1 \leq i \leq n}\left(A_{i}(z)-\overline{A}(z)\right)^2}{\min(n_1, n_0)^2\cdot \mathbb{V}_A}\rightarrow0.\label{eq:lindeberg-feller}
	\end{equation}
\end{condition}

Condition~\ref{strength} is the same as Condition (b) of Proposition 3.1 from \cite{hong2020inference}, and requires the existence of compliers. Condition~\ref{lindeberg-feller} is an extension of the conditions in \cite{li2017general}, and ensures the asymptotic normality of $\overline{A}_1$ and $\overline{A}_0$. Part (3) of Condition~\ref{lindeberg-feller} is weaker than assumption (2.1) of Theorem 2.1 from \cite{hong2020inference} that involves moments higher than order two.

Under Assumption \ref{assu} and Condition \ref{lindeberg-feller}, applying Theorem 1 of \cite{li2017general}, $\widehat{\tau}_A$ converges to a normal distribution with mean $\tau_A=0$ and variance $\mathbb{V}_A$. The next theorem says that, under Assumption \ref{assu}, Conditions \ref{strength}, \ref{lindeberg-feller} and CRE, the Wald estimator is an asymptotically unbiased and normally distributed estimator for $\tau$.

\setcounter{theorem}{0}
\begin{theorem}
	\label{Wald_CRE}
	Under Assumption \ref{assu}, Conditions \ref{strength}, \ref{lindeberg-feller} and CRE, as $n\rightarrow\infty$,
	\begin{equation*}
		\widehat{\tau}_{Wald}-\tau=o_p(1)\quad \text{and}\quad \frac{\widehat{\tau}_{Wald}-\tau}{\sqrt{\mathbb{V}_A}/\tau_W}\stackrel{d}{\rightarrow}N(0,1)
	\end{equation*}
\end{theorem}

Proofs of all theorems and propositions are given in Section C of the Supplementary Material.

To construct an asymptotically conservative confidence interval for $\tau$, we can use $\widehat{\tau}_W$ to estimate $\tau_W$, and use $\widehat{V}_A=S_{\widehat{A}(1)}^2/n_1+S_{\widehat{A}(0)}^2/n_0$ to estimate $\mathbb{V}_A$, where $S_{\widehat{A}(z)}^2$ is the sample variance of $\widehat{A}_i=Y_i-\widehat{\tau}_{Wald}W_i$ for units with $Z_i=z$ ($z=0,1$). Define $V_Y=S_{Y(1)}^2/n_1+S_{Y(0)}^2/n_0$, $V_W=S_{W(1)}^2/n_1+S_{W(0)}^2/n_0$, and $C_{YW}=S_{Y(1),W(1)}/n_1+S_{Y(0),W(0)}/n_0$. It is easy to see that $\widehat{V}_A=V_Y-2\widehat{\tau}_{Wald}C_{YW}+\widehat{\tau}^2_{Wald}V_W$. Let $z_{\alpha/2}$ be the $(1-\alpha/2)$th quantile of the standard normal distribution.

\begin{theorem}
	\label{conservative_ci_CRE}
	Under Assumption \ref{assu}, Conditions \ref{strength}, \ref{lindeberg-feller} and CRE, the confidence interval
	\begin{equation}
		\label{ci_CRE}
		\widehat{\tau}_{Wald}\pm z_{\alpha/2}\sqrt{V_Y-2\widehat{\tau}_{Wald}C_{YW}+\widehat{\tau}^2_{Wald}V_W}/\left|\widehat{\tau}_{W}\right|
	\end{equation}
	has at least $(1-\alpha)$ coverage rate for $\tau$ as $n\rightarrow\infty$.
\end{theorem}

Proposition S1 in the Supplementary Material shows that the confidence interval in Theorem~\ref{conservative_ci_CRE} is the same as the super-population confidence interval for the population LATE obtained by the delta method.  A major concern about the delta method is its poor performance when the instrument is weak. This has been pointed out by, for example, \cite{bound1995problems} and \cite{nelson1990distribution} in the super-population context, and by \cite{kang2018inference} in the finite-population context. 

\subsubsection{The AR-like Confidence Set}\label{sec:FAR_CRE_Noreg}

Assuming homogeneous complier treatment effects, \cite{li2017general} constructed a confidence set for $\tau$, which is identical to that obtained by \cite{fieller1954some} and is asymptotically equivalent to that obtained by \cite{anderson1949estimation} when homoscedastic variance is assumed. Following \cite{ding2023first}, we refer to it as the FAR confidence set. \cite{kang2018inference} shows that, in the presence of heterogeneous complier treatment effects, this confidence set is also valid. We next construct the FAR confidence set under Condition~\ref{lindeberg-feller}, which is weaker than the set of assumptions in \cite{kang2018inference} that keep $\tau_Y$ and $\tau_W$ fixed across all values of $n$.

Recall that under Assumption \ref{assu} and Condition \ref{lindeberg-feller}, $\widehat{\tau}_A$ converges to a normal distribution with mean 0 and variance $\mathbb{V}_A$. It follows that, as $n\rightarrow\infty$, 
\begin{equation}
	\label{finite_CLT}
	\Pr\left(\widehat{\tau}_A^2\leq z_{\alpha/2}^2\mathbb{V}_A\right)\rightarrow 1-\alpha.
\end{equation}
Applying Proposition 1 of \cite{li2017general}, we can replace $\mathbb{V}_A$ in \eqref{finite_CLT} with 
\begin{equation}
	\label{def_V_A}
	V_A=S_{A(1)}^2/n_1+S_{A(0)}^2/n_0
\end{equation}
to obtain
\begin{equation*}
	\liminf_{n\rightarrow\infty}\left(\Pr\left(\widehat{\tau}_A^2\leq z_{\alpha/2}^2V_A\right)-\Pr\left(\widehat{\tau}_A^2\leq z_{\alpha/2}^2\mathbb{V}_A\right)\right)\geq0.
\end{equation*}
It is easy to see that $V_A=V_Y-2\tau C_{YW}+\tau^2 V_W$. With \eqref{def_tauhat_A}, an asymptotic conservative $1-\alpha$ confidence set for $\tau$ is the solution to the following inequality:
\begin{equation}
	\label{confidence_set_cre}
	\left(\widehat{\tau}_Y-\tau\widehat{\tau}_W\right)^2\leq z_{\alpha/2}^2\left(V_Y-2\tau C_{YW}+\tau^2V_W\right).
\end{equation}
\cite{kang2018inference} show that the solution can be a finite interval $[c_1,c_2]$, the whole real line, an empty set, the union of two disjoint infinite intervals $(-\infty,c_1]\cup [c_2,\infty)$, or one of the two infinite intervals, $(-\infty,c_1]$ and $[c_2,\infty)$. While \cite{kang2018inference} excludes the empty set as a solution when $\widehat{\tau}_W^2>z_{\alpha/2}^2V_W$, we can exclude the empty set as a solution as long as $\widehat{\tau}_W\neq 0$. The key observation is that when $\widehat{\tau}_W\neq 0$, $\widehat{\tau}_{Wald}=\widehat{\tau}_Y/\widehat{\tau}_W$ is always included in the solution to \eqref{confidence_set_cre}.  Infinite confidence sets suggest little or no identification due to weak instruments.

It is easy to see that under Condition~\ref{strength}, if we replace $\tau$ in the right hand side of $\eqref{confidence_set_cre}$ by $\widehat{\tau}_{Wald}$, we obtain the confidence interval with the Wald method in~\eqref{ci_CRE}.

\subsection{A Two-stage Procedure}\label{sec:two-stage-CRE}
We propose a two-stage procedure that uses the Wald method when a first-stage test concludes that the instrument is strong, and uses the FAR confidence set when the first-stage test concludes that the instrument is weak.  

The null hypothesis is
\begin{equation}
	\label{f_null}
	H_0:\tau_W\leq p_{+},
\end{equation}
where $p_{+}$ is a pre-specified small value in $(0,1)$. Given that $W_i$ is binary, it follows from Theorem 1 of \cite{li2017general} that under part (1) of Condition~\ref{lindeberg-feller}, $(\widehat{\tau}_W-\tau_W)/\sqrt{V_W}$ asymptotically has a standard normal distribution. Therefore we use the test statistic  $T=(\widehat{\tau}_W-p_{+})/\sqrt{V_W}$ and reject $H_0$ when $T>z_{\gamma}$ for a pre-specified significance level $\gamma\in (0,1)$.

We now formally define the confidence set of $\tau$ for any given values of $p_{+}\in(0,1)$, $\alpha\in(0,1)$ and $\gamma\in(0,1)$:
\begin{equation*}
	\widehat{I}_{CRE}=\begin{cases}
		\widehat{\tau}_{Wald}\pm z_{\alpha/2}\sqrt{V_Y-2\widehat{\tau}_{Wald}C_{YW}+\widehat{\tau}^2_{Wald}V_W}/\left|\widehat{\tau}_{W}\right|,\text{ if }T>z_{\gamma},\\
		\left\{\tau:\left(\widehat{\tau}_Y-\tau\widehat{\tau}_W\right)^2\leq z_{\alpha/2}^2\left(V_Y-2\tau C_{YW}+\tau^2V_W\right)\right\},\text{ if }T\leq z_{\gamma}.
	\end{cases}
\end{equation*}
The next theorem says that, under Assumption \ref{assu}, Condition \ref{lindeberg-feller} and CRE, $\widehat{I}_{CRE}$ is asymptotically conservative regardless of whether Condition~\ref{strength} holds, and that $\widehat{I}_{CRE}$ is more efficient than the FAR confidence set.
\begin{theorem}
	\label{confidence_set_mix}
	Under Assumption \ref{assu}, Condition \ref{lindeberg-feller} and CRE,\\ 
	(i) (Asymptotic Conservativeness) $$\liminf_{n\rightarrow\infty}\Pr\left(\tau\in\widehat{I}_{CRE}\right)\geq1-\alpha.$$
	(ii) (Efficiency) When $T\leq z_{\gamma}$, $\widehat{I}_{CRE}$ is the FAR confidence set \eqref{confidence_set_cre}. When $T>z_{\gamma}$, $\widehat{I}_{CRE}$ is the Wald confidence interval \eqref{ci_CRE}, which is shorter than \eqref{confidence_set_cre} unless \eqref{confidence_set_cre} is empty.
\end{theorem}

Under Assumption \ref{assu} and Condition \ref{lindeberg-feller}, the asymptotic conservativeness of $\widehat{I}_{CRE}$ does not depend on $p_+$ or $\gamma$. However in practice with finite sample we need to specify their values. Higher $p_+$ means that it is less likely to accept an instrument as being strong and more likely to use the FAR confidence set.  This leads to better coverage rates when $\tau_W$ is close to 0, but longer confidence interval/set. In the Monte Carlo simulations in Section~\ref{sec:Monte Carlo}, we set $p_+=0.01$.

To set the value of $\gamma$, we use the boundary case with $\tau_W=0$ and $\tau\in\{-\infty,\infty\}$ as a guidance.  This boundary case has been excluded by Assumption \ref{assu}, but we find that, in setting the value of $\gamma$, extrapolating this boundary case is useful.

First note that the confidence interval for the Wald method is finite as long as $\widehat{\tau}_W\neq 0$. Since the asymptotic distribution of $\widehat{\tau}_W$ is continuous, the asymptotic probability that this confidence interval covers the true value of $\tau$ is 0 when $\tau_W=0$. Therefore to achieve asymptotically conservative coverage rate we need to use the FAR confidence set.

Second note that the two-stage confidence set $\widehat{I}_{CRE}$ equals the FAR confidence set when $T\leq z_{\gamma}$, which is equivalent to
\begin{equation}\label{CI_inf1}
	\widehat{\tau}_W\leq p_+ +z_{\gamma}\sqrt{V_W}.
\end{equation}
According to the Supplementary Material for \cite{kang2018inference}, when $\tau_W=0$, the FAR confidence set includes $-\infty$ and $\infty$ when $\widehat{\tau}_W^2<z_{\alpha/2}^2V_W$,\footnote{Since the asymptotic distribution of $\widehat{\tau}_W$ is continuous, we ignore cases with $\widehat{\tau}_W^2=z_{\alpha/2}^2V_W$.} that is, when
\begin{equation}\label{CI_inf2}
	-z_{\alpha/2}\sqrt{V_W}<\widehat{\tau}_W<z_{\alpha/2}\sqrt{V_W}. 
\end{equation}
With \eqref{CI_inf1} and \eqref{CI_inf2}, it becomes clear that, when $\tau_W=0$, the two-stage confidence set $\widehat{I}_{CRE}$ includes $-\infty$ and $\infty$ when  $-z_{\alpha/2}\sqrt{V_W}<\widehat{\tau}_W\leq p_+ +z_{\gamma}\sqrt{V_W}$ for any $\gamma\geq \alpha/2$. $\widehat{\tau}_W/\sqrt{V_W}$ asymptotically follows a standard normal distribution, and therefore (1) when $\gamma=\alpha/2$, $\widehat{I}_{CRE}$ asymptotically has a coverage rate of at least $(1-\alpha)$, or, in other words, there is no undercoverage; (2) when $\gamma=3\alpha/2$, $\widehat{I}_{CRE}$ asymptotically has a coverage rate of at least $(1-2\alpha)$, or, in other words, there is an undercoverage rate of at most $\alpha$.

In the Monte Carlo simulations in Section~\ref{sec:Monte Carlo}, we set $\alpha=0.05$ and $\gamma\in\{0.025,0.075\}$ (that is, $\gamma\in\{\alpha/2,3\alpha/2\}$).

\section{Inference under ReM without Regression Adjustment}

We introduce a set of notations for later use. For any set of quantities $\{Q_i:i=1,\cdots,n\}$, let $\mathbb{S}_{Q,\bm{x}}=\frac{1}{n-1}\sum_{i=1}^{n}\boldsymbol{x}_i(Q_i-\overline{Q})$ be the finite population covariance between $Q_i$ and $\bm{x}_i$; let $\mathbb{S}_{Q\mid\boldsymbol{x}}^2=\mathbb{S}_{Q,\bm{x}}^{\top}\mathbb{S}_{\bm{x}\bm{x}}^{-1}\mathbb{S}_{Q,\bm{x}}$ be the finite population variance of the linear projection of $Q_i$ on $\bm{x}_i$. For units with $Z_i=z$ ($z=0,1$), let
\begin{equation*}
	\bm{S}_{\bm{x}\bm{x},z}=\frac{1}{n_z-1}\sum_{i:\ Z_{i}=z}(\bm{x}_i-\overline{\bm{x}}_z)(\bm{x}_i-\overline{\bm{x}}_z)^{\top }
\end{equation*}
be the sample covariance matrix of covariates, where $\overline{\bm{x}}_z=\frac{1}{n_z}\sum_{i:\ Z_{i}=z}\bm{x}_i$, let $\boldsymbol{S}_{Q(z),\boldsymbol{x}}=\frac{1}{n_z-1}\sum_{i:\ Z_i=z}(\boldsymbol{x}_i-\overline{\bm{x}}_z)(Q_i-\overline{Q})$ be the sample covariance between $Q_i$ and $\boldsymbol{x}_i$, and let $S_{Q(z)\mid\boldsymbol{x}}^2=\boldsymbol{S}_{Q(z),\boldsymbol{x}}^{\top}\bm{S}_{\bm{x}\bm{x},z}^{-1}\boldsymbol{S}_{Q(z),\boldsymbol{x}}$ be the sample variance of the linear projection of $Q_i$ on $\boldsymbol{x}_i$.

Let $\mathbb{S}_{A(z),\bm{x}}$ ($z=0,1$) be the finite population covariance between $A_i(z)$ and $\bm{x}_i$. We introduce the regularity conditions from \cite{li2020rerandomization} as follows.
\setcounter{theorem}{2}
\begin{condition}
	\label{strict_cond}
	As $n\rightarrow\infty$, (1) the proportion of units under treatment, $n_1/n$, has a limit in $(0,1)$; (2) $\mathbb{S}_{A(z)}^2$, $\mathbb{S}_{A(1)-A(0)}^2$, $\mathbb{S}_{\bm{x}\bm{x}}$ and $\mathbb{S}_{A(z),\bm{x}}$ ($z=0,1$) have finite limiting values, and the limit of $\mathbb{S}_{\bm{x}\bm{x}}$ is non-singular; (3) $\max_{1\leq i\leq n}\left|A_i(z)-\overline{A}(z)\right|^{2} / n \rightarrow 0$ ($z=0,1$) and $\max_{1\leq i\leq n}\left\|\bm{x}_{i}\right\|^{2}/n$ $\rightarrow 0$.
\end{condition}

\subsection{The Wald Method}\label{sec:ReM_Wald}

Applying Proposition 1 from \cite{li2018asymptotic}, the squared multiple correlation coefficient between $\widehat{\tau}_A$ and $\widehat{\tau}_{\boldsymbol{x}}$ under CRE can be expressed as
\begin{equation}
	\mathbb{R}^2_{A}=\mathbb{V}_{A|\bm{x}}/\mathbb{V}_{A},\label{def_bR2_A}
\end{equation}
where 
\begin{equation*}
	\mathbb{V}_{A|\bm{x}}=\left(\frac{\mathbb{S}^2_{A(1)\mid \boldsymbol{x}}}{n_1}+\frac{\mathbb{S}^2_{A(0)\mid \boldsymbol{x}}}{n_0}-\frac{\mathbb{S}^2_{A(1)-A(0)\mid \boldsymbol{x}}}{n}\right).
\end{equation*}
Here, $\mathbb{S}^2_{A(z)\mid \boldsymbol{x}}$ ($z=0,1$) and $\mathbb{S}^2_{A(1)-A(0)\mid \boldsymbol{x}}$ are finite-population variances of the linear projections of $A_i(z)$ ($z=0,1$) and $A_i(1)-A_i(0)$ on $\boldsymbol{x}_i$, and $\mathbb{V}_A$ is defined in~\eqref{def_bV_A}.

Moreover, applying Theorem 1 from \cite{li2018asymptotic}, under Assumption \ref{assu}, Condition \ref{strict_cond} and ReM, the asymptotic distribution of $\widehat{\tau}_A/\sqrt{\mathbb{V}_A}$ can be expressed as a linear combination of two independent random variables,
\begin{equation}
	\label{mix_normal_trunc}
	\sqrt{1-\mathbb{R}^2_{A}} \cdot \varepsilon_0+\sqrt{\mathbb{R}^2_{A}} \cdot L_{K, a},
\end{equation}
where $\varepsilon_0$ follows the standard normal and the distribution of $L_{K,a}$ is symmetric around zero with the following form
\begin{equation}
	\label{truncate_dist}
	L_{K,a}\sim \chi _{K,a}S\sqrt{\beta _{K}}.
\end{equation}%
where $\chi_{K,a}^2=\chi_{K}^{2}|\chi_{K}^{2}\leq a$ is a truncated $\chi^{2}$ random variable with $K$ degrees of freedom, $S$ is a random variable taking values $\pm 1$ with probability 1/2, $\beta _{K}\sim \text{Beta}(1/2,(K-1)/2)$ is a Beta random
variable degenerating to a point mass at 1 when $K=1$, and ($\chi _{K,a}$, $S$ ,$\beta _{K}$) are jointly independent.

The next theorem says that the Wald estimator is an asymptotically unbiased estimator and also gives its asymptotic distribution.
\begin{theorem}
	\label{Wald_ReM}
	Under Assumption \ref{assu}, Conditions \ref{strength}, \ref{strict_cond} and ReM, as $n\rightarrow\infty$, $\widehat{\tau}_{Wald}-\tau=o_p(1)$ and
	\begin{equation*}
		\frac{\widehat{\tau}_{Wald}-\tau}{\sqrt{\mathbb{V}_A}/\tau_W}\stackrel{d}{\rightarrow}\sqrt{1-\mathbb{R}^2_{A}} \cdot \varepsilon_0+\sqrt{\mathbb{R}^2_{A}} \cdot L_{K, a},
	\end{equation*}
	where the distribution in the right hand side is given by~\eqref{mix_normal_trunc}.
\end{theorem}

To construct an asymptotically conservative confidence interval for $\tau$, we need to estimate $\tau_W$, $\mathbb{V}_A$ and the quantile of the distribution in~\eqref{mix_normal_trunc}.
We can use $\widehat{\tau}_W$ to estimate $\tau_W$. Applying the arguments in Appendix A4 of \cite{li2018asymptotic}, we can use
\begin{equation}
	\label{def_VA_ReM} V_A^{ReM}=\frac{S_{A(1)}^2}{n_1}+\frac{S_{A(0)}^2}{n_0}-\frac{\left(\boldsymbol{S}_{A(1),\boldsymbol{x}}-\boldsymbol{S}_{A(0),\boldsymbol{x}}\right)^{\top}\mathbb{S}_{\bm{x}\bm{x}}^{-1}\left(\boldsymbol{S}_{A(1),\boldsymbol{x}}-\boldsymbol{S}_{A(0),\boldsymbol{x}}\right)}{n}
\end{equation}
to estimate $\mathbb{V}_A$, and use 
\begin{equation}
	\label{def_R2A} R^2_{A}=V_{A\mid\bm{x}}/V_A^{ReM}
\end{equation}
to estimate $\mathbb{R}^2_{A}$, where
\begin{equation} V_{A\mid\bm{x}}=\frac{S_{A(1)\mid\bm{x}}^2}{n_1}+\frac{S_{A(0)\mid\bm{x}}^2}{n_0}-\frac{\left(\boldsymbol{S}_{A(1),\boldsymbol{x}}-\boldsymbol{S}_{A(0),\boldsymbol{x}}\right)^{\top}\mathbb{S}_{\bm{x}\bm{x}}^{-1}\left(\boldsymbol{S}_{A(1),\boldsymbol{x}}-\boldsymbol{S}_{A(0),\boldsymbol{x}}\right)}{n}.\label{def_VAx}
\end{equation}

Define
\begin{align} V_Y^{ReM}&=\frac{S_{Y(1)}^2}{n_1}+\frac{S_{Y(0)}^2}{n_0}-\frac{\left(\boldsymbol{S}_{Y(1),\boldsymbol{x}}-\boldsymbol{S}_{Y(0),\boldsymbol{x}}\right)^{\top}\mathbb{S}_{\bm{x}\bm{x}}^{-1}\left(\boldsymbol{S}_{Y(1),\boldsymbol{x}}-\boldsymbol{S}_{Y(0),\boldsymbol{x}}\right)}{n},\notag\\ V_W^{ReM}&=\frac{S_{W(1)}^2}{n_1}+\frac{S_{W(0)}^2}{n_0}-\frac{\left(\boldsymbol{S}_{W(1),\boldsymbol{x}}-\boldsymbol{S}_{W(0),\boldsymbol{x}}\right)^{\top}\mathbb{S}_{\bm{x}\bm{x}}^{-1}\left(\boldsymbol{S}_{W(1),\boldsymbol{x}}-\boldsymbol{S}_{W(0),\boldsymbol{x}}\right)}{n},\label{def V_W_prime} \\ C_{YW}^{ReM}&=\frac{S_{Y(1),W(1)}}{n_1}+\frac{S_{Y(0),W(0)}}{n_0}-\frac{\left(\boldsymbol{S}_{Y(1),\boldsymbol{x}}-\boldsymbol{S}_{Y(0),\boldsymbol{x}}\right)^{\top}\mathbb{S}_{\bm{x}\bm{x}}^{-1}\left(\boldsymbol{S}_{W(1),\boldsymbol{x}}-\boldsymbol{S}_{W(0),\boldsymbol{x}}\right)}{n},\notag
\end{align}
These quantities are observable. It is easy to see that 
\begin{equation}
	V_A^{ReM}=V_Y^{ReM}-2\tau C_{YW}^{ReM}+\tau^2V_{W}^{ReM}.\label{def_VA_ReM2} 
\end{equation}
Similarly, $R_A^2$ also involves $\tau$. As $\tau$ is unknown, we can replace $\tau$ with $\widehat{\tau}_{Wald}$ in $V_A^{ReM}$ and $R_A^2$ to obtain 
\begin{equation}
	\widehat{V}_A^{ReM}=V_Y^{ReM}-2\widehat{\tau}_{Wald} C_{YW}^{ReM}+\widehat{\tau}_{Wald}^2V_{W}^{ReM}\label{hat_VA_ReM2}
\end{equation}
and $\widehat{R}_A^2$. Following \cite{li2018asymptotic}, if $\widehat{R}_A^2$ is negative, we set it be 0. (Asymptotically $\widehat{R}_A^2$ is nonnegative.) Similarly, $\widehat{V}_A^{ReM}$ could also be negative in finite sample. When this happens we set it to $0$ in order to avoid problematic variance estimators.
For $\rho\in[0,1]$, let $\lambda_{\alpha/2}\left(\rho\right)$ be the $(1-\alpha/2)$th quantile of the distribution of
\begin{equation}
	\label{mix_normal_trunc_general}
	\sqrt{1-\rho} \cdot \varepsilon_0+\sqrt{\rho} \cdot L_{K, a}.
\end{equation}
We can use $\lambda_{\alpha/2}\left(\widehat{R}_A^2\right)$ to estimate the $(1-\alpha/2)$th quantile of the distribution in~\eqref{mix_normal_trunc}. 

\begin{theorem}
	\label{conservative_ReM}
	Under Assumption \ref{assu}, Conditions \ref{strength}, \ref{strict_cond} and ReM, as $n\rightarrow\infty$, the confidence interval
	\begin{equation}
		\label{ci_ReM} \widehat{\tau}_{Wald}\pm\lambda_{\alpha/2}\left(\widehat{R}^2_{A}\right)\sqrt{V_Y^{ReM}-2\widehat{\tau}_{Wald}C_{YW}^{ReM}+\widehat{\tau}_{Wald}^2V_W^{ReM}}/\left|\widehat{\tau}_W\right|
	\end{equation}
	has at least $(1-\alpha)$ coverage rate for $\tau$.
\end{theorem}

\subsection{The FAR Confidence Set}
\label{subsec:FAR_rem}

Given the asymptotic distribution of $\widehat{\tau}_A/\sqrt{\mathbb{V}_A}$ in~\eqref{mix_normal_trunc}, as $n\rightarrow\infty$, 
\begin{equation}
	\Pr\left(\widehat{\tau}_A^2\leq\lambda^2_{\alpha/2}\left(\mathbb{R}^2_{A}\right)\mathbb{V}_A\right)\rightarrow 1-\alpha.\label{ci_detail_raw}
\end{equation}
Applying the arguments in Appendix A4 of \cite{li2018asymptotic}, $\lambda_{\alpha/2}^2\left(R^2_{A}\right)V_A^{ReM}$ is asymptotically larger than or equivalent to $\lambda_{\alpha/2}^2\left(\mathbb{R}^2_{A}\right)\mathbb{V}_A$. With \eqref{def_tauhat_A}, the solution to
\begin{equation}
	\label{confidence_set_rem}
	\left(\widehat{\tau}_{Y}-\tau\widehat{\tau}_{W}\right)^2\leq\lambda_{\alpha/2}^2\left(R^2_{A}\right)V_A^{ReM}=\lambda_{\alpha/2}^2\left(R^2_{A}\right)\left(V_Y^{ReM}-2\tau C_{YW}^{ReM}+\tau^2V_W^{ReM}\right)
\end{equation}
is an asymptotically conservative confidence set for $\tau$. 

Given that $R_A^2$ depends on $\tau$ and therefore $\lambda_{\alpha/2}^2\left(R^2_{A}\right)$ depends on $\tau$ in a complicated way, it is impossible to solve \eqref{confidence_set_rem} analytically. We propose an approach that yields FAR confidence sets in closed form with a little sacrifice of efficiency.

Let $R^{*2}_A$ denote the minimum value of $R_A^2$ when $\tau$ varies over $[-\infty,\infty]$. According to Lemma A4 of \cite{li2018asymptotic}, $\lambda_{\alpha/2}(\rho)$ is a non-increasing function of $\rho$, hence for any $\tau$,
\begin{equation*}
	\lambda_{\alpha/2}\left(R^{*2}_{A}\right)\geq\lambda_{\alpha/2}\left(R^2_{A}\right).
\end{equation*}
It follows that the set
\begin{equation}
	\label{confidence_set_rem2}
	\left(\widehat{\tau}_{Y}-\tau\widehat{\tau}_{W}\right)^2\leq\lambda_{\alpha/2}^2\left(R^{*2}_{A}\right)\left(V_Y^{ReM}-2\tau C_{YW}^{ReM}+\tau^2V_W^{ReM}\right)
\end{equation}
contains \eqref{confidence_set_rem}, hence is also an asymptotically conservative confidence set for $\tau$. The solution to \eqref{confidence_set_rem2} is similar to that of \eqref{confidence_set_cre}. Moreover, since $\lambda_{\alpha/2}\left(R^{*2}_{A}\right)\leq\lambda_{\alpha/2}(0)=z_{\alpha/2}$ and $V_A^{ReM}\leq V_A$ according to their definitions in~\eqref{def_V_A} and~\eqref{def_VA_ReM} (i.e., $V_Y^{ReM}-2\tau C_{YW}^{ReM}+\tau^2V_W^{ReM}\leq V_Y-2\tau C_{YW}+\tau^2V_W$), the confidence set \eqref{confidence_set_rem2} is contained by \eqref{confidence_set_cre}. This implies that ReM has an efficiency gain over CRE. A remaining task is to find $R^{*2}_A$, which we outline in Section A of the Supplementary Material.

Similar to Section~\ref{sec:FAR_CRE_Noreg}, we can exclude the empty set as a solution as long as $\widehat{\tau}_W\neq 0$. It is easy to see that under Condition~\ref{strength}, if in the right hand side of $\eqref{confidence_set_rem2}$ we replace $\tau$ by $\widehat{\tau}_{Wald}$ and replace $R^{*2}_{A}$ with $\widehat{R}^2_A$, we obtain the confidence interval with the Wald method in~\eqref{ci_ReM}.

\subsection{The Two-stage Procedure}

In the first stage, we again test the null hypothesis in \eqref{f_null}. Define $R^2_W$ similarly as $R^2_A$ in~\eqref{def_R2A}, with $A_i$ replaced by $W_i$. Applying Corollary 2 from \cite{li2020rerandomization}, $(\widehat{\tau}_W-\tau_W)/\sqrt{V_W^{ReM}}$ asymptotically follows the distribution in~\eqref{mix_normal_trunc_general} with $\rho = R^2_W$, where $V_W^{ReM}$ is defined in~\eqref{def V_W_prime}. Therefore we use the test statistics $T^{ReM}=(\widehat{\tau}_W-p_{+})/\sqrt{V_W^{ReM}}$ and reject $H_0$ when $T>\lambda_{\gamma}\left(R^2_W\right)$ for a prespecified significance level $\gamma\in (0,1)$.

We formally define the confidence set of $\tau$ for any given values of $p_{+}\in(0,1)$, $\alpha\in(0,1)$ and $\gamma\in(0,1)$:
\begin{equation*}
	\widehat{I}_{ReM}=\begin{cases} \widehat{\tau}_{Wald}\pm\lambda_{\alpha/2}\left(\widehat{R}^2_{A}\right)\sqrt{V_Y^{ReM}-2\widehat{\tau}_{Wald}C_{YW}^{ReM}+\widehat{\tau}_{Wald}^2V_W^{ReM}}/\left|\widehat{\tau}_W\right|,\text{ if }T^{ReM}>\lambda_{\gamma}\left(R^2_W\right),\\
		\left\{\tau:\left(\widehat{\tau}_Y-\tau\widehat{\tau}_W\right)^2\leq\lambda_{\alpha/2}^2\left(R^{*2}_{A}\right)\left(V_Y^{ReM}-2\tau C_{YW}^{ReM}+\tau^2V_W^{ReM}\right)\right\},\text{ if }T^{ReM}\leq\lambda_{\gamma}\left(R^2_W\right).
	\end{cases}
\end{equation*}

The next theorem is similar to Theorem~\ref{confidence_set_mix}.
\begin{theorem}
	\label{confidence_set_mix2}
	Under Assumption \ref{assu}, Condition \ref{strict_cond} and ReM,\\
	(i) (Asymptotic Conservativeness) $$\liminf_{n\rightarrow\infty}\Pr\left(\tau\in\widehat{I}_{ReM}\right)\geq 1-\alpha.$$
	(ii) (Efficiency) When $T^{ReM}\leq\lambda_{\gamma}(R^2_W)$, $\widehat{I}_{ReM}$ is the FAR confidence set \eqref{confidence_set_rem2}. When $T^{ReM}>\lambda_{\gamma}(R^2_W)$, $\widehat{I}_{ReM}$ is the Wald confidence interval \eqref{ci_ReM}, which is shorter than \eqref{confidence_set_rem2} unless \eqref{confidence_set_rem2} is empty.
\end{theorem}
The specification of $p_+$ and $\gamma$ is similar to that given in Section~\ref{sec:two-stage-CRE}.

\section{Inference with Regression Adjustment}
\label{sec_reg}

For later use, we introduce a set of notations. For any set of quantities $\{Q_i:i=1,\cdots,n\}$, let $\widehat{\tau}_{Q|\bm{x}}$ be the estimated coefficient on $Z_i$ in the Ordinary Least Squares (OLS) regression of $Q_i$ on $Z_i$, $\bm{x}_i$ and $Z_i\bm{x}_i$. Let $V_Q^{\dagger}$ ($\dagger \in\{EHW, HC2, HC3\}$) be the Eicker-Huber-White (EHW), HC2 and HC3 robust estimators \citep{MacKinnon:2013} for the variance of $\widehat{\tau}_{Q|\bm{x}}$. 

As mentioned at the end of Section~\ref{sec:design}, CRE can be regarded as a special case of ReM with $a=\infty$. In this Section, we will only give theoretical results under ReM, which hold with any value of $a$ and therefore hold under CRE.

\subsection{The Wald Method}

We first define the Wald estimator with regression adjustment as
\begin{equation}
	\label{est_adj} \widehat{\tau}_{Wald|\bm{x}}=\frac{\widehat{\tau}_{Y\mid\boldsymbol{x}}}{\widehat{\tau}_{W\mid\boldsymbol{x}}}.
\end{equation}
In the setting with a binary outcome, \cite{ren2023model} called $\widehat{\tau}_{Wald|\bm{x}}$ the indirect least squares estimator with interaction, and studied its finite-population asymptotic behaviors under CRE. We will study the asymptotic behaviors of $\widehat{\tau}_{Wald|\bm{x}}$ for general outcome and under ReM (with CRE as a special case).

Applying Theorem 3 and Proposition 3 of \cite{li2020rerandomization}, under Assumption \ref{assu} and ReM,
\begin{equation}
	\label{adist_lin}
	\frac{\widehat{\tau}_{A\mid\bm{x}}}{\sqrt{\mathbb{V}_A\left(1-\mathbb{R}^2_{A}\right)}}\stackrel{d}{\rightarrow}N(0,1).
\end{equation}
The next theorem says that, under Conditions \ref{strength}, \ref{strict_cond} and ReM, $\widehat{\tau}_{Wald|\bm{x}}$ is an asymptotically unbiased and normally distributed estimator for $\tau$.
\begin{theorem}
	\label{Wald_Reg}
	Under Assumption \ref{assu}, Conditions \ref{strength}, \ref{strict_cond} and ReM, as $n\rightarrow\infty$,
	\begin{equation}
		\widehat{\tau}_{Wald|\bm{x}}-\tau=o_p(1)\quad \text{and}\quad \frac{\widehat{\tau}_{Wald|\bm{x}}-\tau}{\sqrt{\mathbb{V}_A\left(1-\mathbb{R}^2_{A}\right)}/\tau_W}\stackrel{d}{\rightarrow}N(0,1).
	\end{equation}
\end{theorem}

To construct an asymptotically conservative confidence interval for $\tau$, we need to estimate $\tau_W$ and $\mathbb{V}_A\left(1-\mathbb{R}^2_{A}\right)$. We can use $\widehat{\tau}_{W|\bm{x}}$ to estimate $\tau_W$. The next propositions says that we can use $V_A^{\dagger}$ ($\dagger\in\{EHW,HC2,HC3\}$) to estimate $\mathbb{V}_A\left(1-\mathbb{R}^2_{A}\right)$.
\setcounter{theorem}{0}
\begin{proposition}
	\label{conservative_reg_raw}
	Under Assumption \ref{assu}, Condition \ref{strict_cond} and ReM, $nV_A^{\dagger}$ ($\dagger\in\{EHW,HC2,HC3\}$) is an asymptotically conservative estimator for $n\mathbb{V}_A\left(1-\mathbb{R}^2_{A}\right)$.
\end{proposition}

Because $A_i=Y_i-\tau W_i$, it is easy to see that $\widehat{\tau}_{A|\bm{x}}=\widehat{\tau}_{Y|\bm{x}}-\tau\widehat{\tau}_{W|\bm{x}}$. Hence $V_A^{\dagger}=V_Y^{\dagger}-2\tau C_{YW}^{\dagger}+\tau^2V_W^{\dagger}$, where $C_{YW}^{\dagger}$ ($\dagger \in\{EHW, HC2, HC3\}$) is the covariance between $\widehat{\tau}_{Y|\bm{x}}$ and $\widehat{\tau}_{W|\bm{x}}$ estimated using the EHW, HC2 or HC3 robust estimator. We estimate $V_A^{\dagger}$ by replacing $\tau$ with $\widehat{\tau}_{Wald|\bm{x}}$ to obtain $\widehat{V}_A^{\dagger}=V_Y^{\dagger}-2\widehat{\tau}_{Wald|\bm{x}} C_{YW}^{\dagger}+\widehat{\tau}_{Wald|\bm{x}}^2V_W^{\dagger}$.

\setcounter{theorem}{7}
\begin{theorem}
	\label{conservative_Reg}
	Under Assumption \ref{assu}, Conditions \ref{strength}, \ref{strict_cond} and ReM, as $n\rightarrow\infty$, the confidence interval
	\begin{equation}
		\label{ci_Reg}	\widehat{\tau}_{Wald|\bm{x}}\pm z_{\alpha/2}\sqrt{V_Y^{\dagger}-2\widehat{\tau}_{Wald|\bm{x}}C_{YW}^{\dagger}+\widehat{\tau}_{Wald|\bm{x}}^2V_W^{\dagger}}/\left|\widehat{\tau}_{W\mid\bm{x}}\right|
	\end{equation}
	has at least $(1-\alpha)$ coverage rate for $\tau$  ($\dagger=EHW,HC2$ or $HC3$).
\end{theorem}

\subsection{The FAR Confidence Set}\label{subsec:FAR_reg}

Chapter 21 of \cite{ding2023first} discusses the FAR confidence set with regression adjustment under CRE. \cite{aronow2024randomization} discusses using a studentized FAR-type test statistic to construct confidence sets with regression adjustment under CRE. We will discuss the FAR confidence set with regression adjustment under ReM (with CRE as a special case).

Given~\eqref{adist_lin}, Proposition \ref{conservative_reg_raw}, and  $\widehat{\tau}_{A|\bm{x}}=\widehat{\tau}_{Y|\bm{x}}-\tau\widehat{\tau}_{W|\bm{x}}$, the solution to the inequality
\begin{equation}
	\label{confidence_set_reg}
	\left(\widehat{\tau}_{Y\mid\bm{x}}-\tau\widehat{\tau}_{W\mid\bm{x}}\right)^2\leq z_{\alpha/2}^2V_A^{\dagger}=z_{\alpha/2}^2\left(V_Y^{\dagger}-2\tau C_{YW}^{\dagger}+\tau^2V_W^{\dagger}\right)
\end{equation}
is an asymptotic conservative confidence set for $\tau$.

Similar to Section~\ref{sec:FAR_CRE_Noreg}, we can exclude the empty set as a solution as long as $\widehat{\tau}_{W\mid\bm{x}}\neq 0$. It is easy to see that under Condition \ref{strength}, if we replace $\tau$ in the right hand side of \eqref{confidence_set_reg} by
$\widehat{\tau}_{Wald|\bm{x}}$, we obtain the confidence interval with the Wald method in \eqref{ci_Reg}.

\subsection{The Two-stage Procedure}

In the first stage, we again test the null hypothesis in \eqref{f_null}. Similar to~\eqref{adist_lin}, we have
\begin{equation*}
	\frac{\widehat{\tau}_{W\mid\bm{x}}}{\sqrt{\mathbb{V}_W\left(1-\mathbb{R}^2_{W}\right)}}\stackrel{d}{\rightarrow}N(0,1),
\end{equation*}
where $\mathbb{V}_W$ and $\mathbb{R}^2_W$ are defined similarly as $\mathbb{V}_A$ and $\mathbb{R}^2_A$. Similar to Proposition~\ref{conservative_reg_raw}, $nV_W^{\dagger}$ ($\dagger\in\{EHW,HC2,HC3\}$) is an asymptotically conservative estimator for $n\mathbb{V}_W\left(1-\mathbb{R}^2_{W}\right)$.  Therefore we use the test statistic $T^{\dagger}=(\widehat{\tau}_{W\mid\bm{x}}-p_{+})/\sqrt{V_W^{\dagger}}$ ($\dagger\in\{EHW,HC2,HC3\}$) and reject $H_0$ when $T^{\dagger}>z_{\gamma}$ for a prespecified significance level $\gamma\in(0,1)$.

We formally define the confidence set of $\tau$ for $\dagger\in\{EHW,HC2,HC3\}$ and for any given values of $p_+\in(0,1)$, $\alpha\in(0,1)$ and $\gamma\in(0,1)$: 
\begin{equation*}
	\widehat{I}_{Reg}^{\dagger}=\begin{cases} \widehat{\tau}_{Wald|\bm{x}}\pm z_{\alpha/2}\sqrt{V_Y^{\dagger}-2\widehat{\tau}_{Wald|\bm{x}}C_{YW}^{\dagger}+\widehat{\tau}_{Wald|\bm{x}}^2V_W^{\dagger}}/\left|\widehat{\tau}_{Wald\mid\bm{x}}\right|,\text{ if }T^{\dagger}>z_{\gamma},\\
		\left\{\tau:\left(\widehat{\tau}_{Y\mid\bm{x}}-\tau\widehat{\tau}_{W\mid\bm{x}}\right)^2\leq z_{\alpha/2}^2\left(V_Y^{\dagger}-2\tau C_{YW}^{\dagger}+\tau^2V_W^{\dagger}\right)\right\},\text{ if }T^{\dagger}\leq z_{\gamma}.
	\end{cases}
\end{equation*}

The next theorem is similar to Theorem~\ref{confidence_set_mix}.
\begin{theorem}
	Under Assumption \ref{assu} and Condition \ref{strict_cond},\\ 
	\label{confidence_set_mix3}
	(i) (Asymptotic Conservativeness) 
	$$\liminf_{n\rightarrow\infty}\Pr\left(\tau\in\widehat{I}_{Reg}^{\dagger}\right)\geq1-\alpha.$$
	(ii) (Efficiency) When $T^{\dagger}\leq z_{\gamma}$,  $\widehat{I}_{Reg}^{\dagger}$ is the FAR confidence set \eqref{confidence_set_reg}; when $T^{\dagger}>z_{\gamma}$, $\widehat{I}_{Reg}^{\dagger}$ is the Wald confidence interval \eqref{ci_Reg}, which is shorter than \eqref{confidence_set_reg} unless \eqref{confidence_set_reg} is empty.
\end{theorem}
The specification of $p_+$ and $\gamma$ is similar to that given in Section~\ref{sec:two-stage-CRE}.

\section{Monte Carlo Simulation} 
\label{sec:Monte Carlo}

\subsection{Setup of the Monte Carlo Simulation}\label{sec:setup_MCstudy}
We use {\it Wald} and {\it FAR} to denote the Wald method and the FAR method.We use {\it $TS_{\gamma-0.075}$} and {\it $TS_{\gamma-0.025}$} to denote the two-stage procedure with $\gamma=0.075$ and $\gamma=0.025$. The strategy adopted by practitioners in the super-population setting of using a first-stage F-test to check the strength of the instrument is used as a comparison  (see e.g. \cite{stock2002survey}).  We use {\it $TS_{F-10}$} to denote a two-stage procedure that uses the Wald method when the first stage $F$ statistic is larger than 10 and uses the FAR confidence set when it is no larger than 10, and use {\it $Wald_{F-10}$} to denote the approach that only gives results from the Wald method when the first stage $F$ statistic is larger than 10.

We fix $K=5$, and set $n_1=n_0=n/2$. Each covariate independently follows a $N(0, 1)$ distribution. Noting that potential outcomes only depend on treatment received, we use $Y_i^W(w)$ ($w=0,1$) to denote the potential outcome if receiving treatment $w$. The data generating process is as follows:
\begin{equation*}
	\begin{aligned}
		Y_i^W(0) &=\bm{\xi}^{\top}\bm{x}_i+\epsilon_{0 i}, \\
		Y_i^W(1) &=\bm{\xi}^{\top}\bm{x}_i+\bm{\eta}^{\top}\bm{x}_i+\epsilon_{1 i},\\
		L_i(0) &=\delta_0+\bm{\psi}^{\top}\bm{x}_i+u_i, \\
		L_i(1) &=\delta_1+L_i(0),\\
		W_i(z) &=I(L_i(z)>0),\ z=0,1.
	\end{aligned}
\end{equation*}
Here, $\bm{\xi}$, $\bm{\eta}$ and $\bm{\psi}$ are all $K\times 1$ vectors of $1$'s. We set $\delta_0=0$, and choose the value of $\delta_1$ such that the fraction of compliers equals a pre-specified $\tau_W$. $\epsilon_{i 0}$, $\epsilon_{i 1}$ and $u_i$ are drawn from independent normal distributions with mean 0 and variances chosen such that the squared multiple correlation coefficient is $0.5$ in each of the equations for $Y_i^W(0)$, $Y_i^W(1)$ and $L_i(0)$.

The values of $(\bm{x}_i, L_i(0), L_i(1), W_i(0), W_i(1), Y_i(0), Y_i(1))$ are fixed, and only the values of $Z_i$ are randomly generated $10,000$ times under CRE or ReM. The observed values are $(\bm{x}_i, Z_i, W_i, Y_i)$.  We thus have $10,000$ datasets for each scenario with different combinations of $n$, $\tau_W$ and experimental design.

We consider three performance measures: (1) median absolute error of point estimates ($\widehat{\tau}_{Wald}$ or $\widehat{\tau}_{Wald|\bm{x}}$, (2) coverage rate of 95\% confidence sets, and (3) median length of 95\% confidence sets. When calculating these performance measures for {\it $Wald_{F-10}$}, only data sets with first stage $F$-statistics larger than 10 are included. For {\it $TS_{\gamma-0.075}$}, {\it $TS_{\gamma-0.025}$} and {\it $TS_{F-10}$/$Wald_{F-10}$}, we also calculate the proportion of datasets for which the first stage test rejects the null hypothesis and concludes that the instrument is strong.

\subsection{Results}
Here we fix $n=200$, and focus on the proportion of datasets judged to have strong instruments by the first stage tests, as well as the median length of 95\% confidence sets. For regression adjustment, we focus on the EHW variance estimator. Section D of the Supplementary Material contains more simulation results with $n=200$, as well as results with $n=100$ or $n=400$.

Table~\ref{tab:SRate} presents the proportion of datasets judged to have strong instruments, which equals the size of the first stage test when $\tau_W=0.005$, and the power of the first stage test when $\tau_W\geq 0.05$. The size is slightly larger than the nominal value 0.05 for {\it $TS_{\gamma-0.075}$}, and is smaller than the nominal value for {\it $TS_{\gamma-0.025}$} and {\it $TS_{F-10}$}.  {\it $TS_{\gamma-0.075}$} has the highest power, and {\it $TS_{F-10}$/$Wald_{F-10}$} has the lowest power. For {\it $TS_{\gamma-0.075}$} and {\it $TS_{\gamma-0.025}$}, compared to the case without regression adjustment under CRE, ReM in the design stage or regression adjustment in the analysis stage increases the power, but there is no additional increase to use both. For {\it $TS_{F-10}$/$Wald_{F-10}$}, regression adjustment increases the power, but the experimental design does not make a difference. None of the first-stage tests has high power when $\tau_W=0.05$ or $\tau_W=0.10$.

\begin{table}[htbp]
	\centering
	\caption{Proportion of datasets judged to have strong instruments ($n=200$).}\label{tab:SRate}
	\begin{tabular}{rcrrrrrrr}
		\hline
		&&\multicolumn{7}{c}{$\tau_W$}\\%[-2.5mm]
		Method & Design & 0.005 & 0.05 & 0.10 & 0.15 & 0.2 & 0.3 & 0.5 \\ 
		\hline
		&&\multicolumn{7}{c}{{\it NoReg}}\\%[-2.5mm]
		\multirow{2}{*}{\it $TS_{\gamma-0.075}$} & CRE & 0.06 & 0.16 & 0.40 & 0.71 & 0.93 & 1.00 & 1.00 \\
		& ReM & 0.07 & 0.22 & 0.54 & 0.84 & 0.98 & 1.00 & 1.00 \\
		\multirow{2}{*}{\it $TS_{\gamma-0.025}$} & CRE & 0.02 & 0.08 & 0.24 & 0.53 & 0.81 & 1.00 & 1.00 \\
		& ReM & 0.02 & 0.10 & 0.33 & 0.68 & 0.92 & 1.00 & 1.00 \\
		\multirow{2}{*}{\it $TS_{F-10}$/$Wald_{F-10}$} & CRE & 0.00 & 0.01 & 0.04 & 0.15 & 0.40 & 0.95 & 1.00 \\
		& ReM & 0.00 & 0.00 & 0.02 & 0.11 & 0.39 & 0.98 & 1.00 \\
		&&\multicolumn{7}{c}{{\it Reg}}\\[-2.5mm]
		\multirow{2}{*}{\it $TS_{\gamma-0.075}$} & CRE & 0.07 & 0.22 & 0.54 & 0.84 & 0.98 & 1.00 & 1.00 \\
		& ReM & 0.07 & 0.22 & 0.54 & 0.85 & 0.98 & 1.00 & 1.00 \\
		\multirow{2}{*}{\it $TS_{\gamma-0.025}$} & CRE & 0.02 & 0.10 & 0.33 & 0.67 & 0.92 & 1.00 & 1.00 \\
		& ReM & 0.02 & 0.10 & 0.33 & 0.68 & 0.92 & 1.00 & 1.00 \\
		\multirow{2}{*}{\it $TS_{F-10}$/$Wald_{F-10}$} & CRE & 0.00 & 0.01 & 0.06 & 0.23 & 0.57 & 0.99 & 1.00 \\
		& ReM & 0.00 & 0.01 & 0.06 & 0.24 & 0.58 & 0.99 & 1.00 \\
		\hline
	\end{tabular}
\end{table}

Tables~\ref{tab:Len_NoReg} and~\ref{tab:Len_Reg_EHW} present the median length of 95\% confidence sets without and with regression adjustment. When $\tau_W=0.005$, {\it  Wald} has severe undercoverage; {\it $Wald_{F-10}$} also has severe undercoverage even if it only involves datasets judged to have strong instruments; the other methods do not have severe undercoverage, but their median interval length is all infinity.  We next focus on cases with $\tau_W\geq 0.05$.

{\it FAR} always has a larger median interval length than {\it Wald}. Each of the three two-stage procedures yields either the {\it Wald} confidence interval or the {\it FAR} confidence set, depending on the result of the corresponding first-stage test.  Therefore all of them have median interval length no smaller than that of {\it Wald} and smaller than that of {\it FAR}. When there is difference among the three two-stage procedures, {\it $TS_{\gamma-0.075}$} has the smallest median interval length, whereas {\it $TS_{F-10}$} has the largest median interval length. This is because according to Table~\ref{tab:SRate}, it is most likely for {\it $TS_{\gamma-0.075}$} to use the {\it Wald} confidence interval rather than the {\it FAR} confidence set, and mostly unlikely for {\it $TS_{F-10}$}. {\it $Wald_{F-10}$} appears to have much smaller median interval length than other methods when $0.05\leq \tau_W\leq 0.2$, but this is because it only involves datasets judged to have strong instruments by the first-stage $F$ test, whose proportion is no larger than 58\% according to Table~\ref{tab:SRate}. 

For {\it Wald}, compared to the case without regression adjustment under CRE, ReM or regression adjustment reduces the median interval length, but there is not much additional reduction to use both. For {\it FAR}, when $\tau_W\geq 0.15$, ReM greatly reduces median interval length only when no regression adjustment is made and $\tau_W\in\{0.15,0.2\}$, but regression adjustment greatly reduces median interval length under both CRE and ReM.  The three two-stage procedures exhibit mixtures of the above phenomena for {\it Wald} and {\it FAR}, with {\it $TS_{\gamma-0.075}$} most resembling {\it Wald} and {\it $TS_{F-10}$} most resembling {\it FAR}.

\begin{table}[htbp]
	\centering
	\caption{Median length of 95\% confidence sets without regression adjustment ($n=200$). If the coverage rate is smaller than 85\%, we report ``/"; if the coverage rate is larger than 85\% but smaller than 90\%, we mark the median length with underlining.}\label{tab:Len_NoReg}
	\begin{tabular}{rcrrrrrrr}
		\hline
		&&\multicolumn{7}{c}{$\tau_W$}\\%[-2.5mm]
		Method & Design & 0.005 & 0.05 & 0.10 & 0.15 & 0.2 & 0.3 & 0.5 \\ 
		\hline
		%&\multicolumn{8}{c}{No Regression Adjustment}\\[-2.5mm]
		\multirow{2}{*}{{\it Wald}} & CRE & / & 62.88 & 32.66 & 20.61 & 14.85 & 9.31 & 5.10 \\
		& ReM & / & 48.24 & 23.06 & 14.28 & 10.26 & 6.51 & 3.55 \\
		\multirow{2}{*}{{\it FAR}} & CRE & $\infty$ & $\infty$ & $\infty$ & 74.08 & 21.23 & 10.37 & 5.21 \\
		& ReM & $\infty$ & $\infty$ & $\infty$ & 54.47 & 19.80 & 10.07 & 5.06 \\
		\multirow{2}{*}{{\it $TS_{\gamma-0.075}$}} & CRE & $\infty$ & $\infty$ & $\infty$ & 20.61 & 14.85 & 9.31 & 5.10 \\
		& ReM & $\infty$ & $\infty$ & 23.99 & 14.28 & 10.26 & 6.51 & 3.55 \\
		\multirow{2}{*}{{\it $TS_{\gamma-0.025}$}} & CRE & $\infty$ & $\infty$ & $\infty$ & 22.39 & 14.85 & 9.31 & 5.10 \\
		& ReM & $\infty$ & $\infty$ & $\infty$ & 14.28 & 10.26 & 6.51 & 3.55 \\
		\multirow{2}{*}{{\it $TS_{F-10}$}} & CRE & $\infty$ & $\infty$ & $\infty$ & 74.08 & 20.98 & 9.31 & 5.10 \\
		& ReM & $\infty$ & $\infty$ & $\infty$ & 54.47 & 19.64 & 6.51 & 3.55 \\
		\multirow{2}{*}{{\it $Wald_{F-10}$}} & CRE & / & 11.98 & 12.23 & 11.88 & 11.38 & 9.19 & 5.10 \\
		& ReM & / & 9.27 & 9.16 & 8.69 & 8.27 & 6.48 & 3.55 \\
		\hline
	\end{tabular}
\end{table}

\begin{table}[htbp]
	\centering
	\caption{Median length of 95\% confidence sets with regression adjustment with EHW variance estimator ($n=200$). If the coverage rate is smaller than 85\%, we report ``/"; if the coverage rate is larger than 85\% but smaller than 90\%, we mark the median length with underlining.}\label{tab:Len_Reg_EHW}
	\begin{tabular}{rcrrrrrrr}
		\hline
		&&\multicolumn{7}{c}{$\tau_W$}\\%[-2.5mm]
		Method & Design & 0.005 & 0.05 & 0.10 & 0.15 & 0.2 & 0.3 & 0.5 \\ 
		\hline
		% &\multicolumn{8}{c}{Regression Adjustment (EHW)}\\[-2.5mm]
		\multirow{2}{*}{{\it Wald}}  & CRE & / & 48.25 & 23.10 & 14.32 & 10.28 & 6.52 & 3.56 \\
		& ReM & / & 47.65 & 22.71 & 14.10 & 10.17 & 6.45 & 3.52 \\
		\multirow{2}{*}{{\it FAR}} & CRE & $\infty$ & $\infty$ & $\infty$ & 23.90 & 12.73 & 7.02 & 3.61 \\
		& ReM & $\infty$ & $\infty$ & $\infty$ & 23.11 & 12.48 & 6.92 & 3.57 \\
		\multirow{2}{*}{{\it $TS_{\gamma-0.075}$}} & CRE & $\infty$ & $\infty$ & 23.90 & 14.32 & 10.28 & 6.52 & 3.56 \\
		& ReM & $\infty$ & $\infty$ & 23.40 & 14.10 & 10.17 & 6.45 & 3.52 \\
		\multirow{2}{*}{{\it $TS_{\gamma-0.025}$}} & CRE & $\infty$ & $\infty$ & $\infty$ & 14.32 & 10.28 & 6.52 & 3.56 \\
		& ReM & $\infty$ & $\infty$ & $\infty$ & 14.10 & 10.17 & 6.45 & 3.52 \\
		\multirow{2}{*}{{\it $TS_{F-10}$}} & CRE & $\infty$ & $\infty$ & $\infty$ & 23.90 & 10.33 & 6.52 & 3.56 \\
		& ReM & $\infty$ & $\infty$ & $\infty$ & 23.12 & 10.18 & 6.45 & 3.52 \\
		\multirow{2}{*}{{\it $Wald_{F-10}$}} & CRE & / & 10.49 & 10.21 & 9.64 & 8.84 & 6.51 & 3.56 \\
		& ReM & / & 10.49 & 10.21 & 9.63 & 8.84 & 6.44 & 3.52 \\
		\hline
	\end{tabular}
\end{table}

To summarize, it is not safe to always use {\it Wald}; also, it is not efficient to always use {\it FAR} since the confidence set could be overly wide. The two-stage procedures can combine the advantages of both {\it Wald} and {\it FAR}. Among them, it is recommended to use {\it $TS_{\gamma-0.075}$} as it is the most efficient without sacrificing much coverage. It is recommended to use regression adjustment in the analysis stage. There is not much additional benefit to use rerandomization in the design stage.

\section{Application to Voter Mobilization Experiments in \cite{green2003getting}}
\label{application}
\cite{green2003getting} presented randomized voter mobilization experiments in several cities prior to the US Election Day of 2001. The treatment group was assigned to a face-to-face contact from a coalition of nonpartisan student and community organizations, encouraging them to vote. Citizens in the control group would not be contacted, but some citizens in the treatment group could not be reached. Therefore the monotonicity assumption holds. The outcome of interest is whether a citizen voted in the 2001 election, and hence $\tau$ represents the average treatment effect on voter turnout rate among the compliers in the experimental units, measured in percentage points. Only citizens who were actually contacted would be affected by the treatment, and hence the exclusion restriction holds.

\cite{green2003getting} stratified subjects in Detroit, Minneapolis, and St. Paul into walk lists before random assignment. 
Following \cite{green2003getting}, we removed a few walk lists suspected with falsified records to obtain a dataset with 105 walk lists (with minimum $n=30$ and maximum $n=365$), and include two covariates for regression adjustment: voting in 2000 and family size.

\begin{figure}[htbp]
	\centering
	\includegraphics[height=5in]{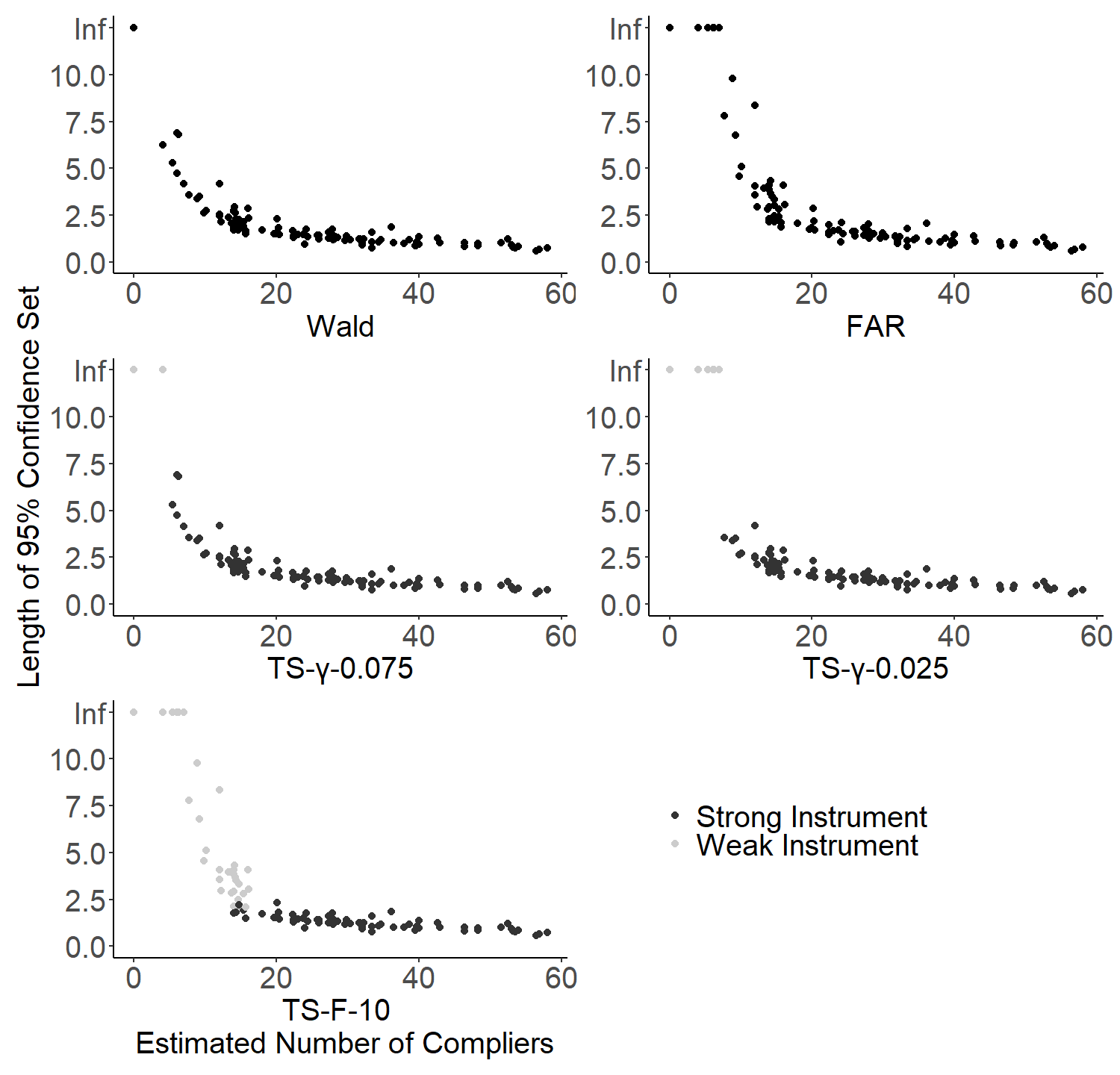}
	\caption{Analysis of the voter mobilization data without regression adjustment.}\label{fig:real_NoReg}
\end{figure}

\begin{figure}[htbp]
	\centering
	\includegraphics[height=5in]{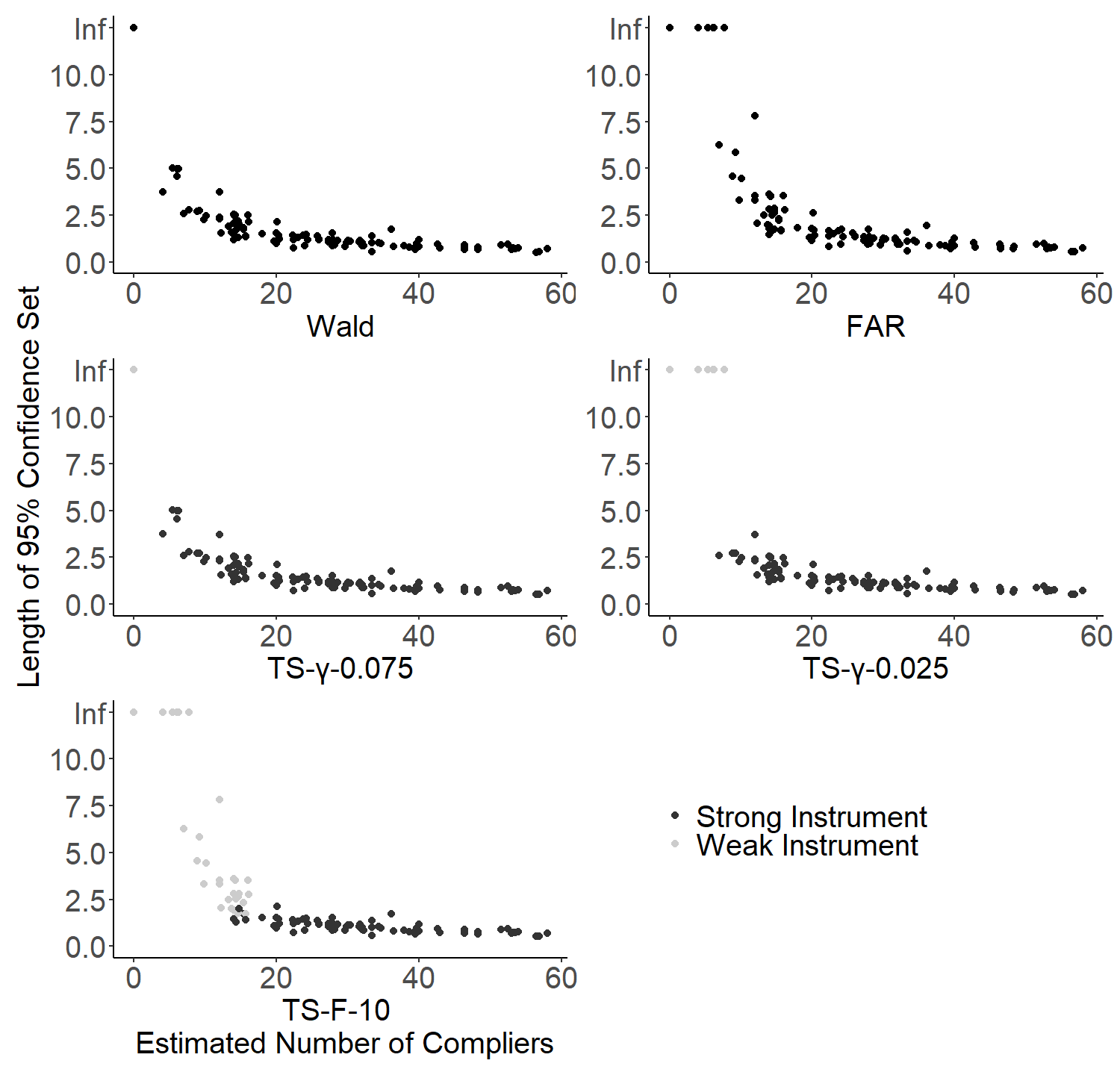}
	\caption{Analysis of the voter mobilization data with regression adjustment.}\label{fig:real_Reg}
\end{figure}

We estimated $\tau$ separately for each walk list using each of the methods in Section~\ref{sec:Monte Carlo}. Figure~\ref{fig:real_NoReg} plots the length of 95\% confidence set/interval against the estimated number of compliers among the experimental units, $n\widehat{\tau}_W$, when no regression adjustment is done.  Figure~\ref{fig:real_Reg} presents a similar plot when regression adjustment is done (with EHW variance estimator). For each two stage procedure, we use black points to indicate strong instruments and gray points to indicate weak instruments, judged by the corresponding first-stage test. 

For one of the walk lists, $\widehat{\tau}_W=0$, and the corresponding 95\% confidence set for {\it Wald} is infinity in both figures.  For the other walk lists, the length of 95\% confidence set for {\it Wald} is always smaller than that for {\it FAR}.  For each method, generally the larger the estimated number of compliers is, the shorter the length of 95\% confidence set is. When there is difference among the three two-stage approaches, {\it $TS_{\gamma-0.075}$} has the highest proportion of walk lists judged to have strong instruments, and the shortest length of 95\% confidence set; conversely, {\it $TS_{F-10}$} has the lowest proportion of strong instruments, and the longest length of 95\% confidence set. Regression adjustment generally reduces length of 95\% confidence set for each method.  

Since results of {\it $Wald_{F-10}$} correspond to those of {\it $TS_{F-10}$} judged to have strong instruments, we do not present a separate panel for {\it $Wald_{F-10}$} in each of the figures. It can be seen that {\it $Wald_{F-10}$} only yields relatively short confidence sets when the estimated number of compliers is relatively large. Therefore, only presenting results of {\it $Wald_{F-10}$} can be misleading.

\section{Discussion}
In pragmatic randomized experiments, incomplete adherence and/or incomplete compliance to the assigned treatment is common. This paper focuses on the experimental units at hand, and presents a two-stage procedure that has asymptotically correct coverage rate of the sample LATE. A first-stage test is used to judge whether the instrument of treatment assignment is strong. If the instrument is judged to be strong, the Wald method is used; if the instrument is judged to be weak, the FAR confidence set is used. The procedure is developed for the situation with and without regression adjustment and for two experimental designs (complete randomization and Mahalaonobis distance based rerandomization). 

Monte Carlo simulations and a real application support using the two-stage procedure with $\gamma=0.075$, and using regression adjustment in the analysis stage.  They also indicate that the typical approach adopted by practitioners to present results only when the first-stage $F$-statistic is larger than 10 can be highly misleading.

\bibliographystyle{apalike}
\bibliography{Refs}
\appendix
\newpage
\makeatletter
\renewcommand \thesection{S\@arabic\c@section}
\renewcommand\thetable{S\@arabic\c@table}
\renewcommand \thefigure{S\@arabic\c@figure}
\renewcommand \theequation{S.\@arabic\c@equation}
\renewcommand \thelemma{S\@arabic\c@lemma}
\makeatother
\setcounter{page}{1}
\setcounter{equation}{0}
\begin{center}
	{\LARGE\bf Supplementary Materials}
\end{center}
\section{Finding $R^{*2}_A$ for The FAR Confidence Set under ReM without Regression Adjustment}
\label{far_detail_rem}

Define 
\begin{align*} V_{Y\mid\bm{x}}=&\frac{S_{Y(1)\mid\bm{x}}^2}{n_1}+\frac{S_{Y(0)\mid\bm{x}}^2}{n_0}-\frac{\left(\boldsymbol{S}_{Y(1),\boldsymbol{x}}-\boldsymbol{S}_{Y(0),\boldsymbol{x}}\right)^{\top}\mathbb{S}_{\bm{x}\bm{x}}^{-1}\left(\boldsymbol{S}_{Y(1),\boldsymbol{x}}-\boldsymbol{S}_{Y(0),\boldsymbol{x}}\right)}{n},\\ V_{W\mid\bm{x}}=&\frac{S_{W(1)\mid\bm{x}}^2}{n_1}+\frac{S_{W(0)\mid\bm{x}}^2}{n_0}-\frac{\left(\boldsymbol{S}_{W(1),\boldsymbol{x}}-\boldsymbol{S}_{W(0),\boldsymbol{x}}\right)^{\top}\mathbb{S}_{\bm{x}\bm{x}}^{-1}\left(\boldsymbol{S}_{W(1),\boldsymbol{x}}-\boldsymbol{S}_{W(0),\boldsymbol{x}}\right)}{n},\\ C_{YW\mid\bm{x}}=&\frac{\boldsymbol{S}_{Y(1),\boldsymbol{x}}^{\top}\bm{S}_{\bm{x}\bm{x},1}^{-1}\boldsymbol{S}_{W(1),\boldsymbol{x}}}{n_1}+\frac{\boldsymbol{S}_{Y(0),\boldsymbol{x}}^{\top}\bm{S}_{\bm{x}\bm{x},0}^{-1}\boldsymbol{S}_{W(0),\boldsymbol{x}}}{n_0}\\	&-\frac{\left(\boldsymbol{S}_{Y(1),\boldsymbol{x}}-\boldsymbol{S}_{Y(0),\boldsymbol{x}}\right)^{\top}\mathbb{S}_{\bm{x}\bm{x}}^{-1}\left(\boldsymbol{S}_{W(1),\boldsymbol{x}}-\boldsymbol{S}_{W(0),\boldsymbol{x}}\right)}{n}.
\end{align*}
It is easy to see that $V_{A\mid\bm{x}}$ in~\eqref{def_VAx} can be written as 
\begin{equation}
V_{A\mid\bm{x}}=V_{Y\mid\bm{x}}-2\tau C_{YW\mid\bm{x}}+\tau^2V_{W\mid\bm{x}}.\label{def_VAx2}
\end{equation}

With \eqref{def_R2A}, \eqref{def_VA_ReM2} and \eqref{def_VAx2}, $R^2_A$ can be expressed as a function of $\tau$,
\begin{equation}
	\label{def_R2A_tau}
	R^2_{A}(\tau)=\frac{V_{Y\mid\bm{x}}-2\tau C_{YW\mid\bm{x}}+\tau^2V_{W\mid\bm{x}}}{V_Y^{ReM}-2\tau C_{YW}^{ReM}+\tau^2V_{W}^{ReM}}.
\end{equation}
To find the minimum value of this function, we need to compare the value of the function when $\tau$ is a solution to $\frac{d}{d\tau}R^2_{A}(\tau)=0$ and the value of the function when $\tau=\pm\infty$. From~\eqref{def_R2A_tau}, it is easy to see that $\frac{d}{d\tau}R^2_{A}(\tau)$ is a ratio with the numerator being a cubic polynomial of $\tau$ and the denominator being non-negative.  Hence there are at most three solutions to $\frac{d}{d\tau}R^2_{A}(\tau)=0$.  Also from~\eqref{def_R2A_tau}, as $\tau\rightarrow \pm\infty$, $R_A^2(\tau)$ converges to $V_{W|\bm{x}}/V_W^{ReM}$.  We can hence compare the values of $R_A^2(\tau)$ at the solutions to $\frac{d}{d\tau}R^2_{A}(\tau)=0$ and $V_{W|\bm{x}}/V_W^{ReM}$ to find $R^{*2}_A$, the global minimum of $R_A^2(\tau)$.

\section{Lemmas and Their Proofs}
We first state and prove a useful lemma that links conclusions under CRE with those under ReM.
\begin{lemma}
	\label{dominated_converge_in_p}
	For any sequence of random variables, $\{U_n\}_{n=1}^{\infty}$, that is $o_p(n^k)$ for some real number $k$ under CRE, it is also $o_p(n^k)$ under ReM as $n\rightarrow\infty$.
\end{lemma}
\textbf{Proof.} Since $\{U_n\}_{n=1}^{\infty}$ is $o_p(n^k)$ for some real number $k$ under CRE,  
for $\forall\epsilon>0$, $$\lim_{n\rightarrow\infty}\Pr\left(\left|\frac{U_n}{n^k}\right|\geq \epsilon\right)=0.$$ Under ReM, the final treatment assignment vector is in the set $\mathcal{A}_a(\bm{X})=\left\{\bm{Z}|M(\bm{Z},\bm{X})\leq a\right\}.$
Recall that $p_a$ is the limit of the probability of accepting a random allocation under ReM. We have
\begin{equation*}
	\begin{aligned}
		&\limsup_{n\rightarrow\infty}\Pr\left(\left.\left|\frac{U_n}{n^k}\right|\geq \epsilon\right|\bm{Z}\in\mathcal{A}_a(\bm{X})\right)\\
		=&\limsup_{n\rightarrow\infty}\Pr\left(\left|U_n\right|\geq n^{k}\epsilon\mid\bm{Z}\in\mathcal{A}_a(\bm{X})\right)\\
		=&\limsup_{n\rightarrow\infty}\frac{\Pr\left(\left|U_n\right|\geq n^{k}\epsilon\ \text{and}\ \bm{Z}\in\mathcal{A}_a(\bm{X})\right)}{\Pr\left(\bm{Z}\in\mathcal{A}_a(\bm{X})\right)}\\
		\leq&\limsup_{n\rightarrow\infty}\frac{\Pr(\left|U_n\right|\geq n^{k}\epsilon)}{p_a}\\
		=&\frac{1}{p_a}\lim_{n\rightarrow\infty}\Pr\left(\left|\frac{U_n}{n^k}\right|\geq \epsilon\right)\\
		=&0,
	\end{aligned}
\end{equation*}
which means $\{U_n\}_{n=1}^{\infty}$ is $o_p(n^k)$ under ReM.\hfill$\square$

The next lemma is about the continuity of the quantile function $\lambda_{\alpha/2}(\cdot)$.
\begin{lemma}
	\label{continuity_lambda}
	For $\rho\in[0,1]$, $\lambda_{\alpha/2}(\rho)$ is continuous in $\rho$.
\end{lemma}
\textbf{Proof.}
Consider a monotonic sequence $\rho_m$, $m=1,2,\cdots$, with limit $\rho_0$. Without loss of generality, assume that $\rho_m$ is non-increasing. It is known from Lemma A4 in \cite{li2018asymptotic} that $\lambda_{\alpha/2}(\rho)$ is a non-increasing function in $\rho$.  Hence $\lambda_{\alpha/2}(\rho_m)$ is a non-decreasing sequence, with 
\begin{equation}
	\label{non-increase1}
	\lambda_{\alpha/2}(\rho_m)\leq\lambda_{\alpha/2}(\rho_0)\text{ for }m=1,2,\cdots.
\end{equation}
Because $\epsilon_0\sim N(0,1)$ and the distribution of $L_{K,a}$ is symmetric around zero and more concentrated than $N(0,1)$, we can easily obtain that $\lambda_{\alpha/2}(\cdot)$ is bounded between 0 and $z_{\alpha/2}$.  Therefore $\lambda_{\alpha/2}(\rho_m)$ has a limit, denoted by $x_0$, and 
\begin{equation}
	\label{non-increase2}
	\lambda_{\alpha/2}(\rho_m)\leq x_0\text{ for } m=1,2,\cdots.
\end{equation}
Combining \eqref{non-increase1} with the statement that $x_0$ is the limit of $\lambda_{\alpha/2}(\rho_m)$, we have 
\begin{equation}
	\label{inequality_x0}
	x_0\leq\lambda_{\alpha/2}(\rho_0).
\end{equation}

Let $F_{\rho}(\cdot)$ denote the cumulative distribution function for $\sqrt{1-\rho} \cdot \varepsilon_0+\sqrt{\rho} \cdot L_{K, a}$. We have $F_{\rho}(\lambda_{\alpha/2}(\rho))=1-\alpha/2$. According to the theorem in \cite{scheffe1947useful}, we have \begin{equation}
	\label{scheffe}
	\lim_{m\rightarrow\infty}F_{\rho_m}(x)=F_{\rho_0}(x)
\end{equation}
for any real number $x$. Because $F_{\rho_m}(x)$ is non-decreasing in $x$, with \eqref{non-increase2} we
have $1-\alpha/2=F_{\rho_m}(\lambda_{\alpha/2}(\rho_m))\leq F_{\rho_m}(x_0)$, and hence $1-\alpha/2\leq\lim_{m\rightarrow\infty}F_{\rho_m}(x_0)$. Furthermore, applying \eqref{scheffe} with $x=x_0$, we have $\lim_{m\rightarrow\infty}F_{\rho_m}(x_0)=F_{\rho_0}(x_0)$, and hence $1-\alpha/2\leq F_{\rho_0}(x_0)$. On the other hand, due to \eqref{inequality_x0}, we have $F_{\rho_0}(x_0)\leq F_{\rho_0}(\lambda_{\alpha/2}(\rho_0))=1-\alpha/2$. Therefore, $F_{\rho_0}(x_0)=1-\alpha/2$ and $x_0=\lambda_{\alpha/2}(\rho_0)$. Hence the limit of $\lambda_{\alpha/2}(\rho_m)$ equals $\lambda_{\alpha/2}(\rho_0)$. 

As the monotonic sequence $\{\rho_m\}$ is chosen arbitrarily, we know that $\lambda_{\alpha/2}(\rho)$ is a continuous function of $\rho$. \hfill$\square$

The next lemma is about the consistency under CRE for point estimators of $\tau_W$, $\widehat{\tau}_W$ and $\widehat{\tau}_{W|\bm{x}}$.  Due to Lemma \ref{dominated_converge_in_p}, these estimators are also consistent under ReM.
\begin{lemma}
	\label{consistency_tauW}
	Under CRE, as $n\rightarrow\infty$, if $n_1/n$ has a limit in $(0,1)$, then $\widehat{\tau}_{W}-\tau_W=o_p(1)$. If in addition $\mathbb{S}_{\bm{x}\bm{x}}$ has a limit that is non-singular and $\lim_{n\rightarrow\infty}\max_{1\leq i\leq n}\Vert \bm{x}_i\Vert^2/n\rightarrow0$, then $\widehat{\tau}_{W\mid\bm{x}}-\tau_W=o_p(1)$.
\end{lemma}
\textbf{Proof.} Let $\mathbb{S}^2_{W(z)}$ ($z=0,1$) denote the finite population variance of $W_i(z)$. Because $W_i(z)$ is binary, we have
\begin{equation*}
	\mathbb{S}_{W(z)}^2=\frac{1}{n-1}\sum_{i=1}^n\left(W_i(z)-\frac{1}{n}\sum_{i=1}^n W_i(z)\right)^2\leq\frac{n}{n-1}.
\end{equation*}
Let $\mathbb{S}^2_{W(1)-W(0)}$ denote the finite population variance of $W_i(1)-W_i(0)$. Applying Theorem 3 in \cite{li2017general}, under CRE, $\widehat{\tau}_W$ has mean $\tau_W$ and variance $\mathbb{V}_{W}=\mathbb{S}^2_{W(1)}/n_1+\mathbb{S}^2_{W(0)}/n_0-\mathbb{S}^2_{W(1)-W(0)}/n\leq\mathbb{S}^2_{W}(1)/n_1+\mathbb{S}^2_{W}(0)/n_0$.
Since $n_1/n$ has a limit in $(0,1)$, we have $\lim_{n\rightarrow\infty} 1/n_1=\lim_{n\rightarrow\infty}1/n_0=0$. By Chebyshev's inequality, for any $\epsilon>0$,
\begin{align*}
	&\lim_{n\rightarrow\infty}\Pr\left(|\widehat{\tau}_W-\tau_W|\geq\epsilon\right)\\ \leq&\lim_{n\rightarrow\infty}\frac{\mathbb{V}_{W}}{\epsilon^2}\leq\lim_{n\rightarrow\infty}\frac{1}{\epsilon^2}\left(\frac{\mathbb{S}^2_{W(1)}}{n_1}+\frac{\mathbb{S}^2_{W(0)}}{n_0}\right)=0,
\end{align*}
i.e., $\widehat{\tau}_W-\tau_W=o_p(1)$.

Let $S_{W(z)}^2$ ($z=0,1$) denote the sample variance of $W_i$ for those with $Z_i=z$. Because $W_i$ is binary, we have
\begin{equation}
	S_{W(z)}^2=\frac{1}{n_z-1}\sum_{i:\ Z_i=z}(W_i-\overline{W}_z)^2\leq\frac{n_z}{n_z-1}.\label{S_Wz_leq}
\end{equation}
For $z=0,1$, let $\bm{S}_{\bm{x}\bm{x},z}=\bm{C}_z^{\top}\bm{C}_z$ be the Cholesky decomposition of $\bm{S}_{\bm{x}\bm{x},z}$. Denote $\widehat{\bm{\beta}}_{W,z}$ as the vector of coefficients on $\bm{x}$ in the OLS regression of $W_i$ on $\bm{x}_i$ using units with $Z_i=z$. For units with $Z_i=z$, the sample variance of the linear projection $\widehat{\bm{\beta}}_{W,z}^{\top}\bm{x}_i$, which equals $\widehat{\bm{\beta}}_{W,z}^{\top}\bm{S}_{\bm{x}\bm{x},z}\widehat{\bm{\beta}}_{W,z}=\widehat{\bm{\beta}}_{W,z}^{\top}\bm{C}_z^{\top}\bm{C}_z\widehat{\bm{\beta}}_{W,z}$, is no larger than the sample variance of $W_i$, $S^2_{W(z)}$. Hence
\begin{equation}
	\label{bound_on_projection}
	\begin{aligned} &\left(\widehat{\bm{\beta}}_{W,z}^{\top}\overline{\bm{x}}_z\right)^2=\left(\left(\widehat{\bm{\beta}}_{W,z}^{\top}\bm{C}_z^{\top}\right)\left(\bm{C}_z\bm{S}_{\bm{x}\bm{x},z}^{-1}\overline{\bm{x}}_z\right)\right)^2\\
		\leq&\left(\widehat{\bm{\beta}}_{W,z}^{\top}\bm{C}_z^{\top}\bm{C}_z\widehat{\bm{\beta}}_{W,z}\right)\left(\overline{\bm{x}}_z^{\top}\bm{S}_{\bm{x}\bm{x},z}^{-1}\bm{C}_z^{\top}\bm{C}_z\bm{S}_{\bm{x}\bm{x},z}^{-1}\overline{\bm{x}}_z\right)\\
		\leq&\ S^2_{W(z)}\left(\overline{\bm{x}}_z^{\top}\bm{S}_{\bm{x}\bm{x},z}^{-1}\overline{\bm{x}}_z\right)\leq\frac{n_z}{n_z-1}\overline{\bm{x}}_z^{\top}\bm{S}_{\bm{x}\bm{x},z}^{-1}\overline{\bm{x}}_z,
	\end{aligned}
\end{equation}
where the first inequality is obtained using Cauchy-Schwarz inequality and the last is obtained using \eqref{S_Wz_leq}.

Let the $k$th covariate of unit $i$ be $x_i^{(k)}$, let $\overline{x}_z^{(k)}$ ($z=0,1$) denote the mean of $x_i^{(k)}$ for units with $Z_i=z$, and let $\mathbb{S}_{\bm{x}\bm{x}}^{(k,k)}$ denote the $k$th diagonal element of $\mathbb{S}_{\bm{x}\bm{x}}$, that is, the finite population variance of $x_i^{(k)}$.  We have
\begin{equation}
	\label{eq:lindeberg-feller-x}
	\begin{aligned}
		\frac{\max_{1\leq i\leq n}\left(x_i^{(k)}\right)^2}{\min(n_1,n_0)\mathbb{S}_{\bm{x}\bm{x}}^{(k,k)}}&=\frac{\max_{1\leq i\leq n}\left(x_i^{(k)}\right)^2}{n}\cdot\frac{n}{\min(n_1,n_0)}\cdot\frac{1}{\mathbb{S}_{\bm{x}\bm{x}}^{(k,k)}}\\
		&\leq \frac{\max_{1\leq i\leq n}\left\|\bm{x}_{i}\right\|^2}{n}\cdot\frac{n}{\min(n_1,n_0)}\cdot\frac{1}{\mathbb{S}_{\bm{x}\bm{x}}^{(k,k)}}\rightarrow 0,
	\end{aligned}
\end{equation}
where the limit is due to the assumptions of the lemma. Applying Theorem 1 of \cite{li2017general}, under \eqref{eq:lindeberg-feller-x},  $\overline{x}_z^{(k)}/\sqrt{(1/n_z-1/n)\mathbb{S}_{\bm{x}\bm{x}}^{(k,k)}}$ converges to a standard normal distribution under CRE. As $n\rightarrow\infty$, because $\sqrt{(1/n_z-1/n)\mathbb{S}_{\bm{x}\bm{x}}^{(k,k)}}$ converges to $0$ according to the assumptions of the lemma, we obtain that $\overline{x}_z^{(k)}=o_p(1)$.

With the fact that $\bm{x}_i$ can be regarded as potential values of the covariates under both $Z_i=0$ and $Z_i=1$,
applying Proposition 3 of \cite{li2017general}, we have $\bm{S}_{\bm{x}\bm{x},z}-\mathbb{S}_{\bm{x}\bm{x}}=o_p(1)$, where $\mathbb{S}_{\bm{x}\bm{x}}$ has a finite non-singular limiting value. Therefore, applying Slusky's theorem together with the continuous mapping theorem to the right hand side of \eqref{bound_on_projection}, we conclude that 
\begin{equation}
\label{beta_Wz_x}
\widehat{\bm{\beta}}_{W,z}^{\top}\overline{\bm{x}}_z=o_p(1).
\end{equation}

Recall that for any set of quantities $\{Q_i:\ i=1,\cdots,n\}$, $\widehat{\tau}_{Q|\bm{x}}$ is the estimated coefficient on $Z_i$ in the OLS regression of $Q_i$ on $Z_i$, $\bm{x}_i$ and $Z_i\bm{x}_i$. As \cite{lin2013agnostic} shows, 
\begin{equation*}
	\widehat{\tau}_{Q\mid\bm{x}}=\frac{1}{n_1}\sum_{i:\ Z_i=1}\left(Q_i-\widehat{\bm{\beta}}_{Q,1}^{\top}\bm{x}_i\right)-\frac{1}{n_0}\sum_{i:\ Z_i=0}\left(Q_i-\widehat{\bm{\beta}}_{Q,0}^{\top}\bm{x}_i\right),
\end{equation*}
where $\widehat{\bm{\beta}}_{Q,z}$ is the vector of coefficients on $\bm{x}_i$ in the OLS regression of $Q_i$ on $\boldsymbol{x}_i$ for units with $Z_i=z$. Therefore
\begin{equation}
\begin{aligned}
	\label{ITThat_W_adj}
	\widehat{\tau}_{W\mid\boldsymbol{x}}&=\left(\frac{1}{n_1}\sum_{i:\ Z_i=1}W_i-\frac{1}{n_0}\sum_{i:\ Z_i=0}W_i\right)-\frac{1}{n_1}\sum_{i:\ Z_i=1}\left(\widehat{\bm{\beta}}_{W,1}^{\top}\bm{x}_i\right)+\frac{1}{n_0}\sum_{i:\ Z_i=0}\left(\widehat{\bm{\beta}}_{W,0}^{\top}\bm{x}_i\right)\\
&=\widehat{\tau}_{W}-\widehat{\bm{\beta}}_{W,1}^{\top}\overline{\bm{x}}_1+\widehat{\bm{\beta}}_{W,0}^{\top}\overline{\bm{x}}_0.
\end{aligned}
\end{equation}
From \eqref{ITThat_W_adj} and \eqref{beta_Wz_x}, we know that $\widehat{\tau}_{W\mid\boldsymbol{x}}-\tau_W=(\widehat{\tau}_{W}-\tau_W)-\widehat{\bm{\beta}}_{W,1}^{\top}\overline{\bm{x}}_1+\widehat{\bm{\beta}}_{W,0}^{\top}\overline{\bm{x}}_0=o_p(1)+o_p(1)+o_p(1)=o_p(1)$. \hfill$\square$

The next lemma says that Condition \ref{strict_cond} implies Condition \ref{lindeberg-feller}, meaning that conclusions obtained under Condition~\ref{lindeberg-feller} automatically hold under Condition~\ref{strict_cond}.
\begin{lemma}
\label{conditions_imply}
If Condition \ref{strict_cond} holds, then Condition \ref{lindeberg-feller} holds.
\end{lemma}
\textbf{Proof.}
First, we have
\begin{equation*}
	\begin{aligned}
		&\frac{\max_{1 \leq i \leq n}\left(A_{i}(z)-\overline{A}(z)\right)^2}{\min(n_1, n_0)^2\cdot \mathbb{V}_A}\\
		=&\left(\frac{n}{\min(n_1, n_0)}\right)^2\cdot\frac{\max_{1\leq i\leq n}\left|A_i(z)-\overline{A}(z)\right|^{2}/n}{n/n_1\cdot\mathbb{S}_{A(1)}+n/n_0\cdot\mathbb{S}_{A(0)}-\mathbb{S}_{A(1)-A(0)}}.
	\end{aligned}
\end{equation*}
Under part (1) of Condition~\ref{strict_cond}, as $n\rightarrow\infty$, $n/\min(n_1,n_0)$, $n/n_1$ and $n/n_0$ all have a finite limit.  This together with parts (2) and (3) of Condition \ref{strict_cond} imply part (3) of Condition \ref{lindeberg-feller}. 

Parts (1) of Condition~\ref{strict_cond} and Condition~\ref{lindeberg-feller} are the same. Combining the definition of $\mathbb{V}_A$ in~\eqref{def_bV_A} and parts (1) and (2) of Condition \ref{strict_cond}, we get part (2) of Condition~\ref{lindeberg-feller}.

So we conclude that Condition \ref{strict_cond} implies Condition \ref{lindeberg-feller}. \hfill$\square$

\section{Proofs of Theorems and Propositions}

\noindent\textbf{Proof of Theorem \ref{Wald_CRE}.}
Since $\tau_W$ has a positive limit inferior under Condition \ref{strength}, we know from Lemma \ref{consistency_tauW} that $\widehat{\tau}_W/\tau_W-1=o_p(1)$. 
Under Assumption \ref{assu} and Condition \ref{lindeberg-feller}, we can apply Theorem 4 of \cite{li2017general} to obtain that $\widehat{\tau}_A/\sqrt{\mathbb{V}_A}\stackrel{d}{\rightarrow}N(0,1)$ under CRE (recall that $\tau_{A}=0$). Since $\widehat{\tau}_{Wald}=\widehat{\tau}_Y/\widehat{\tau}_W$ and $\widehat{\tau}_A=\widehat{\tau}_Y-\tau\widehat{\tau}_W$, we have
\begin{equation}
	\widehat{\tau}_{Wald}-\tau=\frac{\widehat{\tau}_A}{\widehat{\tau}_W}.\label{diff_Wald}
\end{equation}
It follows from Slutsky's theorem that, as $n\rightarrow\infty$,
\begin{equation*}
	\frac{\widehat{\tau}_{Wald}-\tau}{\sqrt{\mathbb{V}_A}/\tau_W}=\frac{\widehat{\tau}_A}{\sqrt{\mathbb{V}_A}}\frac{\tau_W}{\widehat{\tau}_W}\stackrel{d}{\rightarrow}N(0,1).
\end{equation*}
Since $n\mathbb{V}_A$ has a finite (positive) limit superior under Condition \ref{lindeberg-feller} and $\tau_W$ has a positive limit inferior under Condition \ref{strength}, it follows that $\widehat{\tau}_{Wald}-\tau=o_p(1)$.
\hfill$\square$

\noindent\textbf{Proof of Theorem \ref{conservative_ci_CRE}.}
Recall that 
\begin{equation*}
	\label{estimated_SA}
	\widehat{V}_A=\frac{S_{\widehat{A}(1)}^2}{n_1}+\frac{S_{\widehat{A}(0)}^2}{n_0},
\end{equation*}
where $S_{\widehat{A}(z)}^2$ is the sample variance of $\widehat{A}_i=Y_i-\widehat{\tau}_{Wald}W_i$ for units with $Z_i=z$ ($z=0,1$). Define
\begin{equation*}
	\label{SA}
	V_A=\frac{S_{A(1)}^2}{n_1}+\frac{S_{A(0)}^2}{n_0}.
\end{equation*}
We first prove that, with $A_i$ well defined under Assumption \ref{assu}, $|\widehat{V}_A-V_{A}|/\mathbb{V}_A=o_p\left(1\right)$ as $n\rightarrow\infty$.

According to the expressions of $A_i$ and $\widehat{A}_i$,  $\widehat{A}_i=A_i-\left(\widehat{\tau}_{Wald}-\tau\right)W_i$. We have
\begin{equation}
	\label{diff_in_sampvar}
	S_{\widehat{A}(z)}^2=S_{A(z)}^2-2\left(\widehat{\tau}_{Wald}-\tau\right)S_{A(z),W(z)}+\left(\widehat{\tau}_{Wald}-\tau\right)^2S_{W(z)}^2
\end{equation}
for $z=0,1$.
The absolute difference between $\widehat{V}_A$ and $V_{A}$ can be bounded as follows:
\begin{equation}
	\begin{aligned}
		\left|\widehat{V}_A-V_{A}\right|
		=&\left|\frac{-2\left(\widehat{\tau}_{Wald}-\tau\right)S_{A(1),W(1)}
			+\left(\widehat{\tau}_{Wald}-\tau\right)^2S^2_{W(1)}}{n_1}\right.\\
		&\left.+\frac{-2\left(\widehat{\tau}_{Wald}-\tau\right)S_{A(0),W(0)}
			+\left(\widehat{\tau}_{Wald}-\tau\right)^2S^2_{W(0)}}{n_0}\right|\\ \leq&\frac{2\left|\widehat{\tau}_{Wald}-\tau\right|\left|S_{A(1),W(1)}\right|+\left(\widehat{\tau}_{Wald}-\tau\right)^2S^2_{W(1)}}{n_1}\\
		&+\frac{2\left|\widehat{\tau}_{Wald}-\tau\right|\left|S_{A(0),W(0)}\right|+
			\left(\widehat{\tau}_{Wald}-\tau\right)^2S^2_{W(0)}}{n_0}.
	\end{aligned}\label{abs_diff}
\end{equation}
Because $W_i$ is binary, we have
\begin{equation}
	\begin{aligned}
		\left|S_{A(z), W(z)}\right|&=\left|\frac{1}{n_z-1}\sum_{i:\ Z_i=z}(A_i-\overline{A}_z)(W_i-\overline{W}_z)\right|\leq\frac{n_z}{n_z-1}\max_{i:Z_i=z}\left|A_{i}(z)-\overline{A}_z\right|\\
		&\leq\frac{n_z}{n_z-1}\left(\max_{i:Z_i=z}\left|A_{i}(z)-\overline{A}(z)\right|+\left|\overline{A}(z)-\frac{1}{n_z}\sum_{i:\ Z_i=z}A_i(z)\right|\right)\\
		&\leq\frac{n_z}{n_z-1}\left(\max_{i:Z_i=z}\left|A_{i}(z)-\overline{A}(z)\right|+\frac{1}{n_z}\sum_{i:\ Z_{i}=z}\left|\overline{A}(z)-A_i(z)\right|\right)\\
		&\leq\frac{2n_z}{n_z-1}\max_{i:Z_i=z}\left|A_{i}(z)-\overline{A}(z)\right|\leq\frac{2n_z}{n_z-1}\max_{1\leq{}i\leq{}n}\left|A_{i}(z)-\overline{A}(z)\right|\label{S_AzWz_leq}
	\end{aligned}
\end{equation}
for $z=0,1$. Plugging \eqref{S_Wz_leq} and \eqref{S_AzWz_leq} into \eqref{abs_diff}, we have
\begin{equation}
	\begin{aligned}
		\left|\widehat{V}_A-V_{A}\right|
		\leq&\frac{4\left|\widehat{\tau}_{Wald}-\tau\right|\max_{1 \leq i \leq n}\left|A_{i}(1)-\overline{A}(1)\right|}{n_1-1}+\frac{\left(\widehat{\tau}_{Wald}-\tau\right)^2}{n_1-1}\\
		&+\frac{4\left|\widehat{\tau}_{Wald}-\tau\right|\max_{1 \leq i \leq n}\left|A_{i}(0)-\overline{A}(0)\right|}{n_0-1}+\frac{\left(\widehat{\tau}_{Wald}-\tau\right)^2}{n_0-1}.
	\end{aligned}\label{abs_diff_VA}
\end{equation}

Dividing both sides of \eqref{abs_diff_VA} by $\mathbb{V}_{A}$, we have
\begin{equation*}
	\begin{aligned}
		\left|\widehat{V}_A-V_{A}\right|/\mathbb{V}_{A}
		\leq&\frac{4\min(n_1,n_0)}{\left(n_1-1\right)\tau_W}\frac{\left|\widehat{\tau}_{Wald}-\tau\right|}{\sqrt{\mathbb{V}_{A}}/\tau_W}\frac{\max_{1 \leq i \leq n}\left|A_{i}(1)-\overline{A}(1)\right|}{\min(n_1, n_0)\sqrt{\mathbb{V}_{A}}}\\
		&+\frac{4\min(n_1,n_0)}{\left(n_0-1\right)\tau_W}\frac{\left|\widehat{\tau}_{Wald}-\tau\right|}{\sqrt{\mathbb{V}_{A}}/\tau_W}\frac{\max_{1 \leq i \leq n}\left|A_{i}(0)-\overline{A}(0)\right|}{\min(n_1, n_0)\sqrt{\mathbb{V}_{A}}}\\	&+\frac{1}{(n_1-1)\tau_W^2}\frac{\left(\widehat{\tau}_{Wald}-\tau\right)^2}{\mathbb{V}_A/\tau_W^2}+\frac{1}{(n_0-1)\tau_W^2}\frac{\left(\widehat{\tau}_{Wald}-\tau\right)^2}{\mathbb{V}_A/\tau_W^2}.
	\end{aligned}
\end{equation*}
It then follows from Condition \ref{strength}, \ref{lindeberg-feller}, the second part of Theorem \ref{Wald_CRE} and Slutsky's theorem that, as $n\rightarrow \infty$, $|\widehat{V}_A-V_{A}|/\mathbb{V}_A\stackrel{p}{\longrightarrow}0$.

Define 
\begin{equation}
\mathbb{V}_A^+=\frac{\mathbb{S}_{A(1)}^2}{n_1}+\frac{\mathbb{S}_{A(0)}^2}{n_0}.\label{def_VAstar}
\end{equation}
Because $\mathbb{V}_A^+\geq\mathbb{V}_A$ by definition, we know that $|\widehat{V}_A-V_{A}|/\mathbb{V}_A^+\stackrel{p}{\longrightarrow}0$. Under Condition \ref{lindeberg-feller} and CRE, we can apply Proposition 1 of \cite{li2017general} to have $V_{A}/\mathbb{V}_A^+\stackrel{p}{\longrightarrow}1$. It follows that $\widehat{V}_A/\mathbb{V}_A^+$ $\stackrel{p}{\longrightarrow}1$. Therefore, $\widehat{V}_A$ is an asymptotically conservative estimator of $\mathbb{V}_{A}$. 

Recall that under Condition~\ref{lindeberg-feller}, $\widehat{\tau}_A$ converges to a normal distribution with mean 0 and variance $\mathbb{V}_A$.  It follows that, as $n\rightarrow\infty$, $\Pr\left(\widehat{\tau}_A^2\leq z_{\alpha/2}^2\mathbb{V}_A\right)\rightarrow 1-\alpha$. With \eqref{def_tauhat_A}, an asymptotic conservative $1-\alpha$ confidence set for $\tau$ is the solution to the following inequality:
\begin{equation*}
	\left(\widehat{\tau}_Y-\tau\widehat{\tau}_W\right)^2\leq z_{\alpha/2}^2\widehat{V}_A=z_{\alpha/2}^2\left(V_Y-2\widehat{\tau}_{Wald}C_{YW}+\widehat{\tau}_{Wald}^2V_W\right).
\end{equation*}
Solving the inequality by considering both cases of $\widehat{\tau}_W>0$ and $\widehat{\tau}_W<0$, we obtain \eqref{ci_CRE}, which completes the proof.
\hfill$\square$

\noindent\textbf{Proposition S1:} Under Assumption \ref{assu}, Conditions \ref{strength}, \ref{lindeberg-feller} and CRE, the interval estimator in \eqref{ci_CRE} is the same as the super-population
confidence interval for the population LATE obtained by the delta method.

\noindent\textbf{Proof of Proposition S1:}
According to Section 23.7 in \cite{imbens2015causal}, under Assumption \ref{assu}, Condition \ref{strength}, \ref{lindeberg-feller} and CRE and applying the delta method, in large samples, $\widehat{\tau}_{Wald}$, as an estimator of the population LATE, is approximately normally distributed with super-population sampling variance
\begin{equation}
\mathbb{V}_{sp}\left(\widehat{\tau}_{Wald}\right)=\frac{1}{\tau_W^2}\cdot \left[\mathbb{V}_{sp}\left(\widehat{\tau}_Y\right)-2\cdot\frac{\tau_Y}{\tau_W}\cdot \mathbb{C}_{sp}\left(\widehat{\tau}_Y,\widehat{\tau}_W\right)+\left(\frac{\tau_Y}{\tau_W}\right)^2\mathbb{V}_{sp}\left(\widehat{\tau}_W\right)\right],
\end{equation}
where $\mathbb{V}_{sp}(\cdot)$ denotes the super-population variance of a random variable, and $\mathbb{C}_{sp}(\cdot,\cdot)$ denotes the super-population covariance of two random variables. A simple estimator for the super-population sampling variance can be based on substituting estimates for the components, that is, substituting $\widehat{\tau}_W$ for $\tau_W$, $\widehat{\tau}_{Wald}$ for $\tau_Y/\tau_W$, $V_Y$ for $\mathbb{V}_{sp}\left(\widehat{\tau}_Y\right)$, $C_{YW}$ for $\mathbb{C}_{sp}\left(\widehat{\tau}_Y,\widehat{\tau}_W\right)$, and
$V_W$ for $\mathbb{V}_{sp}\left(\widehat{\tau}_W\right)$. This is equivalent to estimating the super-population sampling standard deviation by $\sqrt{V_Y-2\widehat{\tau}_{Wald}C_{YW}+\widehat{\tau}_{Wald}^2 V_W}/|\widehat{\tau}_W|$.
 
Therefore the interval estimator in \eqref{ci_CRE} is the same as the super-population
confidence interval for the population LATE obtained by the delta method. \hfill$\square$

\noindent\textbf{Proof of Theorem \ref{confidence_set_mix}.} 

\noindent\textbf{Proof of Part (a).} The first part of Lemma \ref{consistency_tauW} implies that 
\begin{equation*}
	\lim_{n\rightarrow\infty}\Pr\left(\widehat{\tau}_W<p_{+}\mid \tau_W<p_{+}/2\right)=1.
\end{equation*}
We thus have
\begin{equation}
	\label{conservative_cond1}
	\begin{aligned}
		&\liminf_{n\rightarrow\infty}\Pr\left(\tau\in\widehat{I}_{CRE}\mid \tau_W<p_{+}/2\right)\\
		\geq&\liminf_{n\rightarrow\infty}\Pr\left(\tau\in\widehat{I}_{CRE}\mid\widehat{\tau}_W<p_{+},\tau_W<p_{+}/2\right)\lim_{n\rightarrow\infty}\Pr\left(\widehat{\tau}_W<p_{+}\mid \tau_W<p_{+}/2\right)\\
=&\liminf_{n\rightarrow\infty}\Pr\left(\tau\in\widehat{I}_{CRE}\mid\widehat{\tau}_W<p_{+},\tau_W<p_{+}/2\right)\\
		\geq&1-\alpha.
	\end{aligned}
\end{equation}
The last inequality in~\eqref{conservative_cond1} is because that, given $\widehat{\tau}_W<p_{+}$, we have $\widehat{\tau}_W\leq p_{+}+z_{\gamma}\sqrt{V_W}$ or equivalently $T\leq z_{\gamma}$, and hence $\widehat{I}_{CRE}$ equals the FAR confidence set \eqref{confidence_set_cre} which is asymptotically conservative according to the arguments in Section \ref{sec:FAR_CRE_Noreg}. 

For the cases with $\tau_W\geq p_{+}/2$, Condition~\ref{strength} holds. We have shown in the proof of Theorem \ref{conservative_ci_CRE} that, as $n\rightarrow\infty$, $V_{A}/\mathbb{V}_A^+\stackrel{p}{\longrightarrow}1$ and $\widehat{V}_A/\mathbb{V}_A^+\stackrel{p}{\longrightarrow}1$. Define
\begin{equation*}
	\widetilde{V}_A=\begin{cases}
		\widehat{V}_A, \text{ if } T>z_{\gamma},\\
		V_{A}, \text{ if } T\leq z_{\gamma}.
	\end{cases}
\end{equation*}
It follows that $\widetilde{V}_A/\mathbb{V}_A^+\stackrel{p}{\longrightarrow}1$. Because $\mathbb{V}_A^+\geq\mathbb{V}_A$,we have
\begin{equation*}
	\liminf_{n\rightarrow\infty}\left(\Pr\left(\widehat{\tau}_A^2\leq z_{\alpha/2}^2\widetilde{V}_A\mid \tau_W\geq p_{+}/2\right)-\Pr\left(\widehat{\tau}_A^2\leq z_{\alpha/2}^2\mathbb{V}_A\mid \tau_W\geq p_{+}/2\right)\right)\geq0,
\end{equation*}
Because $\widehat{\tau}_A/\sqrt{\mathbb{V}_A}$ converges to $N(0,1)$, $\lim_{n\rightarrow\infty}\Pr\left(\widehat{\tau}_A^2\leq z_{\alpha/2}^2\mathbb{V}_A\mid \tau_W\geq p_{+}/2\right)=1-\alpha$. It is easy to see that the solution to $\widehat{\tau}_A^2\leq z_{\alpha/2}^2\widetilde{V}_A$ equals $\widehat{I}_{CRE}$, and hence
\begin{equation}
	\label{conservative_cond2}
	\liminf_{n\rightarrow\infty}\Pr\left(\tau\in\widehat{I}_{CRE}\mid \tau_W\geq p_{+}/2\right)\geq1-\alpha.
\end{equation}
It follows from \eqref{conservative_cond1}, \eqref{conservative_cond2} and the law of total probability that $\liminf_{n\rightarrow\infty}\Pr\left(\tau\in\widehat{I}_{CRE}\right)\geq1-\alpha$ under Condition \ref{lindeberg-feller} and CRE.

\noindent\textbf{Proof of Part (b).} Applying Fieller's theorem (\cite{finney1978statistical}, Chapter 4), when \eqref{confidence_set_cre} is not empty, the upper and lower boundaries of \eqref{confidence_set_cre} are
\begin{equation*}
	\label{ci_CRE3}
	\left[\widehat{\tau}_{Wald}-\frac{gC_{YW}}{V_W}\pm \frac{z_{\alpha/2}}{\left|\widehat{\tau}_{W}\right|}\sqrt{V_Y+\widehat{\tau}_{Wald}^2V_W-2\widehat{\tau}_{Wald}C_{YW}-g\left(V_Y-\frac{C_{YW}^2}{V_W}\right)}\right]/(1-g),
\end{equation*}
where $g=z_{\alpha/2}^2V_W/\widehat{\tau}_W^2$. As noted by \cite{finney1978statistical}, when $g\geq1$, the range of $\tau$ satisfying \eqref{confidence_set_cre} is infinite. So we can focus on the case where $g<1$.

Given $g<1$, the length of \eqref{confidence_set_cre} satisfies
\begin{equation}
	\begin{aligned}	&\frac{2z_{\alpha/2}}{\left|\widehat{\tau}_{W}\right|(1-g)}\sqrt{V_Y+\widehat{\tau}_{Wald}^2V_W-2\widehat{\tau}_{Wald}C_{YW}-g\left(V_Y-\frac{C_{YW}^2}{V_W}\right)}\\ =&\frac{2z_{\alpha/2}}{\left|\widehat{\tau}_{W}\right|(1-g)}\sqrt{(1-g)\left(V_Y+\widehat{\tau}_{Wald}^2V_W-2\widehat{\tau}_{Wald}C_{YW}\right)+g\left(\widehat{\tau}_{Wald}^2V_W-2\widehat{\tau}_{Wald}C_{YW}+\frac{C_{YW}^2}{V_W}\right)}\\ =&\frac{2z_{\alpha/2}}{\left|\widehat{\tau}_{W}\right|(1-g)}\sqrt{(1-g)\left(V_Y+\widehat{\tau}_{Wald}^2V_W-2\widehat{\tau}_{Wald}C_{YW}\right)+\frac{g}{V_W}\left(\widehat{\tau}_{Wald}V_W-C_{YW}\right)^2}\\ \geq&\frac{2z_{\alpha/2}}{\left|\widehat{\tau}_{W}\right|(1-g)}\sqrt{(1-g)\left(V_Y+\widehat{\tau}_{Wald}^2V_W-2\widehat{\tau}_{Wald}C_{YW}\right)}\\		>&\frac{2z_{\alpha/2}}{\left|\widehat{\tau}_{W}\right|}\sqrt{V_Y+\widehat{\tau}_{Wald}^2V_W-2\widehat{\tau}_{Wald}C_{YW}}.
	\end{aligned}\label{length_ineq}
\end{equation}
The last line in~\eqref{length_ineq} equals the length of \eqref{ci_CRE}. This completes the proof. \hfill$\square$

\noindent\textbf{Proof of Theorem \ref{Wald_ReM}.}
We have shown in the proof of Theorem \ref{Wald_CRE} that $\widehat{\tau}_W/\tau_W\stackrel{p}{\rightarrow}1$ under CRE. It follows from Lemma \ref{dominated_converge_in_p} that $\widehat{\tau}_W/\tau_W\stackrel{p}{\rightarrow}1$ also under ReM. On the other hand, as $n\rightarrow\infty$, the distribution of $\widehat{\tau}_A/\sqrt{\mathbb{V}_A}$ converges to \eqref{mix_normal_trunc} under Assumption \ref{assu}, Condition \ref{strict_cond} and ReM. It follows from \eqref{diff_Wald} and Slutsky's theorem that, as $n\rightarrow\infty$,
\begin{equation*} \frac{\widehat{\tau}_{Wald}-\tau}{\sqrt{\mathbb{V}_A}/\tau_W}=\frac{\widehat{\tau}_A}{\sqrt{\mathbb{V}_A}}\frac{\tau_W}{\widehat{\tau}_W}
\end{equation*}
converges in distribution to \eqref{mix_normal_trunc} under ReM. $n\mathbb{V}_A$ has a finite (positive) limit superior under Condition \ref{lindeberg-feller}. Due to Lemma~\ref{conditions_imply}, it also has a has a finite (positive) limit superior under Condition~\ref{strict_cond}. $\tau_W$ has a positive limit inferior under Condition \ref{strength}. It follows that $\widehat{\tau}_{Wald}-\tau=o_p(1)$.
\hfill$\square$

\noindent\textbf{Proof of Theorem \ref{conservative_ReM}.}
Under Assumption \ref{assu}, Condition \ref{strict_cond} and ReM, we can apply Lemma A15 of \cite{li2018asymptotic} to have that, for $z=0,1$, 
\begin{align}
S_{A(z)}^2-\mathbb{S}_{A(z)}^2&=o_p(1),\label{eq:converge_SAz}\\ \bm{S}_{A(z),\bm{x}}-\mathbb{S}_{A(z),\bm{x}}&=o_p(1).\label{eq:converge_SAzx} 
\end{align}
We next show that, for $z=0,1$,  $S_{\widehat{A}(z)}^2-\mathbb{S}_{A(z)}^2=o_p(1)$ and $\bm{S}_{\widehat{A}(z),\bm{x}}-\mathbb{S}_{A(z),\bm{x}}=o_p(1)$.

Combining \eqref{S_Wz_leq} with Cauchy-Schwarz inequality, we have
\begin{equation}
	|S_{A(z), W(z)}|\leq\sqrt{S^2_{A(z)}S^2_{W(z)}}\leq\sqrt{\frac{n_z}{n_z-1}S_{A(z)}^2}.\label{S_BzWz_leq}
\end{equation}
Combing \eqref{diff_in_sampvar}, \eqref{S_Wz_leq} and \eqref{S_BzWz_leq}, the absolute difference between $S_{\widehat{A}(z)}^2$ and $S_{A(z)}^2$ can be bounded by
\begin{equation}
	\label{abs_diff_sampvar}
	\begin{aligned}
		\left|S_{\widehat{A}(z)}^2-S_{A(z)}^2\right|
		\leq&2\left|\widehat{\tau}_{Wald}-\tau\right|\left|S_{A(z),W(z)}\right|+\left(\widehat{\tau}_{Wald}-\tau\right)^2S_{W(z)}^2\\
		\leq&2\sqrt{\frac{n_z}{n_z-1}S_{A(z)}^2}\left|\widehat{\tau}_{Wald}-\tau\right|+\frac{n_z}{n_z-1}\left(\widehat{\tau}_{Wald}-\tau\right)^2.
	\end{aligned}
\end{equation}
With \eqref{eq:converge_SAz} and given that $\mathbb{S}_{A(z)}^2$ has finite limiting value according to Condition \ref{strict_cond}, applying Theorem \ref{Wald_ReM} and Slutsky's theorem, the right hand side of \eqref{abs_diff_sampvar} converges to 0 in probability. We conclude that $S_{\widehat{A}(z)}^2-S_{A(z)}^2=o_p(1)$. Combining this with \eqref{eq:converge_SAz}, we have $S_{\widehat{A}(z)}^2-\mathbb{S}_{A(z)}^2=o_p(1)$.

Recall that $x_i^{(k)}$ is the $k$th covariate of unit $i$ and $\overline{x}_z^{(k)}$ ($z=0,1$) is the mean of $x_i^{(k)}$ for units with $Z_i=z$. Applying Cauchy-Schwarz inequality to the $k$th component of $\bm{S}_{W(z),\bm{x}}$, we have
\begin{equation}
	\begin{aligned} &\left(\frac{1}{n_z-1}\sum_{i:\ Z_{i}=z}\left(W_i-\overline{W}_z\right)\left(x_{i}^{(k)}-\overline{x}_z^{(k)}\right)\right)^2\\
		\leq&\left(\frac{1}{n_z-1}\sum_{i:\ Z_{i}=z}\left(W_i-\overline{W}_z\right)^2\right)\left(\frac{1}{n_z-1}\sum_{i:\ Z_{i}=z}\left(x_{i}^{(k)}-\overline{x}_z^{(k)}\right)^2\right)\\
		=&\ S^2_{W(z)}\frac{1}{n_z-1}\sum_{i:\ Z_{i}=z}\left(x_{i}^{(k)}-\overline{x}_z^{(k)}\right)^2.
	\end{aligned}\label{S_Wz_x_com}
\end{equation}
Therefore
\begin{equation*}
	\begin{aligned}
		\left\Vert \bm{S}_{W(z),\bm{x}}\right\Vert\leq&\sqrt{\sum_{k=1}^{K}S^2_{W(z)}\frac{1}{n_z-1}\sum_{i:\ Z_{i}=z}\left(x_{i}^{(k)}-\overline{x}_z^{(k)}\right)^2}\\=&\sqrt{S^2_{W(z)}\operatorname{trace}(\bm{S}_{\bm{x}\bm{x},z})}\leq\sqrt{\frac{n_z}{n_z-1}\operatorname{trace}(\bm{S}_{\bm{x}\bm{x},z})},
	\end{aligned}
\end{equation*}
where the last inequality is due to \eqref{S_Wz_leq}.
Since $A_i-\widehat{A}_i=W_i\left(\widehat{\tau}_{Wald}-\tau\right)$, we have
\begin{equation*}
	\bm{S}_{A(z),\bm{x}}-\bm{S}_{\widehat{A}(z),\bm{x}}=\bm{S}_{W(z),\bm{x}}\left(\widehat{\tau}_{Wald}-\tau\right).
\end{equation*}
It follows that
\begin{equation}
	\label{bound_for_covariance}
	\begin{aligned}
		&\left\Vert \bm{S}_{\widehat{A}(z),\bm{x}}-\bm{S}_{A(z),\bm{x}}\right\Vert=\left\Vert \bm{S}_{W(z),\bm{x}}\right\Vert\left|\widehat{\tau}_{Wald}-\tau\right|\\
		\leq&\sqrt{\frac{n_z}{n_z-1}\operatorname{trace}(\bm{S}_{\bm{x}\bm{x},z})}|\widehat{\tau}_{Wald}-\tau|
	\end{aligned}
\end{equation}
Under Condition~\ref{strict_cond} and ReM, we can apply Lemma A15 of \cite{li2018asymptotic} to have $\bm{S}_{\bm{x}\bm{x},z}-\mathbb{S}_{\bm{x}\bm{x}}=o_p(1)$, where $\mathbb{S}_{\bm{x}\bm{x}}$ has a finite non-singular limiting value according to Condition \ref{strict_cond}. Applying Theorem \ref{Wald_ReM} and Slutsky's theorem, the right hand side of \eqref{bound_for_covariance} converges to 0 in probability. Therefore, we conclude that $\bm{S}_{\widehat{A}(z),\bm{x}}-\bm{S}_{A(z),\bm{x}}=o_p(1)$. Combining this with \eqref{eq:converge_SAzx}, we have $\bm{S}_{\widehat{A}(z),\bm{x}}-\mathbb{S}_{A(z),\bm{x}}=o_p(1)$.

Define
\begin{align}
\mathbb{V}_A^{+,ReM}&=\frac{\mathbb{S}_{A(1)}^2}{n_1}+\frac{\mathbb{S}_{A(0)}^2}{n_0}-\frac{\mathbb{S}_{A(1)-A(0)\mid\bm{x}}^2}{n}\notag\\
&=\frac{\mathbb{S}_{A(1)}^2}{n_1}+\frac{\mathbb{S}_{A(0)}^2}{n_0}-\frac{\left(\mathbb{S}_{A(1),\bm{x}}-\mathbb{S}_{A(0),\bm{x}}\right)^{\top}\mathbb{S}_{\bm{x}\bm{x}}^{-1}\left(\mathbb{S}_{A(1),\bm{x}}-\mathbb{S}_{A(0),\bm{x}}\right)}{n},\label{def_bV_A_ReM_plus}\\	\widehat{V}_A^{ReM}&=\frac{S_{\widehat{A}(1)}^2}{n_1}+\frac{S_{\widehat{A}(0)}^2}{n_0}-\frac{\left(\bm{S}_{\widehat{A}(1),\bm{x}}-\bm{S}_{\widehat{A}(0),\bm{x}}\right)^{\top}\mathbb{S}_{\bm{x}\bm{x}}^{-1}\left(\bm{S}_{\widehat{A}(1),\bm{x}}-\bm{S}_{\widehat{A}(0),\bm{x}}\right)}{n}.\notag
\end{align}
According to part (1) in Condition~\ref{strict_cond}, $n_1/n$ has a limit in $(0,1)$, which implies that $n_0/n=1-n_1/n$ also has a limit in $(0,1)$. Applying the continuous mapping theorem, we have $\widehat{V}_A^{ReM}-\mathbb{V}_A^{+,ReM}=o_p(n^{-1})$.
Define
\begin{equation*}
\begin{aligned}
	\label{estimated_SAx'}
\mathbb{V}_{A\mid\bm{x}}&=\frac{\mathbb{S}_{A(1)\mid\bm{x}}^2}{n_1}+\frac{\mathbb{S}_{A(0)\mid\bm{x}}^2}{n_0}-\frac{\mathbb{S}_{A(1)-A(0)\mid\bm{x}}^2}{n},\\
&=\frac{\mathbb{S}_{A(1)\mid\bm{x}}^2}{n_1}+\frac{\mathbb{S}_{A(0)\mid\bm{x}}^2}{n_0}-\frac{\left(\mathbb{S}_{A(1),\bm{x}}-\mathbb{S}_{A(0),\bm{x}}\right)^{\top}\mathbb{S}_{\bm{x}\bm{x}}^{-1}\left(\mathbb{S}_{A(1),\bm{x}}-\mathbb{S}_{A(0),\bm{x}}\right)}{n},\\	\widehat{V}_{A\mid\bm{x}}^{ReM}&=\frac{S_{\widehat{A}(1)\mid\bm{x}}^2}{n_1}+\frac{S_{\widehat{A}(0)\mid\bm{x}}^2}{n_0}-\frac{\left(\bm{S}_{\widehat{A}(1),\bm{x}}-\bm{S}_{\widehat{A}(0),\bm{x}}\right)^{\top}\mathbb{S}_{\bm{x}\bm{x}}^{-1}\left(\bm{S}_{\widehat{A}(1),\bm{x}}-\bm{S}_{\widehat{A}(0),\bm{x}}\right)}{n}.
\end{aligned}
\end{equation*}
Similarly, we have $\widehat{V}_{A\mid\bm{x}}^{ReM}-\mathbb{V}_{A\mid\bm{x}}=o_p(n^{-1})$.

Define 
\begin{equation}
\mathbb{R}^{2-}_{A}=\frac{\frac{\mathbb{S}_{A(1)|\bm{x}}^2}{n_1}+\frac{\mathbb{S}_{A(0)|\bm{x}}^2}{n_0}-\frac{\mathbb{S}_{A(1)-A(0)\mid\bm{x}}^2}{n}}{\frac{\mathbb{S}_{A(1)}^2}{n_1}+\frac{\mathbb{S}_{A(0)}^2}{n_0}-\frac{\mathbb{S}_{A(1)-A(0)\mid\bm{x}}^2}{n}}.\label{def_bR2_A_minus}
\end{equation}
Recall from Section~\ref{sec:ReM_Wald} that $\widehat{R}_A^2$ is obtained by replacing $\tau$ in $R_A^2$ defined in \eqref{def_R2A} with $\widehat{\tau}_{Wald}$. Applying the continuous mapping theorem, 
we have $\widehat{R}_A^2-\mathbb{R}^{2-}_{A}=o_p(1)$. Furthermore, with Lemma \ref{continuity_lambda}, we can apply the continuous mapping theorem again to have $\lambda_{\alpha/2}\left(\widehat{R}^2_{A}\right)-\lambda_{\alpha/2}\left(\mathbb{R}^{2-}_{A}\right)=o_p(1)$. 

Since $\mathbb{S}_{A(1)-A(0)\mid\bm{x}}$ is the finite population variance of the linear projection of $A_i(1)-A_i(0)$ on $\bm{x}_i$, we have $\mathbb{S}_{A(1)-A(0)\mid\bm{x}}^2\leq\mathbb{S}_{A(1)-A(0)}^2$. Combining this with \eqref{def_bV_A_ReM_plus}, \eqref{def_bV_A}, \eqref{def_bR2_A} and \eqref{def_bR2_A_minus}, we have $\mathbb{V}_A^{+,ReM}\geq \mathbb{V}_{A}$ and $\mathbb{R}^{2-}_{A}\leq{}\mathbb{R}^2_{A}$. Since we have proved that $\widehat{V}_A^{ReM}-\mathbb{V}_A^{+,ReM}=o_p(n^{-1})$, it follows that $\mathbb{V}_{A}$ is asymptotically conservatively estimated by $\widehat{V}_A^{ReM}$. Recall that $\lambda_{\alpha/2}(\rho)$ is a non-increasing function of $\rho$. We further know that $\lambda_{\alpha/2}\left(\mathbb{R}_A^{2}\right)\mathbb{V}_A^2$ is asymptotically conservatively estimated by $\lambda_{\alpha/2}\left(\widehat{R}^2_{A}\right)\widehat{V}_A^{ReM}$. 

Under ReM, $\widehat{\tau}_A^2\leq\lambda_{\alpha/2}^2\left(\mathbb{R}^2_{A}\right)\mathbb{V}_A$ holds with probability $1-\alpha$ asymptotically. With \eqref{def_tauhat_A} and
\eqref{hat_VA_ReM2},
\begin{equation*}
\begin{aligned} \left(\widehat{\tau}_Y-\tau\widehat{\tau}_W\right)^2&\leq\lambda_{\alpha/2}^2\left(\widehat{R}^2_A\right)\widehat{V}_A^{ReM}\\
&=\lambda_{\alpha/2}^2\left(\widehat{R}^2_A\right)\left(V_{Y}^{ReM}-2\widehat{\tau}_{Wald}C_{YW}^{ReM}+\widehat{\tau}_{Wald}^2V_{W}^{ReM}\right)
\end{aligned}
\end{equation*}
has at least a coverage rate of $1-\alpha$ for $\tau$ as $n\rightarrow\infty$. Solving the inequality by considering both cases of $\widehat{\tau}_W>0$ and $\widehat{\tau}_W<0$, we obtain \eqref{ci_ReM}, which completes the proof.
\hfill$\square$

\noindent\textbf{Proof of Theorem \ref{confidence_set_mix2}.}
As the proof is similar to that of Theorem \ref{confidence_set_mix}, we only present key points that are different.

\noindent\textbf{Proof of Part (a).} Conditional on $\tau_W<p_+/2$, similar to the arguments for \eqref{conservative_cond1}, $\widehat{I}_{ReM}$ is asymptotically conservative because the FAR confidence set in Section \ref{subsec:FAR_rem} is asymptotically conservative. 

Now consider conditioning on $\tau_W\geq p_{+}/2$. Theorem \ref{conservative_ReM} implies that the Wald confidence interval is convervative; the arguments in Section \ref{subsec:FAR_rem} imply that the FAR confidence set is asymptotically conservative. Hence $\widehat{I}_{ReM}$ is asymptotically conservative.

Overall, $\widehat{I}_{ReM}$ is asymptotically conservative.

\noindent\textbf{Proof of Part (b).}  Applying Fieller's theorem (\cite{finney1978statistical}, Chapter 4), when \eqref{confidence_set_rem2} is not empty, the upper and lower boundaries of \eqref{confidence_set_rem2} are
\begin{equation*}
	\frac{\widehat{\tau}_{Wald}-\frac{g'C_{YW}^{ReM}}{V_W^{ReM}}\pm \frac{\lambda_{\alpha/2}\left(R^{*2}_A\right)}{\left|\widehat{\tau}_{W}\right|}\sqrt{V_Y^{ReM}+\widehat{\tau}_{Wald}^2V_W^{ReM}-2\widehat{\tau}_{Wald}C_{YW}^{ReM}-g'\left(V_Y^{ReM}-\frac{\left(C_{YW}^{ReM}\right)^2}{V_W^{ReM}}\right)}}{1-g'},
\end{equation*}
where $g'=\lambda_{\alpha/2}^2\left(R^{*2}_A\right)V_W^{ReM}/\widehat{\tau}_W^2$. Replacing $z_{\alpha/2}$, $V_Y$, $V_W$, $C_{YW}$ and $g$ in \eqref{length_ineq} with  $\lambda_{\alpha/2}\left(R^{*2}_A\right)$, $V_Y^{ReM}$, $V_W^{ReM}$, $C_{YW}^{ReM}$ and $g'$, and taking into account  $\lambda_{\alpha/2}\left(R^{*2}_A\right)\geq\lambda_{\alpha/2}^2\left(\widehat{R}^2_{A}\right)$, we prove part (b).
\hfill$\square$

\noindent\textbf{Proof of Theorem \ref{Wald_Reg}.}
From Lemma \ref{consistency_tauW}, under Condition~\ref{strict_cond}, $\widehat{\tau}_{W\mid\bm{x}}/\tau_W-1=o_p(1)$.
Since $A_i=Y_i-\tau W_i$, which is well defined under Assumption \ref{assu}, we have $\widehat{\tau}_{A|\bm{x}}=\widehat{\tau}_{Y|\bm{x}}-\tau \widehat{\tau}_{W|\bm{x}}$. 
Combining this with \eqref{est_adj}, we have
\begin{equation}
	\widehat{\tau}_{Wald|\bm{x}}-\tau=\frac{\widehat{\tau}_{A\mid\bm{x}}}{\widehat{\tau}_{W\mid\bm{x}}}.\label{diff_Reg}
\end{equation}

It follows from Slutsky's theorem and \eqref{adist_lin} that, under ReM and as $n\rightarrow\infty$,
\begin{equation*}	\frac{\widehat{\tau}_{Wald|\bm{x}}-\tau}{\sqrt{\mathbb{V}_A(1-\mathbb{R}^2_{A})}/\tau_W}=\frac{\widehat{\tau}_{A\mid\bm{x}}}{\sqrt{\mathbb{V}_A(1-\mathbb{R}^2_{A})}}\frac{\tau_W}{\widehat{\tau}_{W\mid\bm{x}}}\stackrel{d}{\rightarrow}N(0,1).
\end{equation*}
According to Lemma \ref{conditions_imply}, Condition \ref{lindeberg-feller} holds and hence $n\mathbb{V}_A$ has a finite (positive) limit superior. Also, $\tau_W$ has a positive
limit inferior under Condition \ref{strength}. It follows that $\widehat{\tau}_{Wald|\bm{x}}-\tau=o_p(1)$.
\hfill$\square$

\noindent\textbf{Proof of Proposition \ref{conservative_reg_raw}.}
Recall that for any set of quantities $\{Q_i:\ i=1,\ldots,n\}$, $\widehat{\bm{\beta}}_{Q,z}$ is the vector of coefficients on $\bm{x}_i$ in the OLS regression of $Q_i$ on $\bm{x}_i$ for units with $Z_i=z$ ($z=0,1$). Let $D_i=A_i-\widehat{\bm{\beta}}_{A,Z_i}^{\top}\bm{x}_i$. We also define $\widetilde{D}_i=A_i-\widetilde{\bm{\beta}}_{A,Z_i}^{\top}\bm{x}_i$, where $\widetilde{\bm{\beta}}_{A,z}$ is the vector of coefficients on $\bm{x}_i$ in the OLS regression of $A_i(z)$ on $\bm{x}_i$ using all units. $D_i$ and $\widetilde{D}_i$ are well defined under Assumption \ref{assu}. For units with $Z_i=z$ ($z=0,1$), let $S_{D(z)}^2$ and $S_{\widetilde{D}(z)}^2$ be the sample variances of $D_i$ and $\widetilde{D}_i$. Define $V_D=S_{D(1)}^2/n_1+S_{D(0)}^2/n_0$ and $V_{\widetilde{D}}=S_{\widetilde{D}(1)}^2/n_1+S_{\widetilde{D}(0)}^2/n_0$. Under Condition \ref{strict_cond} and ReM, we can apply Theorem 8 and Proposition 3 of \cite{li2020rerandomization} to obtain that, $nV_A^{EHW}\stackrel{p}{\rightarrow}nV_D$ and $nV_{\widetilde{D}}\stackrel{p}{\rightarrow}nV_D$. Furthermore, Corollary 7 of \cite{li2020rerandomization} indicates that $nV_{\widetilde{D}}$ has a limit in probability that is no less than $n\mathbb{V}_A\left(1-\mathbb{R}^2_{A}\right)$. Therefore $nV_A^{EHW}$ is a conservative estimator of $n\mathbb{V}_A\left(1-\mathbb{R}^2_{A}\right)$. The remaining task is to show that $V_A^{HCj}-V_A^{EHW}=o_p(n^{-1})$ for $j=2,3$.

Let $\bm{\Omega}$ denote the $n\times(2(K+1))$ design matrix in which the $i$th row is $\bm{\Omega}_i^{\top}=(1,Z_i,\bm{x}_i, Z_i\bm{x}_i)^{\top}$.  Let $h_i$ denote the $i$th diagonal element of $\bm{\Omega}(\bm{\Omega}^{\top}\bm{\Omega})^{-1}\bm{\Omega}^{\top}$. Apparently $h_i\geq 0$. Let $u_i$ denote the residual in the OLS regression of $A_i$ on $Z_i$, $\bm{x}_i$ and $Z_i\bm{x}_i$ using all units. The HC$j$ ($j=2,3$) estimator for the covariance matrix of the coefficients in the OLS regression of $A_i$ on $Z_i$, $\bm{x}_i$ and $Z_i\bm{x}_i$ using all units is
\begin{equation}
	\bm{\Sigma}^{HCj}_{A}=\left(\sum_{i=1}^n \bm{\Omega}_i\bm{\Omega}_i^{\top}\right)^{-1}\left(\sum_{i=1}^nu_i^2\left(1-h_i\right)^{-(j-1)}\bm{\Omega}_i\bm{\Omega}_i^{\top}\right)\left(\sum_{i=1}^n \bm{\Omega}_i\bm{\Omega}_i^{\top}\right)^{-1},\label{eq:VA_HCj}
\end{equation}
and $V^{HCj}_{A}$ is the diagonal element of $\bm{\Sigma}^{HCj}_{A}$ that corresponds to the coefficient for $Z_i$, that is, the second diagonal element of $\bm{\Sigma}^{HCj}_{A}$. It can be seen that when $j=1$, $\bm{\Sigma}^{HC1}_{A}=\bm{\Sigma}^{HC1}_{EHW}$, and hence $V_A^{HC1}=V_A^{EHW}$.

Define
\begin{equation}\label{def:G}
	\begin{aligned}
		\bm{G}&\equiv n^{-1} \sum_{i=1}^n \bm{\Omega}_i\bm{\Omega}_i^{\top}\\
		&=
		\left(\begin{array}{cccc}1 &\sum_{i=1}^n Z_i/n &\sum_{i=1}^n \bm{x}_i^{\top}/n & \sum_{i=1}^n Z_i\bm{x}_i^{\top}/n\\
			\sum_{i=1}^n Z_i/n & \sum_{i=1}^n Z_i^2/n & \sum_{i=1}^n Z_i\bm{x}_i^{\top}/n & \sum_{i=1}^n Z_i^2\bm{x}_i^{\top}/n\\
			\sum_{i=1}^n \bm{x}_i/n & \sum_{i=1}^n Z_i\bm{x}_i/n & \sum_{i=1}^n \bm{x}_i\bm{x}_i^{\top}/n & \sum_{i=1}^n Z_i\bm{x}_i\bm{x}_i^{\top}/n\\
			\sum_{i=1}^n Z_i\bm{x}_i/n & \sum_{i=1}^n Z_i^2\bm{x}_i/n & \sum_{i=1}^n Z_i\bm{x}_i\bm{x}_i^{\top}/n & \sum_{i=1}^n Z_i^2\bm{x}_i\bm{x}_i^{\top}/n\end{array}\right)\\
		&=\left(\begin{array}{cc}
			\bm{G}_{11} & \bm{G}_{12} \\
			\bm{G}_{21} & \bm{G}_{22}
		\end{array}\right),
	\end{aligned}
\end{equation}
where
\begin{equation*}
	\begin{aligned}
		&\bm{G}_{11}=\left(\begin{array}{cc}
			1 & n_1/n \\
			n_1/n & n_1/n
		\end{array}\right),\quad
		\bm{G}_{12}=\left(\begin{array}{cc}
			\bm{0} & n_1/n\cdot\overline{\bm{x}}_1^{\top} \\
			n_1/n\cdot\overline{\bm{x}}_1^{\top} & n_1/n\cdot\overline{\bm{x}}_1^{\top}
		\end{array}\right)\\
		&\bm{G}_{21}=\bm{G}_{12}^{\top},\quad
		\bm{G}_{22}=n^{-1}\sum_{i=1}^{n}\left(\begin{array}{cc}
			\bm{x}_i\bm{x}_i^{\top} & Z_i\bm{x}_i\bm{x}_i^{\top} \\
			Z_i\bm{x}_i\bm{x}_i^{\top} & Z_i\bm{x}_i\bm{x}_i^{\top}
		\end{array}\right),
	\end{aligned}
\end{equation*}
taking into account that $\sum_{i=1}^n Z_i=n_1$, $\sum_{i=1}^n \bm{x}_i/n=\overline{\bm{x}}=\bm{0}$, $\sum_{i=1}^n Z_i\bm{x}_i/n=n_1/n\cdot\sum_{i:\ Z_i=1}\bm{x}_i/n_1=n_1/n\cdot\overline{\bm{x}}_1$ and $Z_i^2=Z_i$.

Define for $j=2,3$,
\begin{equation*}
	\bm{H}^{(j)}\equiv n^{-2} \sum_{i=1}^nu_i^2\left(1-h_i\right)^{-(j-1)}\bm{\Omega}_i \bm{\Omega}_i^{\top}.
\end{equation*}
According to \eqref{eq:VA_HCj}, for $j=2,3$, we have
\begin{equation}
	\bm{\Sigma}^{HCj}_{A}=\bm{G}^{-1}\bm{H}^{(j)}\bm{G}^{-1}.\label{eq:VA_HCj_short}
\end{equation}

Applying Lemma A8 and A11 of \cite{li2020rerandomization}, we obtain that under Condition \ref{strict_cond} and ReM, both $\bm{G}_{11}$ and $\bm{G}_{22}$ converge in probability to non-singular matrices and $\bm{G}_{12}=\bm{G}_{21}^{\top}=O_p(n^{-1/2})$. Therefore, $\bm{G}$ converges in probability to a positive definite matrix denoted as $\bm{G}^{\infty}$. Let $\delta$ be a positive number that is smaller than all of the eigenvalues of $\bm{G}^{\infty}$, and let $\bm{\Delta}=\delta\bm{I}$ be an $n\times n$ diagonal matrix with all of the diagonal elements being $\delta$. Therefore $\bm{G}^{-1}-\bm{\Delta}^{-1}$ converges in probability to a negative definite matrix. 

According to the definition of $\bm{G}$, $\bm{\Omega}^{\top}\bm{\Omega}=n\bm{G}$. Hence $h_i$ (the $i$th diagonal element of $\bm{\Omega}(\bm{\Omega}^{\top}\bm{\Omega})^{-1}\bm{\Omega}^{\top}$ with $h_i\geq 0$) is $h_i=n^{-1}\bm{\Omega}_i^{\top}\bm{G}^{-1}\bm{\Omega}_i$.  Also, it is easy to see that $\bm{\Omega}_i^{\top}\bm{\Delta}^{-1}\bm{\Omega}_i=(1+Z_i+\bm{x}_i^{\top}\bm{x}_i+Z_i\bm{x}_i^{\top}\bm{x}_i)/\delta$.
Therefore, $nh_i-(1+Z_i+\bm{x}_i^{\top}\bm{x}_i+Z_i\bm{x}_i^{\top}\bm{x}_i)/\delta=\bm{\Omega}_i^{\top}\left(\bm{G}^{-1}-\bm{\Delta}^{-1}\right)\bm{\Omega}_i$.
Because $\bm{G}^{-1}-\bm{\Delta}^{-1}$ converges in probability to a negative definite matrix, $\bm{\Omega}_i(\bm{G}^{-1}-\bm{\Delta}^{-1})\bm{\Omega}_i$ has a non-positive limit superior in probability.  Also, from part (3) of Condition \ref{strict_cond}, we know that $\lim_{n\rightarrow\infty}\left(\max_{1 \leq i \leq n}\bm{x}_i^{\top}\bm{x}_i\right)/n=0$. Hence $\max_{1 \leq i \leq n}h_i=o_p(1)$. 

For $j=2,3$, according to the definitions of $\bm{\Sigma}^{HCj}_{A}$ and $\bm{\Sigma}^{EHW}_{A}$, we have
\begin{equation*}
	\bm{\Sigma}^{HCj}_{A}-\bm{\Sigma}^{EHW}_{A}\leq\left(\max_{1 \leq i \leq n}\left(1-h_i\right)^{-(j-1)}-1\right)\bm{\Sigma}^{EHW}_{A}.
\end{equation*}
Since $\max_{1 \leq i \leq n}h_i=o_p(1)$ implies that $\left(\max_{1 \leq i \leq n}\left(1-h_i\right)^{-(j-1)}-1\right)=o_p(1)$ and recall that $V^{EHW}_{A}=o_p(n^{-1})$, we have $V^{HCj}_{A}-V^{EHW}_{A}\leq o_p(1)\cdot V^{EHW}_{A}=o_p(n^{-1})$. This completes the proof.
\hfill$\square$

\noindent\textbf{Proof of Theorem \ref{conservative_Reg}.} Recall that for any set of quantities $\{Q_i:\ i=1,\ldots,n\}$, $\widehat{\bm{\beta}}_{Q,z}$ is the vector of coefficients on $\bm{x}_i$ in the OLS regression of $Q_i$ on $\bm{x}_i$ for units with $Z_i=z$ ($z=0,1$). With slight abuse of notation, in this proof we let $\widehat{A}_i=Y_i-\widehat{\tau}_{Wald|\bm{x}}W_i$, $D_i=A_i-\widehat{\beta}_{A,Z_i}\bm{x}_i$ and $\widehat{D}_i=\widehat{A}_i-\widehat{\bm{\beta}}_{\widehat{A},Z_i}^{\top}\bm{x}_i$. $D_i$ is well defined under Assumption \ref{assu}. For units with $Z_i=z$ ($z=0,1$), let $S_{D(z)}^2$ and $S_{\widehat{D}(z)}^2$ be the sample variance of $D_i$ and  $\widehat{D}_i$. Define $V_D=S_{D(1)}^2/n_1+S_{D(0)}^2/n_0$ and $\widehat{V}_D=S_{\widehat{D}(1)}^2/n_1+S_{\widehat{D}(0)}^2/n_0$. We first show that 
\begin{equation}
	\label{consistency_VD}
	\widehat{V}_D-V_D=o_p(n^{-1}).
\end{equation}

Due to the linear relationship between $\widehat{A}_i$ and $(Y_i,W_i)$, we have $\widehat{\bm{\beta}}_{\widehat{A},z}=\widehat{\bm{\beta}}_{Y,z}-\widehat{\tau}_{Wald|\bm{x}}\widehat{\bm{\beta}}_{W,z}$. Define $B_i=W_i-\widehat{\beta}_{W,Z_i}^{\top}\bm{x}_i$. For units with $Z_i=z$ ($z=0,1$), let $\overline{B}_z$ be the sample mean of $B_i$, $S_{B(z)}^2$ be the sample variance of $B_i$, and $S_{D(z),B(z)}$ be the sample covariance between $D_i$ and $B_i$.

Recall that $A_i=Y_i-\tau W_i$, and hence $\widehat{\bm{\beta}}_{A,z}=\widehat{\bm{\beta}}_{Y,z}-\tau\widehat{\bm{\beta}}_{W,z}$.
For $z=0,1$,
\begin{equation*}
	\begin{aligned}
		\widehat{D}_i-D_i&=\left(\widehat{A}_i-A_i\right)-\left(\widehat{\bm{\beta}}_{\widehat{A},Z_i}-\widehat{\bm{\beta}}_{A,Z_i}\right)^{\top}\bm{x}_i\\
		&=-\left(\widehat{\tau}_{Wald|\bm{x}}-\tau\right)W_i+\left(\widehat{\tau}_{Wald|\bm{x}}-\tau\right)\widehat{\bm{\beta}}_{W,Z_i}^{\top}\bm{x}_i\\
		&=-\left(\widehat{\tau}_{Wald|\bm{x}}-\tau\right)B_i.
	\end{aligned}\label{diff_Dhat_D}
\end{equation*}
Hence we have
\begin{align*} S_{\widehat{D}(z)}^2=S_{D(z)}^2-2\left(\widehat{\tau}_{Wald|\bm{x}}-\tau\right)S_{D(z),B(z)}+\left(\widehat{\tau}_{Wald|\bm{x}}-\tau\right)^2S_{B(z)}^2.
\end{align*}
The absolute difference between $S_{\widehat{D}(z)}^2$ and $S_{D(z)}^2$ can be bounded by
\begin{equation}
	\begin{aligned}
		&\left|S_{\widehat{D}(z)}^2-S_{D(z)}^2\right|\\
		=&\left|-2\left(\widehat{\tau}_{Wald|\bm{x}}-\tau\right)S_{D(z),B(z)}+\left(\widehat{\tau}_{Wald|\bm{x}}-\tau\right)^2S_{B(z)}^2\right|\\
		\leq&2\left|\widehat{\tau}_{Wald|\bm{x}}-\tau\right|\left|S_{D(z),B(z)}\right|+\left(\widehat{\tau}_{Wald|\bm{x}}-\tau\right)^2S_{B(z)}^2.
	\end{aligned}\label{abs_diff_adj}
\end{equation}
Because $B_i$ is the residual of the OLS regression of $W_i$ on $\bm{x}_i$, we have
\begin{equation}
	S_{B(z)}^2\leq{}S_{W(z)}^2\leq\frac{n_z}{n_z-1},\label{S_Vz_leq}
\end{equation}
where the second inequality is due to \eqref{S_Wz_leq}.
Because $D_i$ is the residual of the OLS regression of $A_i$ on $\bm{x}_i$, we have
\begin{equation}
	S_{D(z)}^2\leq{}S_{A(z)}^2.\label{S_Dz_leq}
\end{equation}
Combining \eqref{S_Vz_leq} and \eqref{S_Dz_leq} with Cauchy-Schwarz inequality, we have
\begin{equation}
	|S_{D(z), B(z)}|\leq\sqrt{S^2_{D(z)}S^2_{B(z)}}\leq\sqrt{\frac{n_z}{n_z-1}S_{A(z)}^2}.\label{S_DzVz_leq}
\end{equation}
Substituting \eqref{S_Vz_leq} and \eqref{S_DzVz_leq} into \eqref{abs_diff_adj}, we obtain the following inequality
\begin{equation}	
	\label{bound_for_variance}
	\begin{aligned}
		&\left|S_{\widehat{D}(z)}^2-S_{D(z)}^2\right|\\
		\leq&2\sqrt{\frac{n_z}{n_z-1}S_{A(z)}^2}\left|\widehat{\tau}_{Wald|\bm{x}}-\tau\right|+\frac{n_z}{n_z-1}\left(\widehat{\tau}_{Wald|\bm{x}}-\tau\right)^2.
	\end{aligned}
\end{equation}
According to Lemma A5 of \cite{li2020rerandomization}, we have that under Condition \ref{strict_cond} and ReM,  $S_{A(z)}^2-\mathbb{S}_{A(z)}^2=o_p(1)$, where $\mathbb{S}_{A(z)}^2$ has a finite limiting value. According to Lemma \ref{consistency_tauW}, $\widehat{\tau}_{Wald|\bm{x}}-\tau=o_p(1)$. Applying the continuous mapping theorem, the right hand side of \eqref{bound_for_variance} converges to 0 in probability, which means $S_{\widehat{D}(z)}^2-S_{D(z)}^2=o_p(1)$. Under part (1) of Condition \ref{strict_cond}, $n_1/n$ and $n_0/n$ both have a limit in (0,1).  Hence, $\widehat{V}_D-V_D=o_p(n^{-1})$.

Consider the OLS regression of $\widehat{A}_i$ on $Z_i$, $\bm{x}_i$ and $Z_i\bm{x}_i$ using all units. It is easy to see that $\widehat{V}^{EHW}_A$ equals the EHW variance estimator of the coefficient on $Z_i$ in this OLS regression. We next show that $\widehat{V}^{EHW}_A$ is asymptotically equivalent to $\widehat{V}_D$. 

Let $\widehat{u}_i$ denote the residual in the above OLS regression. The EHW estimator for the covariance matrix of the coefficients in this OLS regression is
\begin{equation}
	\widehat{\bm{\Sigma}}_A^{EHW}=\left(\sum_{i=1}^n \bm{\Omega}_i\bm{\Omega}_i^{\top}\right)^{-1}\left(\sum_{i=1}^n\widehat{u}_i^{2}\bm{\Omega}_i\bm{\Omega}_i^{\top}\right)\left(\sum_{i=1}^n \bm{\Omega}_i\bm{\Omega}_i^{\top}\right)^{-1},\label{eq:V_EHW}
\end{equation}
where $\bm{\Omega}_i$ is defined in the proof of Proposition 1. $\widehat{V}^{EHW}_A$ is the diagonal element of $\widehat{\bm{\Sigma}}_A^{EHW}$ that corresponds to the coefficient for $Z_i$, that is, the second diagonal element of $\widehat{\bm{\Sigma}}_A^{EHW}$.

Define
\begin{equation*}
	\bm{H}\equiv n^{-2} \sum_{i=1}^n\widehat{u}_i^{2}\bm{\Omega}_i \bm{\Omega}_i^{\top}=\left(\begin{array}{cc}
		\bm{H}_{11} & \bm{H}_{12} \\
		\bm{H}_{21} & \bm{H}_{22}
	\end{array}\right),
\end{equation*}
where
\begin{equation*}
	\begin{aligned}
		&\bm{H}_{11}=n^{-2}\sum_{i=1}^{n}\left(\begin{array}{cc}
			\widehat{u}_i^{2} & Z_i\widehat{u}_i^{2} \\
			Z_i\widehat{u}_i^{2} & Z_i\widehat{u}_i^{2}
		\end{array}\right),\quad
		\bm{H}_{12}=n^{-2}\sum_{i=1}^{n}\left(\begin{array}{cc}
			\widehat{u}_i^{2}\bm{x}_i^{\top} & Z_i\widehat{u}_i^{2}\bm{x}_i^{\top} \\
			Z_i\widehat{u}_i^{2}\bm{x}_i^{\top} & Z_i\widehat{u}_i^{2}\bm{x}_i^{\top}
		\end{array}\right)\\
		&\bm{H}_{21}=\bm{H}_{12}^{\top},\quad
		\bm{H}_{22}=n^{-2}\sum_{i=1}^{n}\left(\begin{array}{cc}
			\widehat{u}_i^{2}\bm{x}_i\bm{x}_i^{\top} & Z_i\widehat{u}_i^{2}\bm{x}_i\bm{x}_i^{\top} \\
			Z_i\widehat{u}_i^{2}\bm{x}_i\bm{x}_i^{\top} & Z_i\widehat{u}_i^{2}\bm{x}_i\bm{x}_i^{\top}
		\end{array}\right)
	\end{aligned}
\end{equation*}
According to \eqref{eq:V_EHW} and the definition of $\bm{G}$ in the proof of Proposition 1, we have
\begin{equation}
	\widehat{\bm{\Sigma}}_A^{EHW}=\bm{G}^{-1}\bm{H}\bm{G}^{-1}.\label{eq:V_EHW_short}
\end{equation}

Let $\widehat{\beta}_0^*$, $\widehat{\beta}_Z^*$, $\widehat{\bm{\beta}}_{\bm{x}}^*$ and $\widehat{\bm{\beta}}_{Z\bm{x}}^*$ denote respectively the intercept, the coefficient on $Z_i$, the vector of coefficients on $\bm{x}_i$ and the vector of coefficients on $Z_i\bm{x}_i$ in the above OLS regression. Recall that $\widehat{\bm{\beta}}_{\widehat{A},z}$ is the vector of coefficients on $\bm{x}_i$ in the OLS regression of $\widehat{A}_i$ on $\bm{x}_i$ using units with $Z_i=z$ ($z=0,1$). We have
$\widehat{\bm{\beta}}_{\widehat{A},0}=\widehat{\bm{\beta}}_{\bm{x}}^*$, $\widehat{\bm{\beta}}_{\widehat{A},1}=\widehat{\bm{\beta}}_{\bm{x}}^*+\widehat{\bm{\beta}}_{Z\bm{x}}^*$.  According to the definition of $\widehat{u}_i$, we have that for units with $Z_i=0$, $$\widehat{u}_i=\widehat{A}_i-\widehat{\beta}_0^*-\widehat{\bm{\beta}}_{\bm{x}}^{*\top}\bm{x}_i=\widehat{A}_i-\widehat{\beta}_0^*-\widehat{\bm{\beta}}_{\widehat{A},0}^{\top}\bm{x}_i=\widehat{D}_i-\beta_0^*,$$ and that for units with $Z_i=1$, $$\widehat{u}_i=\widehat{A}_i-\widehat{\beta}_0^*-\widehat{\beta}_Z^*-\widehat{\bm{\beta}}_{\bm{x}}^{*\top}\bm{x}_i-\widehat{\bm{\beta}}_{Z\bm{x}}^{*\top}\bm{x}_i=\widehat{A}_i-\left(\widehat{\beta}_0^*+\widehat{\beta}_Z^*\right)-\widehat{\bm{\beta}}_{\widehat{A},1}^{\top}\bm{x}_i=\widehat{D}_i-\beta_0^*-\beta_Z^*.$$ It is easy to see that, for $z=0,1$,
\begin{equation}
	\frac{1}{n_z}\sum_{i:\ Z_i=z}\widehat{u}_i^{2}=\frac{n_z-1}{n_z}S^2_{\widehat{D}(z)}. \label{eq:hatu2}
\end{equation}

Applying Lemma A9 of \cite{li2020rerandomization}, we have $S^2_{D(z)}-\mathbb{S}^2_{A(z)\backslash\bm{x}}=o_p(1)$, where $\mathbb{S}^2_{A(z)\backslash\bm{x}}$ is the finite population variance of the residual in the OLS regression of $A_i(z)$ on $\bm{x}_i$ using all units. Combining this with \eqref{eq:hatu2} and $S_{\widehat{D}(z)}^2-S_{D(z)}^2=o_p(1)$, we have
\begin{equation*}
	\frac{1}{n_z}\sum_{i:\ Z_i=z}\widehat{u}_i^{2}=S^2_{D(z)}+o_p(1)=\mathbb{S}^2_{A(z)\backslash\bm{x}}+o_p(1),\quad{}z=0,1.
\end{equation*}
Since $\mathbb{S}^2_{A(z)\backslash\bm{x}}$ is the variance of the projection residual of $A_i(z)$, we have $\mathbb{S}^2_{A(z)\backslash\bm{x}}\leq\mathbb{S}^2_{A(z)}$, with $\mathbb{S}^2_{A(z)}$ having a finite limit for $z=0,1$ according to part (2) of Condition \ref{strict_cond}. Under part (1) of Condition \ref{strict_cond}, $n_1/n$ and $n_0/n$ both have a limit in $(0,1)$.  Thus, $\sum_{i:\ Z_i=z}\widehat{u}_i^{2}=O_p(n)$ for $z=0,1$. It follows that $\sum_{i=1}^n\widehat{u}_i^{2}=\sum_{i:\ Z_i=1}\widehat{u}_i^{2}+\sum_{i:\ Z_i=0}\widehat{u}_i^{2}=O_p(n)$, and $\sum_{i=1}^n Z_i\widehat{u}_i^{2}=\sum_{i:\ Z_i=1}\widehat{u}_i^{2}=O_p(n)$.  Combining this with the definition of $\bm{H}_{11}$, we have $\bm{H}_{11}=O_p(n^{-1})$.

Part (3) of Condition \ref{strict_cond} implies that $\max_{1\leq{}i\leq{}n}|x_i^{(k)}|=o_p(n^{1/2})$.  According to the definition of $\bm{H}_{12}$, each element of $\bm{H}_{12}$ is either in the form of $\sum_{i=1}^n \widehat{u}_i^{2}x_i^{(k)}/n^2=\sum_{i:\ Z_i=1} \widehat{u}_i^{2}x_i^{(k)}/n^2+\sum_{i:\ Z_i=0} \widehat{u}_i^{2}x_i^{(k)}/n^2$, or in the form of $\sum_{i=1}^n Z_i\widehat{u}_i^{2}x_i^{(k)}/n^2=\sum_{i:\ Z_i=1} \widehat{u}_i^{2}x_i^{(k)}/n^2$.  For $z=0,1$, we have
\begin{equation*}
	\begin{aligned}
		&\left|\frac{1}{n^2}\sum_{i:\ Z_i=z}\widehat{u}_i^{2}x_i^{(k)}\right|\\
		\leq & \frac{1}{n^2}\left(\sum_{i:\ Z_i=z} \widehat{u}_i^{2}\right)\left(\max_{1\leq{}i\leq{}n}|x_i^{(k)}|\right)\\
		=&\frac{1}{n^2}\cdot O_p(n)\cdot o_p(n^{1/2})=o_p(n^{-1/2}).
	\end{aligned}
\end{equation*}
Hence $\sum_{i=1}^n \widehat{u}_i^{2}x_i^{(k)}/n^2=o_p(n^{-1/2})$, and $\sum_{i=1}^n Z_i\widehat{u}_i^{2}x_i^{(k)}/n^2=o_p(n^{-1/2})$.  Therefore, $\bm{H}_{12}=o_p(n^{-1/2})$.

According to the definition of $\bm{H}_{22}$, each element of $\bm{H}_{22}$ is either in the form of $\sum_{i=1}^n \widehat{u}_i^{2}x_i^{(k)}x_i^{(l)}/n^2=\sum_{i:\ Z_i=1} \widehat{u}_i^{2}x_i^{(k)}x_i^{(l)}/n^2+\sum_{i:\ Z_i=1} \widehat{u}_i^{2}x_i^{(k)}x_i^{(l)}/n^2$, or in the form of \\$\sum_{i=1}^n Z_i\widehat{u}_i^{2}x_i^{(k)}x_i^{(l)}/n^2=\sum_{i:\ Z_i=1} \widehat{u}_i^{2}x_i^{(k)}x_i^{(l)}/n^2$.  For $z=0,1$, we have
\begin{equation*}
	\begin{aligned}
		&\left|\frac{1}{n^2}\sum_{i:\ Z_i=z}\widehat{u}_i^{2}x_i^{(k)}x_i^{(l)}\right|\\
		\leq & \frac{1}{n^2}\left(\sum_{i:\ Z_i=z} \widehat{u}_i^{2}\right)\left(\max_{1\leq{}i\leq{}n}|x_i^{(k)}|\right)\left(\max_{1\leq{}i\leq{}n}|x_i^{(l)}|\right)\\
		=&\frac{1}{n^2}\cdot O_p(n)\cdot o_p(n^{1/2})\cdot o_p(n^{1/2})=o_p(1).
	\end{aligned}
\end{equation*}
Hence $\sum_{i=1}^n \widehat{u}_i^{2}x_i^{(k)}x_i^{(l)}/n^2=o_p(1)$, and $\sum_{i=1}^n Z_i\widehat{u}_i^{2}x_i^{(k)}x_i^{(l)}/n^2=o_p(1)$.  Therefore, $\bm{H}_{22}=o_p(1)$.  In summary, we have $\bm{H}=o_p(1)$.

Recall the definitions of $\bm{G}_{11}$, $\bm{G}_{12}$, $\bm{G}_{21}$ and $\bm{G}_{22}$ in the proof of Proposition 1. Let
\begin{equation*}
	\bm{G}_{diag}=\left(\begin{array}{cc}
		\bm{G}_{11} & \bm{0} \\
		\bm{0} & \bm{G}_{22}
	\end{array}\right),\quad
	\bm{G}_{diff}=\left(\begin{array}{cc}
		\bm{0} & \bm{G}_{12} \\
		\bm{G}_{21} & \bm{0}
	\end{array}\right).
\end{equation*}
We have $\bm{G}=\bm{G}_{diag}+\bm{G}_{diff}$.  According to the results in the proof of Proposition \ref{conservative_reg_raw}, $\bm{G}_{diff}=O_p(n^{-1/2})$ and $\bm{G}_{diag}$ converges in probability to a non-singular matrix. According to Hua's identity, if two square matrices $\bm{M}$ and $\bm{N}$ are non-singular and $\bm{M}-\bm{N}^{-1}$ is also non-singular, then $(\bm{M}-\bm{N}^{-1})^{-1}-\bm{M}^{-1}$ is non-singular, with $(\bm{M}-\bm{N}^{-1})^{-1}-\bm{M}^{-1}=(\bm{M}\bm{N}\bm{M}-\bm{M})^{-1}$.  Plugging in $\bm{M}=\bm{G}_{diag}$ and $\bm{N}=-\bm{G}_{diff}^{-1}$, we have
\begin{equation}	\bm{G}_{diag}^{-1}-\bm{G}^{-1}=(\bm{G}_{diag}+\bm{G}_{diag}\bm{G}_{diff}^{-1}\bm{G}_{diag})^{-1}=O_p(n^{-1/2}).\label{eq:Ginv_diff}
\end{equation}
With \eqref{eq:V_EHW_short}, we have
\begin{equation}
\label{eq:diffmat}
	\begin{aligned}
		&\widehat{\bm{\Sigma}}_A^{EHW}-\bm{G}_{diag}^{-1}\bm{H}\bm{G}_{diag}^{-1}\\
		=&\bm{G}^{-1}\bm{H}\bm{G}^{-1}-\bm{G}_{diag}^{-1}\bm{H}\bm{G}_{diag}^{-1}\\
		=&(\bm{G}^{-1}-\bm{G}_{diag}^{-1})\bm{H}\bm{G}^{-1}+\bm{G}_{diag}^{-1}\bm{H}(\bm{G}^{-1}-\bm{G}_{diag}^{-1})\\
		=&(\bm{G}^{-1}-\bm{G}_{diag}^{-1})\bm{H}\bm{G}_{diag}^{-1}+\bm{G}_{diag}^{-1}\bm{H}(\bm{G}^{-1}-\bm{G}_{diag}^{-1})\\
		&+(\bm{G}^{-1}-\bm{G}_{diag}^{-1})\bm{H}(\bm{G}^{-1}-\bm{G}_{diag}^{-1}).
	\end{aligned}
\end{equation}
With \eqref{eq:Ginv_diff} and $\bm{H}=o_p(1)$, we have $(\bm{G}^{-1}-\bm{G}_{diag}^{-1})\bm{H}(\bm{G}^{-1}-\bm{G}_{diag}^{-1})=O_p(n^{-1/2})\cdot o_p(1) \cdot O_p(n^{-1/2})=o_p(n^{-1})$. Since $\bm{G}_{diag}$ is a block-diagonal matrix, we have
\begin{equation*}
	\bm{G}_{diag}^{-1}=\left(\begin{array}{cc}
		\bm{G}_{11}^{-1} & \bm{0} \\
		\bm{0} & \bm{G}_{22}^{-1}
	\end{array}\right).
\end{equation*}
With~\eqref{eq:diffmat}, we obtain that
\begin{equation}\label{eq:V_EHW_diff}
	\begin{aligned}
		\widehat{\bm{\Sigma}}_A^{EHW}-\bm{G}_{diag}^{-1}\bm{H}\bm{G}_{diag}^{-1}=&O_p(n^{-1/2})\left(\begin{array}{cc}
			\bm{H}_{11}\bm{G}_{11}^{-1} & \bm{H}_{12}\bm{G}_{22}^{-1} \\
			\bm{H}_{21}\bm{G}_{11}^{-1} & \bm{H}_{22}\bm{G}_{22}^{-1}
		\end{array}\right)\\
		&+\left(\begin{array}{cc}
			\bm{G}_{11}^{-1}\bm{H}_{11} & \bm{G}_{11}^{-1}\bm{H}_{12} \\
			\bm{G}_{22}^{-1}\bm{H}_{21} & \bm{G}_{22}^{-1}\bm{H}_{22}
		\end{array}\right)O_p(n^{-1/2})+o_p(n^{-1})
	\end{aligned}
\end{equation}

Let $\widehat{\bm{\Sigma}}_A^{EHW}[2,2]$ denote the submatrix of $\widehat{\bm{\Sigma}}_A^{EHW}$ that consists of the first two rows and the first two columns. It is easy to see that the corresponding submatrix of $\bm{G}_{diag}^{-1}\bm{H}\bm{G}_{diag}^{-1}$ equals $\bm{G}_{11}^{-1}\bm{H}_{11}\bm{G}_{11}^{-1}$. Taking the submatrix of \eqref{eq:V_EHW_diff}, we have
\begin{equation*}
	\begin{aligned}
		\widehat{\bm{\Sigma}}_A^{EHW}[2,2]-\bm{G}_{11}^{-1}\bm{H}_{11}\bm{G}_{11}^{-1}=O_p(n^{-1/2})\cdot \bm{H}_{11}\bm{G}_{11}^{-1}+\bm{G}_{11}^{-1}\bm{H}_{11}\cdot O_p(n^{-1/2}) + o_p(n^{-1})
	\end{aligned}
\end{equation*}
Since $\bm{G}_{11}$ converges in probability to a non-singular matrix and $\bm{H}_{11}=O_p(n^{-1})$, we obtain that $\widehat{\bm{\Sigma}}_A^{EHW}[2,2]-\bm{G}_{11}^{-1}\bm{H}_{11}\bm{G}_{11}^{-1}=o_p(n^{-1})$.

Plugging in the expressions for $\bm{G}_{11}$ and $\bm{H}_{11}$, we have
\begin{align*}
	&\bm{G}_{11}^{-1}\bm{H}_{11}\bm{G}_{11}^{-1}\\=&\left(\begin{array}{cc}
		1/n_0 & -1/n_0 \\
		-1/n_0 & n/(n_1n_0)
	\end{array}\right)
	\sum_{i=1}^{n}\left(\begin{array}{cc}
		\widehat{u}_i^{2} & Z_i\widehat{u}_i^{2} \\
		Z_i\widehat{u}_i^{2} & Z_i\widehat{u}_i^{2}
	\end{array}\right)
	\left(\begin{array}{cc}
		1/n_0 & -1/n_0 \\
		-1/n_0 & n/(n_1n_0)
	\end{array}\right)\\
	=&\left(\begin{array}{cc}
		n_0^{-2}\sum_{i:Z_{i}=0}\widehat{u}_i^{2} & -n_0^{-2}\sum_{i:Z_{i}=0}\widehat{u}_i^{2} \\
		-n_0^{-2}\sum_{i:Z_{i}=0}\widehat{u}_i^{2} & n_1^{-2}\sum_{i:Z_{i}=1}\widehat{u}_i^{2}+n_0^{-2}\sum_{i:Z_{i}=0}\widehat{u}_i^{2}
	\end{array}\right).
\end{align*}
By definition, $\widehat{V}^{EHW}_A$ is the element of $\widehat{\bm{\Sigma}}_A^{EHW}[2,2]$ that is in the second row and second column. Hence we have
\begin{equation} \widehat{V}^{EHW}_A=\frac{1}{n_1^2}\sum_{i:Z_{i}=1}\widehat{u}_i^{2}+\frac{1}{n_0^2}\sum_{i:Z_{i}=0}\widehat{u}_i^{2}+o_p(n^{-1}).\label{eq:varhat_EHW_Dhat}
\end{equation}

By definition, $\widehat{V}_D=S_{\widehat{D}(1)}^2/n_1+S_{\widehat{D}(0)}^2/n_0$. Combining this with \eqref{eq:hatu2}, we have
\begin{equation*} \widehat{V}_D=\frac{\sum_{i:Z_{i}=1}\widehat{u}_i^{2}}{n_1(n_1-1)}+\frac{\sum_{i:Z_{i}=0}\widehat{u}_i^{2}}{n_0(n_0-1)}.
\end{equation*}
Combining this with \eqref{eq:varhat_EHW_Dhat}, we have
\begin{equation}
	\widehat{V}^{EHW}_A =\widehat{V}_D-\frac{\sum_{i:Z_{i}=1}\widehat{u}_i^{2}}{n_1^2(n_1-1)}-\frac{\sum_{i:Z_{i}=0}\widehat{u}_i^{2}}{n_0^2(n_0-1)}+o_p(n^{-1}).
\end{equation}
Recall that $\sum_{Z_i=z}\widehat{u}_i^{2}=O_p(n)$ for $z=0,1$.  Also, $n_1/n$ and $n_0/n$ both have a limit in $(0,1)$.  We conclude that $\widehat{V}^{EHW}_A=\widehat{V}_D+o_p(n^{-1})$. We can show $\widehat{V}^{HCj}_A=\widehat{V}^{EMW}_A+o_p(n^{-1})$ for $j=2,3$ similarly to the proof of Proposition \ref{conservative_reg_raw}. Therefore $\widehat{V}^{\dag}_A=\widehat{V}_D+o_p(n^{-1})$ for $\dag=HC2$ or $HC3$.

According to Corollary 7 and Proposition 3 of \cite{li2020rerandomization}, $V_D$ is an asymptotically conservative estimator of the variance of $\widehat{\tau}_{A\mid\bm{x}}$. Therefore, asymptotically $\widehat{\tau}_{A\mid\bm{x}}^2\leq z_{\alpha/2}^2V_D$ has a probability of at least $1-\alpha$ to hold. With $\widehat{V}_D=V_D+o_p(1)$ and $\widehat{V}^{\dag}_A=\widehat{V}_D+o_p(1)$ ($\dag\in\{EHW,HC2,HC3\}$), $\widehat{\tau}_{A\mid\bm{x}}^2\leq z_{\alpha/2}^2\widehat{V}^{\dag}_A$, or equivalently
\begin{equation*} \left(\widehat{\tau}_{Y\mid\bm{x}}-\tau\widehat{\tau}_{W\mid\bm{x}}\right)^2\leq z_{\alpha/2}^2\left(V^{\dag}_{Y}+\widehat{\tau}_{Wald|\bm{x}}^2V^{\dag}_{W}-2\widehat{\tau}_{Wald|\bm{x}}C^{\dag}_{YW}\right),
\end{equation*}
asymptotically has at least a coverage rate of $1-\alpha$ for $\tau$. Solving the inequality by considering
both cases of $\widehat{\tau}_{W\mid\bm{x}}>0$ and $\widehat{\tau}_{W\mid\bm{x}} < 0$, we obtain \eqref{ci_Reg}, which completes the proof.
\hfill$\square$

\noindent\textbf{Proof of Theorem \ref{confidence_set_mix3}.}
As the proof procedures here are almost the same as those of Theorem \ref{confidence_set_mix}, we just provide key steps that are different. Consider $\dag\in\{EHW,HC2,HC3\}$.

\noindent\textbf{Proof of Theorem Part (a).} Conditional on $\tau_W<p_+/2$, similar to the arguments for \eqref{conservative_cond1}, $\widehat{I}_{Reg}^{\dag}$ is asymptotically conservative because the FAR confidence set in Section \ref{subsec:FAR_reg} is asymptotically conservative. 

Now consider conditioning on $\tau_W\geq p_{+}/2$. Theorem \ref{conservative_Reg} implies that the Wald confidence interval is convervative; the arguments in Section \ref{subsec:FAR_reg} imply that the FAR confidence set is asymptotically conservative. Hence $\widehat{I}_{Reg}^{\dag}$ is asymptotically conservative.

Overall, $\widehat{I}_{Reg}^{\dag}$ is asymptotically conservative.

\noindent\textbf{Proof of Theorem Part (b).} Applying Fieller's theorem (\cite{finney1978statistical}, Chapter 4), when \eqref{confidence_set_reg} is not empty, the upper and lower boundaries of \eqref{confidence_set_reg} are
\begin{equation*}
	\left[\widehat{\tau}_{Wald|\bm{x}}-\frac{g^{\dag}C_{YW}^{\dag}}{V_W^{\dag}}\pm \frac{z_{\alpha/2}}{\left|\widehat{\tau}_{W\mid\bm{x}}\right|}\sqrt{V_Y^{\dag}+\widehat{\tau}_{Wald|\bm{x}}^2V_W^{\dag}-2\widehat{\tau}_{Wald|\bm{x}}C_{YW}^{\dag}-g^{\dag}\left(V_Y^{\dag}-\frac{C_{YW}^{^{\dag}2}}{V_W^{\dag}}\right)}\right]/(1-g^{\dag}),
\end{equation*}
where $g^{\dag}=z_{\alpha/2}^2V_W^{\dag}/\widehat{\tau}_{W\mid\bm{x}}^2$.  Replacing $V_Y$, $V_W$, $C_{YW}$ and $g$ in \eqref{length_ineq} with $V_Y^{\dag}$, $V_W^{\dag}$, $C_{YW}^{\dag}$ and $g^{\dag}$, we prove part (b).
\hfill$\square$

\section{More simulation results}
\label{simulation_results}

\subsection{More simulation results with $n=200$} 
Table~\ref{tab:MAE} reports the mean absolute error.  It can be seen that when $\tau_W\geq 0.05$, compared to the case without regression adjustment under CRE, 
ReM or regression adjustment reduces the mean absolute error, but there is no additional reduction to use both.

\begin{table}[htbp]
	\centering
	\caption{Mean absolute error ($n=200$).}\label{tab:MAE}
	\begin{tabular}{rcrrrrrrr}
		\hline
		&&\multicolumn{7}{c}{$\tau_W$}\\%[-2.5mm]
		Case & Design & 0.005 & 0.05 & 0.10 & 0.15 & 0.2 & 0.3 & 0.5 \\ 
		\hline
		%&\multicolumn{8}{c}{No Regression Adjustment}\\[-2.5mm]
		\multirow{2}{*}{{\it NoReg}} & CRE & 20.14 & 8.35 & 4.98 & 3.31 & 2.37 & 1.48 & 0.78 \\
		& ReM & 23.24 & 6.52 & 3.72 & 2.34 & 1.66 & 1.03 & 0.51 \\
		\multirow{2}{*}{{\it Reg}} & CRE & 23.10 & 6.59 & 3.66 & 2.27 & 1.63 & 1.02 & 0.51 \\
		& ReM & 23.25 & 6.36 & 3.66 & 2.30 & 1.63 & 1.03 & 0.50 \\
		\hline
	\end{tabular}
\end{table}

For cases without and with regression adjustment, respectively, Tables~\ref{tab:CRate_NoReg} and~\ref{tab:CRate_Reg} present the coverage rate of 95\% confidence sets, measured in percentage difference from the nominal level 0.95. (If $CRate$ denotes the coverage rate, then Tables~\ref{tab:CRate_NoReg} and~\ref{tab:CRate_Reg} present $100\times (CRate-0.95)$.) It can be seen that when $\tau_W = 0.005$, {\it Wald} has severe undercoverage, and {\it $Wald_{F-10}$} has extremely severe undercoverage.

\begin{table}[htbp]
	\centering
	\caption{Coverage rate of 95\% confidence set without regression adjustment ($n=200$).}\label{tab:CRate_NoReg}
	\begin{tabular}{rcrrrrrrr}
		\hline
		&&\multicolumn{7}{c}{$\tau_W$}\\%[-2.5mm]
		Method & Design & 0.005 & 0.05 & 0.10 & 0.15 & 0.2 & 0.3 & 0.5 \\ 
		\hline
		%&\multicolumn{8}{c}{No Regression Adjustment}\\[-2.5mm]
		\multirow{2}{*}{{\it Wald}} & CRE & -14.72 & 4.52 & 4.32 & 3.32 & 2.74 & 2.45 & 2.24 \\
		& ReM & -25.91 & 4.93 & 4.61 & 4.35 & 3.66 & 2.58 & 3.19 \\
		\multirow{2}{*}{{\it FAR}} & CRE & -0.16 & 0.26 & 0.36 & 0.76 & 0.88 & 1.16 & 2.08 \\
		& ReM & -0.05 & 4.33 & 4.35 & 4.61 & 4.67 & 4.63 & 4.87 \\
		\multirow{2}{*}{{\it $TS_{\gamma-0.075}$}} & CRE & -3.18 & 0.88 & 1.69 & 1.99 & 2.29 & 2.44 & 2.24 \\
		& ReM & -3.91 & 4.39 & 4.26 & 4.29 & 3.66 & 2.58 & 3.19 \\
		\multirow{2}{*}{{\it $TS_{\gamma-0.025}$}} & CRE & -1.03 & 0.48 & 1.06 & 1.35 & 1.72 & 2.40 & 2.24 \\
		& ReM & -0.62 & 4.33 & 4.14 & 4.19 & 3.65 & 2.58 & 3.19 \\
		\multirow{2}{*}{{\it $TS_{F-10}$}} & CRE & -0.16 & 0.26 & 0.39 & 0.69 & 0.72 & 2.16 & 2.24 \\
		& ReM & -0.05 & 4.33 & 4.28 & 4.37 & 3.88 & 2.59 & 3.19 \\
		\multirow{2}{*}{{\it $Wald_{F-10}$}} & CRE & -95.00 & -4.68 & -1.35 & -1.78 & 0.29 & 2.33 & 2.24 \\
		& ReM & -95.00 & 5.00 & 0.02 & 2.52 & 2.70 & 2.47 & 3.12 \\
		\hline
	\end{tabular}
\end{table}

\begin{table}[htbp]
	\centering
	\caption{Coverage rate of 95\% confidence set with regression adjustment ($n=200$).}\label{tab:CRate_Reg}
	\begin{tabular}{rcrrrrrrr}
		\hline
		&&\multicolumn{7}{c}{$\tau_W$}\\%[-2.5mm]
		Method & Design & 0.005 & 0.05 & 0.10 & 0.15 & 0.2 & 0.3 & 0.5 \\ 
		\hline
		%&\multicolumn{8}{c}{No Regression Adjustment}\\[-2.5mm]
		\multirow{2}{*}{{\it Wald}} & CRE & -27.40 & 4.96 & 4.64 & 4.35 & 3.83 & 2.61 & 3.03 \\
		& ReM & -26.57 & 4.93 & 4.58 & 4.39 & 3.74 & 2.64 & 3.18 \\
		\multirow{2}{*}{{\it FAR}} & CRE & -0.82 & -0.18 & 0.16 & 0.43 & 0.77 & 1.33 & 3.05 \\
		& ReM & -0.55 & -0.17 & 0.18 & 0.47 & 0.84 & 1.41 & 3.13 \\
		\multirow{2}{*}{{\it $TS_{\gamma-0.075}$}} & CRE & -4.66 & 0.94 & 2.42 & 3.79 & 3.77 & 2.61 & 3.03 \\
		& ReM & -4.31 & 0.86 & 2.38 & 3.77 & 3.68 & 2.64 & 3.18 \\
		\multirow{2}{*}{{\it $TS_{\gamma-0.025}$}} & CRE & -1.45 & 0.28 & 1.42 & 2.94 & 3.57 & 2.61 & 3.03 \\
		& ReM & -1.19 & 0.28 & 1.31 & 2.86 & 3.46 & 2.64 & 3.18 \\
		\multirow{2}{*}{{\it $TS_{F-10}$}} & CRE & -0.82 & -0.17 & 0.35 & 1.02 & 2.13 & 2.58 & 3.03 \\
		& ReM & -0.55 & -0.14 & 0.34 & 0.95 & 2.19 & 2.60 & 3.18 \\
		\multirow{2}{*}{{\it $Wald_{F-10}$}} & CRE & -95.00 & 5.00 & 2.26 & 3.15 & 3.12 & 2.59 & 3.03 \\
		& ReM & -95.00 & 1.67 & 2.11 & 3.46 & 3.16 & 2.62 & 3.18 \\
		\hline
	\end{tabular}
\end{table}

We also tried regression adjustment with HC2 or HC3 variance estimator.  The results are quite similar to regression adjustment with EHW variance estimator, except that the median interval length is longer.  We omit these results.

\subsection{Simulation results with $n=100$ and $n=400$}  

Tables~\ref{tab:MAE_100}-\ref{tab:Len_Reg_100} present simulation results with $n=100$. The smallest possible positive value of $\tau_W$ is 0.01 (instead of 0.005) when there is 1 complier. Tables~\ref{tab:MAE_400}-\ref{tab:Len_Reg_400} present simulation results with $n=400$. The relative performance of different methods is similar to that for $n=200$.

\begin{table}[htbp]
	\centering
	\caption{Median absolute error ($n=100$).}\label{tab:MAE_100}
	\begin{tabular}{rcrrrrrrr}
		\hline
		&&\multicolumn{7}{c}{$\tau_W$}\\%[-2.5mm]
		Case & Design & 0.01 & 0.05 & 0.10 & 0.15 & 0.2 & 0.3 & 0.5 \\ 
		\hline
		%&\multicolumn{8}{c}{No Regression Adjustment}\\[-2.5mm]
		\multirow{2}{*}{\it NoReg} & CRE & 15.50 & 8.54 & 6.45 & 4.49 & 3.28 & 2.02 & 1.06 \\
		& ReM & 18.60 & 8.08 & 5.33 & 3.49 & 2.47 & 1.60 & 0.74 \\
		\multirow{2}{*}{\it Reg} & CRE & 18.81 & 8.18 & 5.33 & 3.52 & 2.51 & 1.60 & 0.75 \\
		& ReM & 18.45 & 7.99 & 5.25 & 3.43 & 2.44 & 1.58 & 0.72 \\
		\hline
	\end{tabular}
\end{table}

\begin{table}[htbp]
	\centering
	\caption{Proportion of datasets judged to have strong instruments ($n=100$).}\label{tab:SRate_100}
	\begin{tabular}{rcrrrrrrr}
		\hline
		&&\multicolumn{7}{c}{$\tau_W$}\\%[-2.5mm]
		Method & Design & 0.01 & 0.05 & 0.10 & 0.15 & 0.2 & 0.3 & 0.5 \\ 
		\hline
		&&\multicolumn{7}{c}{{\it NoReg}}\\%[-2.5mm]
		\multirow{2}{*}{\it $TS_{\gamma-0.075}$} & CRE & 0.08 & 0.15 & 0.29 & 0.50 & 0.71 & 0.97 & 1.00 \\
		& ReM & 0.07 & 0.16 & 0.35 & 0.61 & 0.86 & 1.00 & 1.00 \\
		\multirow{2}{*}{\it $TS_{\gamma-0.025}$} & CRE & 0.02 & 0.05 & 0.12 & 0.25 & 0.45 & 0.90 & 1.00 \\
		& ReM & 0.03 & 0.07 & 0.19 & 0.41 & 0.70 & 0.98 & 1.00 \\
		\multirow{2}{*}{\it $TS_{F-10}$/$Wald_{F-10}$} & CRE & 0.00 & 0.00 & 0.01 & 0.04 & 0.11 & 0.53 & 1.00 \\
		& ReM & 0.00 & 0.00 & 0.00 & 0.02 & 0.06 & 0.54 & 1.00 \\
		&&\multicolumn{7}{c}{{\it Reg}}\\[-2.5mm]
		\multirow{2}{*}{\it $TS_{\gamma-0.075}$} & CRE & 0.09 & 0.18 & 0.36 & 0.63 & 0.87 & 1.00 & 1.00 \\
		& ReM & 0.08 & 0.18 & 0.37 & 0.64 & 0.87 & 1.00 & 1.00 \\
		\multirow{2}{*}{\it $TS_{\gamma-0.025}$} & CRE & 0.03 & 0.08 & 0.20 & 0.41 & 0.70 & 0.98 & 1.00 \\
		& ReM & 0.03 & 0.08 & 0.20 & 0.42 & 0.71 & 0.98 & 1.00 \\
		\multirow{2}{*}{\it $TS_{F-10}$/$Wald_{F-10}$} & CRE & 0.00 & 0.00 & 0.02 & 0.07 & 0.21 & 0.73 & 1.00 \\
		& ReM & 0.00 & 0.01 & 0.02 & 0.08 & 0.23 & 0.75 & 1.00 \\
		\hline
	\end{tabular}
\end{table}

\begin{table}[htbp]
	\centering
	\caption{Coverage rate of 95\% confidence set without regression adjustment ($n=100$).}\label{tab:CRate_NoReg_100}
	\begin{tabular}{rcrrrrrrr}
		\hline
		&&\multicolumn{7}{c}{$\tau_W$}\\%[-2.5mm]
		Method & Design & 0.01 & 0.05 & 0.10 & 0.15 & 0.2 & 0.3 & 0.5 \\ 
		\hline
		%&\multicolumn{8}{c}{No Regression Adjustment}\\[-2.5mm]
		\multirow{2}{*}{\it Wald} & CRE & -7.81 & 4.98 & 4.66 & 4.00 & 3.24 & 3.54 & 2.28 \\
		& ReM & -19.65 & 4.81 & 4.77 & 4.64 & 4.25 & 2.54 & 2.58 \\
		\multirow{2}{*}{\it FAR} & CRE & 0.08 & 0.42 & 0.38 & 0.41 & 0.62 & 0.72 & 1.78 \\
		& ReM & -1.56 & 2.69 & 3.44 & 3.62 & 3.75 & 3.29 & 4.69 \\
		\multirow{2}{*}{\it $TS_{\gamma-0.075}$} & CRE & -3.24 & 1.14 & 1.47 & 1.70 & 1.61 & 3.38 & 2.28 \\
		& ReM & -5.16 & 2.98 & 3.79 & 4.09 & 4.03 & 2.54 & 2.58 \\
		\multirow{2}{*}{\it $TS_{\gamma-0.025}$} & CRE & -0.71 & 0.66 & 0.82 & 0.79 & 0.89 & 3.04 & 2.28 \\
		& ReM & -2.62 & 2.71 & 3.53 & 3.81 & 3.78 & 2.48 & 2.58 \\
		\multirow{2}{*}{\it $TS_{F-10}$} & CRE & 0.07 & 0.43 & 0.40 & 0.41 & 0.54 & 1.79 & 2.28 \\
		& ReM & -1.56 & 2.69 & 3.42 & 3.60 & 3.64 & 2.11 & 2.58 \\
		\multirow{2}{*}{\it $Wald_{F-10}$} & CRE & -95.00 & 5.00 & -1.25 & -4.59 & -3.42 & 2.45 & 2.28 \\
		& ReM & NA$^1$ & 5.00 & -4.52 & 2.63 & 2.38 & 1.33 & 2.58 \\
		\hline
        \multicolumn{9}{l}{$^1$No dataset is judged to have strong instrument.}
	\end{tabular}
\end{table}
\begin{table}[htbp]
	\centering
	\caption{Coverage rate of 95\% confidence set with regression adjustment ($n=100$).}\label{tab:CRate_Reg_100}
	\begin{tabular}{rcrrrrrrr}
		\hline
		&\multicolumn{8}{c}{$\tau_W$}\\[-2.5mm]
		Method & Design & 0.01 & 0.05 & 0.10 & 0.15 & 0.2 & 0.3 & 0.5 \\ 
		\hline
		%&\multicolumn{8}{c}{No Regression Adjustment}\\[-2.5mm]
		\multirow{2}{*}{\it Wald} & CRE & -18.29 & 4.62 & 4.77 & 4.55 & 4.18 & 2.36 & 2.54 \\
		& ReM & -19.54 & 4.76 & 4.75 & 4.56 & 4.09 & 2.43 & 2.44 \\
		\multirow{2}{*}{\it FAR} & CRE & -1.67 & -0.76 & -0.54 & -0.36 & -0.20 & 0.09 & 2.28 \\
		& ReM & -2.22 & -1.19 & -1.12 & -1.16 & -0.63 & -0.35 & 2.28 \\
		\multirow{2}{*}{\it $TS_{\gamma-0.075}$} & CRE & -6.01 & 0.06 & 1.15 & 2.50 & 3.52 & 2.34 & 2.54 \\
		& ReM & -6.23 & -0.16 & 0.97 & 2.27 & 3.26 & 2.42 & 2.44 \\
		\multirow{2}{*}{\it $TS_{\gamma-0.025}$} & CRE & -2.94 & -0.44 & 0.33 & 1.37 & 2.65 & 2.19 & 2.54 \\
		& ReM & -3.33 & -0.81 & -0.15 & 0.91 & 2.26 & 2.24 & 2.44 \\
		\multirow{2}{*}{\it $TS_{F-10}$} & CRE & -1.67 & -0.76 & -0.44 & -0.13 & 0.27 & 1.07 & 2.54 \\
		& ReM & -2.22 & -1.18 & -1.07 & -0.92 & -0.12 & 0.93 & 2.44 \\
		\multirow{2}{*}{\it $Wald_{F-10}$} & CRE & -95.00 & -4.52 & 2.22 & 1.79 & 2.63 & 1.55 & 2.54 \\
		& ReM & -95.00 & 0.16 & 1.44 & 2.50 & 2.50 & 1.82 & 2.44 \\
		\hline
	\end{tabular}
\end{table}

\begin{table}[htbp]
	\centering
	\caption{Median length of 95\% confidence sets without regression adjustment ($n=100$). If the coverage rate is smaller than 85\%, we report ``/"; if the coverage rate is larger than 85\% but smaller than 90\%, we mark the median length with underlining.}\label{tab:Len_NoReg_100}
	\begin{tabular}{rcrrrrrrr}
		\hline
		&\multicolumn{8}{c}{$\tau_W$}\\[-2.5mm]
		Method & Design & 0.01 & 0.05 & 0.10 & 0.15 & 0.2 & 0.3 & 0.5 \\ 
		\hline
		%&\multicolumn{8}{c}{No Regression Adjustment}\\[-2.5mm]
		\multirow{2}{*}{{\it Wald}} & CRE & \underline{80.62} & 67.76 & 45.68 & 28.57 & 20.06 & 12.46 & 6.68 \\
		& ReM & / & 60.47 & 35.20 & 20.75 & 14.75 & 9.43 & 4.83 \\
		\multirow{2}{*}{{\it FAR}} & CRE & $\infty$ & $\infty$ & $\infty$ & $\infty$ & 143.42 & 16.34 & 7.04 \\
		& ReM & $\infty$ & $\infty$ & $\infty$ & $\infty$ & 68.18 & 14.68 & 6.68 \\
		\multirow{2}{*}{{\it $TS_{\gamma-0.075}$}} & CRE & $\infty$ & $\infty$ & $\infty$ & $\infty$ & 20.06 & 12.46 & 6.68 \\
		& ReM & $\infty$ & $\infty$ & $\infty$ & 20.75 & 14.75 & 9.43 & 4.83 \\
		\multirow{2}{*}{{\it $TS_{\gamma-0.025}$}} & CRE & $\infty$ & $\infty$ & $\infty$ & $\infty$ & 143.02 & 12.46 & 6.68 \\
		& ReM & $\infty$ & $\infty$ & $\infty$ & $\infty$ & 14.75 & 9.43 & 4.83 \\
		\multirow{2}{*}{{\it $TS_{F-10}$}} & CRE & $\infty$ & $\infty$ & $\infty$ & $\infty$ & 143.42 & 12.84 & 6.68 \\
		& ReM & $\infty$ & $\infty$ & $\infty$ & $\infty$ & 68.18 & 9.67 & 4.83 \\
		\multirow{2}{*}{{\it $Wald_{F-10}$}} & CRE & / & 11.31 & 11.41 & 11.13 & 10.75 & 10.43 & 6.68 \\
		& ReM & NA$^1$ & 8.69 & 8.80 & 8.60 & 8.40 & 8.19 & 4.83 \\
		\hline
        \multicolumn{9}{l}{$^1$No dataset is judged to have strong instrument.}
	\end{tabular}
\end{table}

\begin{table}[htbp]
	\centering
	\caption{Median length of 95\% confidence sets with regression adjustment ($n=100$). If the coverage rate is smaller than 85\%, we report ``/"; if the coverage rate is larger than 85\% but smaller than 90\%, we mark the median length with underlining.}\label{tab:Len_Reg_100}
	\begin{tabular}{rcrrrrrrr}
		\hline
		&\multicolumn{8}{c}{$\tau_W$}\\[-2.5mm]
		Method & Design & 0.01 & 0.05 & 0.10 & 0.15 & 0.2 & 0.3 & 0.5 \\ 
		\hline
		%&\multicolumn{8}{c}{No Regression Adjustment}\\[-2.5mm]
		\multirow{2}{*}{{\it Wald}} & CRE & / & 62.62 & 35.04 & 21.12 & 14.90 & 9.48 & 4.90 \\
		& ReM & / & 59.36 & 34.44 & 20.57 & 14.55 & 9.31 & 4.77 \\
		\multirow{2}{*}{{\it FAR}} & CRE & $\infty$ & $\infty$ & $\infty$ & $\infty$ & 24.51 & 11.13 & 5.07 \\
		& ReM & $\infty$ & $\infty$ & $\infty$ & $\infty$ & 23.06 & 10.84 & 4.93 \\
		\multirow{2}{*}{{\it $TS_{\gamma-0.075}$}} & CRE & \underline{$\infty$} & $\infty$ & $\infty$ & 21.12 & 14.90 & 9.48 & 4.90 \\
		& ReM & \underline{$\infty$} & $\infty$ & $\infty$ & 20.57 & 14.55 & 9.31 & 4.77 \\
		\multirow{2}{*}{{\it $TS_{\gamma-0.025}$}} & CRE & $\infty$ & $\infty$ & $\infty$ & $\infty$ & 14.90 & 9.48 & 4.90 \\
		& ReM & $\infty$ & $\infty$ & $\infty$ & $\infty$ & 14.55 & 9.31 & 4.77 \\
		\multirow{2}{*}{{\it $TS_{F-10}$}} & CRE & $\infty$ & $\infty$ & $\infty$ & $\infty$ & 24.51 & 9.48 & 4.90 \\
		& ReM & $\infty$ & $\infty$ & $\infty$ & $\infty$ & 23.06 & 9.31 & 4.77 \\
		\multirow{2}{*}{{\it $Wald_{F-10}$}} & CRE & / & 10.61 & 10.44 & 10.20 & 9.91 & 8.72 & 4.90 \\
		& ReM & / & 10.68 & 10.20 & 10.23 & 9.93 & 8.64 & 4.77 \\
		\hline
        \multicolumn{9}{l}{$^1$ No dataset is judged to have strong instrument.}
	\end{tabular}
\end{table}

\begin{table}[htbp]
	\centering
	\caption{Median absolute error ($n=400$).}\label{tab:MAE_400}
	\begin{tabular}{rcrrrrrrr}
		\hline
		&&\multicolumn{7}{c}{$\tau_W$}\\%[-2.5mm]
		Case & Design & 0.005 & 0.05 & 0.10 & 0.15 & 0.2 & 0.3 & 0.5 \\ 
		\hline
		%&\multicolumn{8}{c}{No Regression Adjustment}\\[-2.5mm]
		\multirow{2}{*}{\it NoReg} & CRE & 24.27 & 7.21 & 3.58 & 2.19 & 1.60 & 1.02 & 0.54 \\
		& ReM & 20.40 & 5.30 & 2.48 & 1.54 & 1.12 & 0.72 & 0.36 \\
		\multirow{2}{*}{\it Reg} & CRE & 20.08 & 5.11 & 2.46 & 1.52 & 1.12 & 0.71 & 0.35 \\
		& ReM & 20.23 & 5.25 & 2.46 & 1.50 & 1.11 & 0.70 & 0.36 \\
		\hline
	\end{tabular}
\end{table}

\begin{table}[htbp]
	\centering
	\caption{Proportion of datasets judged to have strong instruments ($n=400$).}\label{tab:SRate_400}
	\begin{tabular}{rcrrrrrrr}
		\hline
		&&\multicolumn{7}{c}{$\tau_W$}\\%[-2.5mm]
		Method & Design & 0.005 & 0.05 & 0.10 & 0.15 & 0.2 & 0.3 & 0.5 \\ 
		\hline
		&&\multicolumn{7}{c}{{\it NoReg}}\\%[-2.5mm]
		\multirow{2}{*}{\it $TS_{\gamma-0.075}$} & CRE & 0.06 & 0.24 & 0.65 & 0.93 & 0.99 & 1.00 & 1.00 \\
		& ReM & 0.06 & 0.33 & 0.80 & 0.99 & 1.00 & 1.00 & 1.00 \\
		\multirow{2}{*}{\it $TS_{\gamma-0.025}$} & CRE & 0.02 & 0.11 & 0.43 & 0.82 & 0.98 & 1.00 & 1.00 \\
		& ReM & 0.02 & 0.17 & 0.62 & 0.95 & 1.00 & 1.00 & 1.00 \\
		\multirow{2}{*}{\it $TS_{F-10}$/$Wald_{F-10}$} & CRE & 0.00 & 0.02 & 0.11 & 0.43 & 0.83 & 1.00 & 1.00 \\
		& ReM & 0.00 & 0.00 & 0.06 & 0.41 & 0.89 & 1.00 & 1.00 \\
		&&\multicolumn{7}{c}{{\it Reg}}\\[-2.5mm]
		\multirow{2}{*}{\it $TS_{\gamma-0.075}$} & CRE & 0.06 & 0.33 & 0.80 & 0.99 & 1.00 & 1.00 & 1.00 \\
		& ReM & 0.06 & 0.33 & 0.80 & 0.99 & 1.00 & 1.00 & 1.00 \\
		\multirow{2}{*}{\it $TS_{\gamma-0.025}$} & CRE & 0.02 & 0.16 & 0.61 & 0.95 & 1.00 & 1.00 & 1.00 \\
		& ReM & 0.02 & 0.17 & 0.62 & 0.95 & 1.00 & 1.00 & 1.00 \\
		\multirow{2}{*}{\it $TS_{F-10}$/$Wald_{F-10}$} & CRE & 0.00 & 0.03 & 0.22 & 0.71 & 0.98 & 1.00 & 1.00 \\
		& ReM & 0.00 & 0.03 & 0.23 & 0.71 & 0.98 & 1.00 & 1.00 \\
		\hline
	\end{tabular}
\end{table}

\begin{table}[htbp]
	\centering
	\caption{Coverage rate of 95\% confidence set without regression adjustment ($n=400$).}\label{tab:CRate_NoReg_400}
	\begin{tabular}{rcrrrrrrr}
		\hline
		&\multicolumn{8}{c}{$\tau_W$}\\[-2.5mm]
		Method & Design & 0.005 & 0.05 & 0.10 & 0.15 & 0.2 & 0.3 & 0.5 \\ 
		\hline
		%&\multicolumn{8}{c}{No Regression Adjustment}\\[-2.5mm]
		\multirow{2}{*}{\it Wald} & CRE & -25.71 & 1.08 & 0.98 & 1.46 & 1.52 & 1.67 & 1.95 \\
		& ReM & -19.93 & 4.28 & 3.92 & 3.43 & 2.58 & 2.22 & 3.32 \\
		\multirow{2}{*}{\it FAR} & CRE & -0.24 & -0.26 & 0.06 & 0.45 & 0.44 & 0.93 & 1.87 \\
		& ReM & 4.61 & 4.71 & 4.60 & 4.60 & 4.58 & 4.60 & 4.86 \\
		\multirow{2}{*}{\it $TS_{\gamma-0.075}$} & CRE & -3.31 & -1.72 & -1.26 & 0.47 & 1.35 & 1.67 & 1.95 \\
		& ReM & -0.61 & 4.15 & 3.87 & 3.43 & 2.58 & 2.22 & 3.32 \\
		\multirow{2}{*}{\it $TS_{\gamma-0.025}$} & CRE & -0.74 & -1.23 & -1.44 & -0.03 & 1.07 & 1.67 & 1.95 \\
		& ReM & 2.93 & 4.12 & 3.75 & 3.42 & 2.58 & 2.22 & 3.32 \\
		\multirow{2}{*}{\it $TS_{F-10}$} & CRE & -0.24 & -0.40 & -0.62 & -0.58 & 0.13 & 1.67 & 1.95 \\
		& ReM & 4.61 & 4.68 & 4.41 & 3.77 & 2.65 & 2.22 & 3.32 \\
		\multirow{2}{*}{\it $Wald_{F-10}$} & CRE & -95.00 & -39.81 & -13.87 & -2.17 & 0.84 & 1.67 & 1.95 \\
		& ReM & NA$^1$ & -7.50 & 0.93 & 2.30 & 2.42 & 2.20 & 3.32 \\
		\hline
        \multicolumn{9}{l}{$^1$ No dataset is judged to have strong instrument.}
	\end{tabular}
\end{table}
\begin{table}[htbp]
	\centering
	\caption{Coverage rate of 95\% confidence set with regression adjustment ($n=400$).}\label{tab:CRate_Reg_400}
	\begin{tabular}{rcrrrrrrr}
		\hline
		&\multicolumn{8}{c}{$\tau_W$}\\[-2.5mm]
		Method & Design & 0.005 & 0.05 & 0.10 & 0.15 & 0.2 & 0.3 & 0.5 \\ 
		\hline
		%&\multicolumn{8}{c}{No Regression Adjustment}\\[-2.5mm]
		\multirow{2}{*}{\it Wald} & CRE & -19.48 & 4.53 & 4.07 & 3.61 & 2.79 & 2.37 & 3.41 \\
		& ReM & -19.94 & 4.33 & 4.01 & 3.47 & 2.70 & 2.37 & 3.45 \\
		\multirow{2}{*}{\it FAR} & CRE & -0.16 & 0.21 & 0.62 & 1.09 & 1.31 & 1.81 & 3.28 \\
		& ReM & -0.36 & 0.12 & 0.73 & 1.05 & 1.21 & 1.71 & 3.34 \\
		\multirow{2}{*}{\it $TS_{\gamma-0.075}$} & CRE & -3.40 & 1.38 & 3.08 & 3.56 & 2.79 & 2.37 & 3.41 \\
		& ReM & -3.55 & 1.42 & 3.10 & 3.46 & 2.70 & 2.37 & 3.45 \\
		\multirow{2}{*}{\it $TS_{\gamma-0.025}$} & CRE & -0.80 & 0.63 & 2.31 & 3.40 & 2.79 & 2.37 & 3.41 \\
		& ReM & -1.03 & 0.67 & 2.44 & 3.28 & 2.70 & 2.37 & 3.45 \\
		\multirow{2}{*}{\it $TS_{F-10}$} & CRE & -0.17 & 0.28 & 1.01 & 2.53 & 2.74 & 2.37 & 3.41 \\
		& ReM & -0.37 & 0.15 & 1.17 & 2.54 & 2.64 & 2.37 & 3.45 \\
		\multirow{2}{*}{\it $Wald_{F-10}$} & CRE & -95.00 & -1.64 & 2.19 & 3.22 & 2.74 & 2.37 & 3.41 \\
		& ReM & -95.00 & -5.42 & 1.97 & 3.01 & 2.66 & 2.37 & 3.45 \\
		\hline
	\end{tabular}
\end{table}

\begin{table}[htbp]
	\centering
	\caption{Median length of 95\% confidence sets without regression adjustment ($n=400$). If the coverage rate is smaller than 85\%, we report ``/"; if the coverage rate is larger than 85\% but smaller than 90\%, we mark the median length with underlining.}\label{tab:Len_NoReg_400}
	\begin{tabular}{rcrrrrrrr}
		\hline
		&\multicolumn{8}{c}{$\tau_W$}\\[-2.5mm]
		Method & Design & 0.005 & 0.05 & 0.10 & 0.15 & 0.2 & 0.3 & 0.5 \\ 
		\hline
		%&\multicolumn{8}{c}{No Regression Adjustment}\\[-2.5mm]
		\multirow{2}{*}{{\it Wald}} & CRE & / & 46.78 & 21.76 & 13.20 & 9.70 & 6.20 & 3.44 \\
		& ReM & / & 34.53 & 15.08 & 9.47 & 6.98 & 4.53 & 2.49 \\
		\multirow{2}{*}{{\it FAR}} & CRE & $\infty$ & $\infty$ & 308.96 & 18.65 & 11.39 & 6.57 & 3.49 \\
		& ReM & $\infty$ & $\infty$ & 233.39 & 18.21 & 11.18 & 6.49 & 3.44 \\
		\multirow{2}{*}{{\it $TS_{\gamma-0.075}$}} & CRE & $\infty$ & $\infty$ & 21.77 & 13.20 & 9.70 & 6.20 & 3.44 \\
		& ReM & $\infty$ & $\infty$ & 15.08 & 9.47 & 6.98 & 4.53 & 2.49 \\
		\multirow{2}{*}{{\it $TS_{\gamma-0.025}$}} & CRE & $\infty$ & $\infty$ & 308.96 & 13.20 & 9.70 & 6.20 & 3.44 \\
		& ReM & $\infty$ & $\infty$ & 15.08 & 9.47 & 6.98 & 4.53 & 2.49 \\
		\multirow{2}{*}{{\it $TS_{F-10}$}} & CRE & $\infty$ & $\infty$ & 308.96 & 18.12 & 9.70 & 6.20 & 3.44 \\
		& ReM & $\infty$ & $\infty$ & 233.39 & 17.83 & 6.98 & 4.53 & 2.49 \\
		\multirow{2}{*}{{\it $Wald_{F-10}$}} & CRE & / & / & / & 10.15 & 9.14 & 6.20 & 3.44 \\
		& ReM & NA$^1$ & \underline{8.69} & 8.41 & 7.80 & 6.81 & 4.52 & 2.49 \\
		\hline
        \multicolumn{9}{l}{$^1$ No dataset is judged to have strong instrument.}
	\end{tabular}
\end{table}
\begin{table}[htbp]
	\centering
	\caption{Median length of 95\% confidence sets with regression adjustment ($n=400$). If the coverage rate is smaller than 85\%, we report ``/"; if the coverage rate is larger than 85\% but smaller than 90\%, we mark the median length with underlining.}\label{tab:Len_Reg_400}
	\begin{tabular}{rcrrrrrrr}
		\hline
		&\multicolumn{8}{c}{$\tau_W$}\\[-2.5mm]
		Method & Design & 0.005 & 0.05 & 0.10 & 0.15 & 0.2 & 0.3 & 0.5 \\ 
		\hline
		%&\multicolumn{8}{c}{No Regression Adjustment}\\[-2.5mm]
		\multirow{2}{*}{{\it Wald}} & CRE & / & 34.36 & 15.00 & 9.41 & 6.96 & 4.52 & 2.48 \\
		& ReM & / & 34.36 & 15.01 & 9.41 & 6.95 & 4.50 & 2.47 \\
		\multirow{2}{*}{{\it FAR}} & CRE & $\infty$ & $\infty$ & 26.62 & 11.19 & 7.58 & 4.67 & 2.50 \\
		& ReM & $\infty$ & $\infty$ & 26.50 & 11.17 & 7.55 & 4.65 & 2.49 \\
		\multirow{2}{*}{{\it $TS_{\gamma-0.075}$}} & CRE & $\infty$ & $\infty$ & 15.00 & 9.41 & 6.96 & 4.52 & 2.48 \\
		& ReM & $\infty$ & $\infty$ & 15.01 & 9.41 & 6.95 & 4.50 & 2.47 \\
		\multirow{2}{*}{{\it $TS_{\gamma-0.025}$}} & CRE & $\infty$ & $\infty$ & 15.00 & 9.41 & 6.96 & 4.52 & 2.48 \\
		& ReM & $\infty$ & $\infty$ & 15.01 & 9.41 & 6.95 & 4.50 & 2.47 \\
		\multirow{2}{*}{{\it $TS_{F-10}$}} & CRE & $\infty$ & $\infty$ & 26.62 & 9.41 & 6.96 & 4.52 & 2.48 \\
		& ReM & $\infty$ & $\infty$ & 26.50 & 9.41 & 6.95 & 4.50 & 2.47 \\
		\multirow{2}{*}{{\it $Wald_{F-10}$}} & CRE & / & 10.64 & 9.87 & 8.61 & 6.94 & 4.52 & 2.48 \\
		& ReM & / & \underline{10.76} & 9.91 & 8.58 & 6.92 & 4.50 & 2.47 \\
		\hline
	\end{tabular}
\end{table}

We now compare the simulation results across three values of $n$. In terms of the median absolute error, when $\tau_W\geq 0.05$, the larger $n$ is, the smaller the mean absolute error is. In terms of the proportion of datasets judged to have strong instruments, generally the larger $n$ is, the larger the proportion is. 

In terms of coverage rate, we consider three ranges: severe undercoverage (less than 85\%), moderate undercoverage (larger than 85\% but smaller than 90\%), slight or no undercoverage (larger than 90\%). When $\tau_W=0.005$ or $\tau_W=0.01$, {\it Wald} has severe undercoverage across all cases except for the case with $n=100$, CRE and no regression adjustment, in which {\it Wald} has moderate undercoverage; {\it $Wald_{F-10}$} has extremely severe undercoverage or does not yield result for any dataset. When $n=100$ and $\tau_W=0.01$, {\it $TS_{\gamma-0.075}$} can have moderate undercoverage. In the case with $n=400$, ReM and $\tau_W=0.05$, {\it $Wald_{F-10}$} has moderate coverage.

In terms of median interval length, for all methods other than {\it $Wald_{F-10}$}, generally the smaller $n$ is, the shorter median interval length is. The median interval length of {\it $Wald_{F-10}$} does not exhibit a consistent pattern when $n$ increases. This is because when $n$ increases, median interval length is affected by both an increasing number of datasets judged to have strong instruments, and increasing precision due to larger sample size.

\end{document}